\documentclass[rmp,aps]{revtex4}
\newcommand{\beq}{\begin{equation}}
\newcommand{\eeq}{\end{equation}}
\newcommand{\eq}{\begin{equation}}
\newcommand{\ee}{\end{equation}}
\newcommand{\s}{{\sigma}}

\newcommand{\etab}{\mbox{\boldmath $\eta $}}
\newcommand{\Pib}{\mbox{\boldmath $\Pi $}}

\newcommand{\sig}{\sigma_{xy}}

\newcommand{\p}{{\partial}}

\newcommand{\barro}{\bar{\bar{\rho}}({\bf q})}
\newcommand{\barrop}{\bar{\bar{\rho}}^p({\bf q})}
\newcommand{\bachi}{\bar{\bar{ \chi}}}
\newcommand{\baro}{\bar{\bar{\rho}}}
\newcommand{\barH}{\bar{\bar{H}}}
\newcommand{\bareS}{\bar{\bar{S}}}
\newcommand{\barchi}{\bar{\bar{ \chi}}({\bf q})}
\newcommand{\n}{{p \over 2ps+1}}
\def\bA{{\mathbf A}}

\def\bB{{\mathbf B}}
\def\bp{{\mathbf p}}
\def\be{{\mathbf e}}
\def\bC{{\mathbf C
}}
\def\bE{{\mathbf E}}
\def\bn{{\mathbf n}}
\def\bG{{\mathbf G}}
\def\bg{{\mathbf g}}
\def\bq{{\mathbf q}}
\def\br{{\mathbf r}}
\def\bs{{\mathbf s}}

\def\bQ{{\mathbf Q}}
\def\bs{{\mathbf s}}
\def\bB{{\mathbf B}}
\def\bH{{\mathbf H}}
\def\bPi{{\mathbf \Pi}}
\def\bJ{{\mathbf J}}
\def\bj{{\mathbf j}}
\def\bR{{\mathbf R}}
\def\bz{{\mathbf z}}
\def\ba{{\mathbf a}}
\def\bk{{\mathbf k}}
\def\bP{{\mathbf P}}
\def\bg{{\mathbf g}}

\def\cH{{\cal H}}

\def\ddt{{\partial\over \partial t}}

\def\rhot{{\tilde{\rho}}}

\def\a{\alpha}
\def\b{\beta}
\def\e{\epsilon}

\def\o{{\omega}}

\def\dd{d^{\dagger}}

\def\half{{1\over2}}
\def\third{{1\over3}}
\def\twof{{2\over5}}

\def\ints{{\int {d^2s\over(2\pi)^2}}}

\def\trho{\tilde{\rho}}
\def\tchi{\tilde{\chi}}
\def\hz{{\hat \bz}}
\def\ua{\uparrow}
\def\da{\downarrow}
\def\eqa{\begin{eqnarray}}
\def\eea{\end{eqnarray}}
\def\pr{{Phys. Rev.}}
\def\prl{{Phys. Rev. Lett.}}
\def\prb{{Phys. Rev. {\bf B}}}
\def\jpc{{Jour. Phys. {\bf C}}}

\usepackage{graphicx}
\usepackage{dcolumn}
\usepackage{bm}

\begin{document}


\title{Hamiltonian Theories of the FQHE}

\author{Ganpathy Murthy}
 \email{murthy@pa.uky.edu}
\affiliation{ Physics Department\\ University of Kentucky,
Lexington  KY 40506}

\author{R.Shankar}
\email{r.shankar@yale.edu}
 \homepage{http://www.yale.edu/~r.shankar}
\affiliation{ Sloane Physics Lab\\ Yale University  New Haven CT
06520 }

\date{\today}

\begin{abstract}
This paper reviews progress on the Fractional Quantum Hall Effect
(FQHE) based on  what we term hamiltonian theories, i.e., theories
that proceed from the microscopic electronic hamiltonian to the
final solution via a sequence of transformations and
approximations, either in the hamiltonian or path integral
approach, as compared to theories  based on  exact diagonalization
or trial wavefunctions. We focus on the Chern-Simons (CS) approach
in which electrons are converted to CS fermions or bosons that
carry along flux tubes and our Extended Hamiltonian Theory (EHT)
in which electrons are paired with pseudo-vortices to form
composite fermions (CF) whose properties are a lot closer to the
ultimate low-energy quasiparticles. We address a variety of
qualitative and quantitative questions: In what sense do electrons
really bind to vortices? What is the internal structure of the
Composite Fermion and what does it mean? What exactly is the
dipole picture? What degree of freedom carries the Hall current
when the quasiparticles are localized or neutral or both? How
exactly is the kinetic energy quenched in the lowest Landau level
and resurrected by interactions? How well does the CF picture work
at and near $\nu =1/2$? Is the system compressible at $\nu =1/2$?
If so, how can Composite Fermions be dipolar at $\nu=1/2$ and
still be compressible?  How is compressibility demonstrated
experimentally? How does the charge of the excitation get
renormalized from that of the electron to that of the CF in an
operator treatment?  Why do Composite Fermions sometimes appear to
be free when they are not? How are (approximate) transport gaps,
zero-temperature magnetic transitions, the temperature dependent
polarizations of gapped and gapless states, the NMR relaxation
rate $1/T_1$ in gapless states, and gaps in inhomogeneous states
computed? It is seen that though the CS and EHT  approaches agree
whenever a comparison is possible, results that are transparent in
one approach are typically opaque in the other, making them truly
complementary.

\end{abstract}

\maketitle
\tableofcontents
\section{\label{intro}\ \ Introduction}

Twenty years ago, von Klitzing, Dorda, and Pepper (1980) made the
first discovery in what has proved to be a vital and exciting
subfield of condensed matter physics to the present day, that of
the quantum Hall effects (QHE). Their discovery that the Hall
conductance $\s_{xy}$ of a two-dimensional electron gas is
quantized at integer multiples of $e^2/2\pi \hbar $, the quantum
unit of conductance, is known as the integer quantum Hall effect
(IQHE). Soon after, Tsui, St\"ormer, and Gossard (1982) discovered
the even more puzzling    fractional quantum Hall effect (FQHE).
Since the first observation of the fraction ${1\over 3}$, many
more have been seen (see, for example, St\"ormer, Tsui, and
Gossard 1999).

This review focuses on what we term the {\em hamiltonian theories}
 of the FQHE, by which we mean theoretical
approaches that begin with the microscopic hamiltonian for
interacting electrons and try to obtain a satisfactory description
of the underlying physics through a sequence of transformations
and approximations in the operator or path integral formalism. We
bother to give a special name to what appears to be business as
usual, because an alternate approach, pioneered by Laughlin
(1983a,b) and based on writing trial wavefunctions, has proven so
extraordinarily successful. While we will of course discuss the
wavefunction approach here because the hamiltonian approach is
inspired by it, the discussion (of this and any other topic) will
be aimed at serving our primary goal: Providing a cogent and
critical description of the hamiltonian approach in one place,
including all the hindsight and insight the intervening years have
provided. Of necessity, many topics will have to be omitted or
treated summarily. Fortunately many excellent reviews
exist\footnote{The earliest comprehensive introduction to the full
range of the quantum Hall effects is Prange and Girvin (1987).
Huckestein (1995) concentrates on the integer quantum Hall effect.
A more recent comprehensive review is Das Sarma and Pinczuk
(1997), while a more focused treatment of the enormous body of
research on Composite Fermions is summarized in Heinonen (1998).}
and the reader is directed to them.

The hamiltonian theories described here address both qualitative
and quantitative issues.  They give a concrete operator
realization of many heuristic pictures that have been espoused and
make precise under what conditions and in what sense these
pictures are valid. They  allow one  to compute (within reasonable
approximations)  a large number of quantities at zero and nonzero
temperatures at equal and unequal times. They open the door for a
treatment of disorder.

The hamiltonian theories themselves fall into two categories. The
first, which we call the Chern-Simons (CS) approach, consists of
making a singular gauge transformation on the electronic wavefunction
that leads to a composite particle which is the union of an electron
and some number of point flux tubes. The composite particles have a
nondegenerate ground state at mean-field level.  They are coupled to a
gauge field that is fully determined by the particle coordinates. This
formalism is best suited for computation of a variety of response
functions. It is particularly effective at $\nu =1/2$, where adherence
to gauge invariance is of paramount importance if certain low-energy
phenomena pertaining to the overdamped mode, coupling to surface
acoustic waves, or the compressibility are to be properly described.
On the other hand, the composite particles do not exhibit in any
transparent way the quasiparticle properties (such as charge $e^*$ or
effective mass $m^*$) deduced from trial wavefunctions, though in
principle they would surface after considerable work. It is also very
hard to obtain a smooth limit as the electron mass $m\to 0$ in this
approach. The Extended Hamiltonian Theory (EHT), which we developed
over the years addresses some of these issues.\footnote{ We used to
refer to this as THE hamiltonian formalism, but were persuaded to
consider a name change that better reflected the state of affairs. Our
compliance could not have been more total- the new acronym is the
exact reverse of the old one.} By the adjective {\em extended} we
signify both that our work is an extension of older CS work and that
our hamiltonian is defined in an extended or enlarged Hilbert space
with additional degrees of freedom. The enlarged space allows us to
introduce a quasiparticle that is a much better approximation to what
we expect, to easily disentangle low-energy (Lowest Landau Level) and
high-energy physics, to compute a variety of gaps, finite temperature
properties, to describe inhomogeneous states and so on. On the other
hand, the proper treatment of the additional degrees of freedom is
very difficult to ensure in the computation of certain very low-energy
quantities. Both the nonzero compressibility and an overdamped mode at
$\nu =1/2$ (and its experimental consequences) are all but invisible
in this approach, though they can be extracted with some effort. No
contradictions exist between the predictions of the CS and EHT
approaches, which by and large tend to make predictions in
complementary regions. Their predictions can be shown to agree in
overlapping regions, but only with some effort.

\subsection{The experiment}
Figure 1 shows in schematic form the experiment in a rectangular
geometry, with the current $j_x$, magnetic field $B_z$ and
electric field $E_y$ in mutually perpendicular directions.By
definition \beq j_x=\s_{xy}E_y. \eeq If we multiply both sides by
the width of the sample we obtain \beq I_x=\s_{xy}V_y \eeq as the
relation between current and voltage, reminding us that in this
$d=2$ problem the Hall conductance and conductivity are the
same.\footnote{The longitudinal conductance and conductivity are
related by the (aspect) ratio of length to width.}

\begin{figure}
\includegraphics[width=3in]{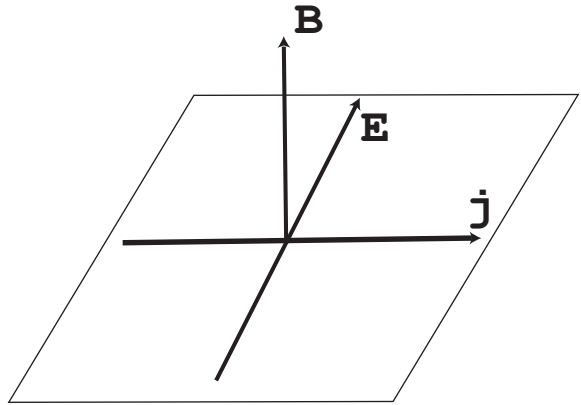}
\caption{\label{sigmaxyz} The Hall experiment in a rectangular
geometry.}
\end{figure}
Let us ask what one expects for $\s_{xy}$ based on the simplest
ideas, so as to place the experimental discoveries in perspective.
Let us ignore interactions and disorder. Then we can assert that
in the steady state, the electric and magnetic forces balance:
\beq eE=evB\eeq where $e$ is the charge of the electrons, and $v$
is their velocity. (Vector indices are omitted when obvious.)
Therefore \beq j=nev = {neE\over B}\eeq where $n$ is the number
density. Thus  \beq \s_{xy}={ne\over B}.\label{hall} \eeq This
result is unaffected by interactions and relies only on Galilean
invariance (relativity). To see this, let us perform a boost to a
frame in which the electric field (to leading order in $v/c$) \beq
{\bf E}'= {\bf E}-{\bf v \times B} \eeq
 vanishes, that is, to a frame with $v=E/B$. In this frame $j=0$.
 Boosting back to the lab frame we obtain $j=neE/B$ and regain
 Eqn.(\ref{hall}).

 Figure 2 is a caricature of what is  measured at $T=0$ in
 the thermodynamic limit. The key feature is
 that $\s_{xy}$, instead of varying linearly with $ne/B$, is
 quantized at steps or plateaus. This article  focuses on steps given by

 \beq \s_{xy}= {e^2\over 2\pi \hbar}\ \nu ,\eeq where \beq \nu = {p \over
2ps+1} \ \ \ \ p=1,2,  \  s=0, 1,2,.. \label{fqhe} \eeq The IQHE
corresponds to $s=0, \ \nu =p$, while FQHE refers to $s>0$.
\footnote{Fractions with denominator $2ps-1$ require only minor
modifications  and are left as an exercise.} We explicitly display
the $\hbar$ which will soon be set equal to unity.

On each step  the longitudinal conductances vanish. Thus the
conductivity tensor assumes the form
  \beq \sigma_{ij}={e^2\over 2\pi \hbar} \left(
  \begin{array}{cc}
  0 & \nu \\
  -\nu & 0
  \end{array}\right).
  \eeq
Consequently the diagonal part of the resistivity tensor is also
zero  and the transport is dissipationless. In the
 real world, due to nonzero  temperature $T$ and finite sample size,
  the transitions between steps acquire a finite width, with  $\s_{xx}>0$
 therein.

\begin{figure}
\includegraphics[width=3in]{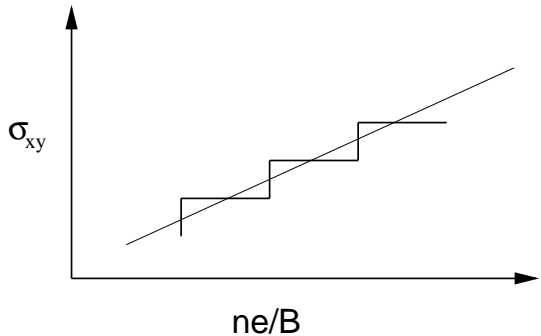}
\caption{\label{sigmaxyplot} Schematic form of measured Hall
conductance at $T=0$. The straight line is the  Galilean invariant
result, while the steps describe the data.}
\end{figure}

Why is the quantization of conductance so surprising, given that
 we have seen so many instances of quantization of observables before?
  The answer is
 that the observable in this case does not refer to an atom or
 molecule, but to a {\em macroscopic} sample, with sample specific
 disorder. The role of disorder is truly paradoxical in the QHE:
 on the one hand, without disorder, we cannot escape the Galilean
 invariant relation  $\s_{xy}=ne/B$; on
 the other hand, despite disorder (which surely varies from sample
 to sample),  the conductance in any step is constant  to better
 than a few  parts in $10^{10}$! A fairly detailed explanation of this  now
 exists and will be described soon. An integral part  of this resolution
 is also why $\s_{xx}=0$, for only then
can we  understand why
 $\sigma_{xy}$ is so well quantized even though  the leads used to
 measure $V_y$ may not be perfectly aligned--  there will be no
 unwanted contributions from the longitudinal voltage drop.

\subsection{What is special about the steps?}
In searching for the physics underlying any one step or plateau,
it makes
 sense to begin with the points where the straight line (the Galilean
 invariant result) intercepts the steps. Not only are these points
 singled out by experiment, but they also have a Hall conductance that
 we could hope to understand without including disorder.  \footnote{In
 the absence of disorder any state would have the desired Hall
 conductance. We are looking for a nontrivial correlated state that is
 robust under the inclusion of disorder and capable of dominating the
 plateau it belongs to.}

 At these points we
 have
 \beq
 {ne\over B} ={e^2\over 2\pi \hbar}{p\over 2ps+1}
 \eeq
 which we rewrite as
 \beq
 {{B\over n (2\pi \hbar /e)}}=2s+ {1\over p}.\label{D}
 \eeq
Thus the ratio of $B$, the flux density, and  $n $,  the particle
density,    has some special rational values at these points. For
example in the case of the fraction $1/3$ there are three quanta
of flux per electron.

To understand what is so special about these values, recall the
following textbook results (Shankar 1994) about a single particle
 of mass $m$ and charge $e$ moving in two dimensions in a perpendicular magnetic
field.
\begin{itemize}
\item The energy is quantized into Landau Levels (LL's) located at
$E= (e\hbar B/m) (n+{1\over 2})\equiv \hbar \omega_0 (n+{1\over
2})$, where $\omega_0$ is called the {\em cyclotron frequency}.
\item Each LL has a degeneracy equal to $\Phi /\Phi_0$, the flux in units of the
flux quantum $\Phi_0=2\pi \hbar / e$, or  :\beq \mbox{Degeneracy
per unit area of each LL}={{B\over (2\pi \hbar
/e)}}.\label{deg}\eeq
\item The wavefunctions of the Lowest Landau Level (LLL) are, in
the symmetric gauge, \beq \psi_{LLL}= z^m e^{-|z|^2/4l^2}\ \ \
z=x+iy\ \ \ \  m=0,1,... \eeq where $l=\sqrt{\hbar /eB}$ is the
magnetic length. We will often drop the gaussian factor in LLL
wavefunctions. Higher LL wavefunctions depend on both $z$ and
$\bar{z}$.
\end{itemize}

Thus  Eqns. (\ref{D}) and Eqn. (\ref{deg} )imply that at the
special points \beq \frac{\mbox{ Number of  states in a LL}
}{\mbox {Number of particles}} = \frac{\mbox{ Flux
 quanta }}{\mbox{  particles}} = {{B\over n (2\pi \hbar /e)}}=2s+{1\over p} =\nu^{-1}.\eeq
 Note that $\nu$, which stood for the dimensionless conductance,
  is thus the number
 of occupied LL's. It is  called the
 {\em filling factor}.  In the FQHE $\nu $ is
 a fraction,  restricted in this article   to be $<\half$ unless otherwise mentioned.

   If  $\nu
<1$, there are
 more LLL states than
particles,  and in the noninteracting limit  they can all be fit
into the LLL with room to spare. For example at $\nu ={1\over 3}$,
there are 3 LLL states per electron. {\em This macroscopic
degeneracy in the noninteracting ground state means  we cannot
even get started with a perturbative treatment of disorder and
interactions.}
 This is the central problem. While this
problem exists for any $\nu$ that is not an integer, experiments
suggest that  $\nu^{-1}=2s+{1\over p}$ is somehow preferred by
nature. Indeed at such points there is a way out, but it is not
simple perturbation theory. This approach is what this article is
all about.

An alternative to  perturbation theory, the Hartree-Fock (HF)
approximation, also does not work, if applied directly to the
electronic hamiltonian. Let us see why. In HF, one takes the
interaction (quartic in fermion operators) and obtains an
expression quadratic in the fermion operators by replacing various
bilinears by their ground-state averages, initially taken as free
parameters, and dropping some higher-order fluctuations. The
quadratic hamiltonian is then solved and the ground state energy
is then minimized as a function of the assumed averages. At the
minimum, the actual averages will self-consistently come out equal
to the assumed averages. Here is a toy model that illustrates the
main ideas. The model has just two Fermion operators $c$ and $d$:
\beq H=\varepsilon_c \ c^{\dag}c+\varepsilon_d \ d^{\dag}d+ u_0 \
c^{\dag}c d^{\dag}d\eeq Let us separate out the bilinears into
averages $\langle c^{\dag}c \rangle =\lambda_c$ and $\langle
d^{\dag}d \rangle =\lambda_d$ and fluctuating parts $:c^{\dag}c :$
and $:d^{\dag}d :$
\begin{eqnarray} c^{\dag}c&=&:c^{\dag}c:+\langle
c^{\dag}c \rangle \equiv :c^{\dag}c:+\lambda_c\\
d^{\dag}d&=&:d^{\dag}d:+\langle d^{\dag}d \rangle \equiv
:d^{\dag}d:+\lambda_d.
\end{eqnarray}
 We now rewrite $H$ as  \beq H=\varepsilon_c \
c^{\dag}c+\varepsilon_d \ d^{\dag}d+ u_0 \left[ \lambda_d
c^{\dag}c+\lambda_c  d^{\dag}d-
\lambda_c\lambda_d+:c^{\dag}c::d^{\dag}d:\right] \eeq and neglect
the last term, quadratic in the fluctuations, to obtain the HF
hamiltonian.

Let  $E_0(\lambda_c , \lambda_d)$ be the energy of the ground
state $|0\rangle$  of the HF hamiltonian. From   the Feynman-
Hellman theorem \beq {\p E_0\over \p \lambda}=\langle 0|{\p H
\over \p \lambda }|0\rangle \eeq which you can prove using the
fact that $\langle 0|0\rangle =1 $  has no $\lambda$ derivative,
it follows  that at the minimum, $\lambda_c = \langle c^{\dag}c
\rangle $ and $\lambda_d = \langle d^{\dag}d \rangle $ .

Why does this fail in the FQHE problem? For integer filling this
procedure gives the expected state of an integer number of LL's filled
(at least for not too strong interactions). However, it turns out that
for generic fractional filling translational symmetry is spontaneously
broken in HF, and the solution corresponds to a crystalline (usually a
Wigner Crystal) state. Due to the breaking of translational symmetry,
the LL and angular momentum indices $(n,m)$ are no longer good quantum
numbers. The single-particle states are now bands in the crystal,
parametrized by a Bloch quasimomentum $\bk$ and a band index
$n_b$. The HF hamiltonian for FQHE will of the form
$H=\sum_{n_b\bk}\varepsilon_{n_b}(\bk)\
c^{\dag}_{n_b}(\bk)c_{n_b}(\bk)$ up to c-number terms (the analogs of
$\lambda_c\lambda_a$). The self-consistent solutions were worked out
by Yoshioka and Fukuyama (1979), Fukuyama and Platzman (1982), and
Yoshioka and Lee(1982).  The HF solution is compressible,
translationally non-invariant, and has no preference for any
particular density. Thus, it does not describe the FQHE phenomenology.

\subsection {The IQHE - A warm-up}
We begin with a brief look at  the IQHE case $s=0$, or $\nu = p$,
which  paves the way for Jain's (1989) view of FQHE as the IQHE of
entities called Composite Fermions.

In the IQHE, exactly $p$ LL's are filled in the noninteracting
limit. There is exactly one totally antisymmetric ground state
which we denote by $\chi_p$. The simplest example is $p=1$, with
just the LLL filled up. The corresponding wavefunction is \beq
\chi_1 = Det \left|
\begin{array}{cccc}
  z_{1}^{0} & z_{1}^{1} & z_{1}^{2} & .... \\
  z_{2}^{0} & z_{2}^{1} & z_{2}^{2} & ...\\
  .... & .... & .... &....
\end{array}\right| \cdot Gaussian
= \prod \limits_{i<j}(z_i-z_j) \exp \left[{-\sum_{i}{|z_i|^2\over
4l^2}}\right]\label{chi1} \eeq

This nondegenerate ground state is separated from excited states
by a gap equal to  the cyclotron energy $\omega_0$.  Though now we
have a starting point for perturbation theory,  we still need to
actually carry out  the perturbative calculations and in
particular understand why the Hall conductance is {\em unaffected}
by these perturbations and stays at the Galilean invariant value.
We also need to understand why the conductance is unchanged as we
make small changes in density, i.e., why there are steps.

The explanation of the IQHE, discovered over the years, can be most
easily understood in the noninteracting limit. The single-electron
problem with certain special types of disorder can be exactly solved
(Prange, 1981, Aoki and Ando, 1981, Prange and Joynt, 1982), and the
exact solution shows that the Hall current is independent of disorder.
Moving beyond this, Trugman (1983) showed in a seminal paper that for
generic disorder in very high magnetic fields the electronic states
follow the equipotentials of the disorder potential, and each LL
gets broadened into a band. The problem has many fruitful analogies
with percolation. In particular, there is exactly one energy at the
band center at which extended states can exist, while states at all
other energies are localized. When the chemical potential $\mu$
crosses this energy, the Hall conductance changes by an integer.
Contrariwise, when $\mu$ lies in energy range corresponding to
localized states, changes of $\mu$ (changes of filling) produce no
change in $\sigma_{xy}$, which explains the steps.

From a somewhat different point of view, Laughlin (1981) and
Halperin (1982) showed by a gauge argument that to have a
quantized Hall conductance in the noninteracting limit all one
needs is $\sigma_{xx}=0$, which is assured if the chemical
potential lies in the region of localized states. This argument
shows that the Hall conductance (in units of the quantum unit of
conductance $e^2/2\pi\hbar$) is a topological integer. There is
also a field-theoretic analysis of the noninteracting quantum Hall
problem in the presence of disorder which yields similar results
(Pruisken, 1984, 1985a,b, and Levine, Libby, and Pruisken, 1983,
1984a,b,c).

The IQHE can also be understood in the other limit, where
interactions dominate, as follows: In this case if one is not
exactly at $\nu=$\ integer, the excess (or deficit of) particles
form a Wigner Crystal due to interactions. First turn off the
interactions and impose an external periodic potential of the same
period as the Wigner Crystal. In this case Thouless, Kohmoto,
Nightingale, and den Nijs (1982) showed that the dimensionless
Hall conductance has to be an integer, and that this integer is
topological. Later work by Thouless (1983), Niu and Thouless
(1984), and Niu, Thouless, Wu (1985) showed that since this
integer is topological, it is robust to adiabatically introducing
disorder and interactions.  As long as the charge gap does not
close, and there are no ground state transitions, the Hall
conductance will remain unchanged\footnote{By definition, an
integer cannot change continuously.}. This allows one to turn off
the external periodic potential as the interactions are being
turned on. For further details on the IQHE the reader is referred
to an excellent review by Huckestein (1995).

\subsection{The FQHE}
We turn now to fractions, such as ${1\over 3}$. There is nothing
special about disordered single-particle states at such a filling:
They are expected to be localized except right near the middle of the
Landau band (Trugman, 1983). If we ignore disorder and interactions,
we are left with the macroscopically degenerate manifold of many-body
states. Presumably interactions will select the ground state. How is
one to find it?

Here the story  branches into two trails. The one blazed by Laughlin (1983a,b)
consists of writing down inspired trial wavefunctions; the other
is  the hamiltonian approach, which starts  with an assault on the
degeneracy problem.

\subsubsection{Laughlin's answer}

After some experimentation, Laughlin wrote down the following
celebrated trial wavefunction for $\nu =1/(2s+1)$ : \beq \Psi_{
{1\over (2s+1)}}= \prod_{j< i}(z_i-z_j)^{2s+1}\exp (- \sum_i
|z_i|^2/4l^2)\eeq which was shown to have nearly unit overlap with
the numerical solution for small systems for generic repulsive
interactions (Laughlin, 1983a,b, 1987). We will refer to
$\nu=1/(2s+1)$ as Laughlin fractions, as per convention.

Of the many remarkable properties of this function, we list those
that have the greatest bearing on what follows. First, it is from
the LLL and obeys the Pauli principle under particle exchange,
since $2s+1$ is odd and spin is assumed polarized. Halperin
(1983,1984) zeroed in on one of its central features: {\em it has
no wasted zeros}, by which he meant the following.  Consider $\psi
(z_1)$, which is $ \Psi (z_1, \ldots, z_N) $ as a function of any
one variable, randomly chosen to be $z_1$, with all others held
fixed.\footnote{The discussion is independent of this  choice.}
Given that the sample is penetrated by $N/\nu=N(2s+1)$ quanta of
flux, the phase of $\psi (z_1)$ has to  have an Aharonov-Bohm
phase change of  $2\pi N (2s+1)$  per particle, or $N(2s+1)$ zeros
given the LLL condition of analyticity. Only $N$ of these {\em
had} to lie on other electrons by the Pauli principle. But they
all {\em do} lie on other electrons, thereby keeping the electrons
away from each other very effectively, producing a low potential
energy. (The kinetic energy is of course the same for any function
in the LLL.)

By showing that quantum averages in these ground states are
precisely statistical averages in a one-component plasma (Baus and
Hansen, 1980, Caillol, Levesque, Weis, and Hansen, 1982), Laughlin
(1983a,b) showed that the system is an {\em incompressible fluid},
which means a fluid that abhors density changes. Unlike a Fermi
gas which increases (decreases) its density globally when
compressed (decompressed), an incompressible fluid is wedded to a
certain density and will first show no response to any applied
pressure, and then suddenly nucleate a localized region of
different density (just the way a Type II superconductor, in which
a magnetic field is not welcome, will allow it to enter in
quantized units in a region that turns normal).

Laughlin (1983a,b) also provided the wavefunction for a state with
such a localized charge deficit. If one imagines inserting a tiny
solenoid at a point $z_0$ and slowly increasing the flux to one
quantum, one must, by gauge invariance, return to an eigenstate of
$H$, and each particle must undergo a $2\pi$ phase shift as it
goes around $z_0$. \footnote{The flux must be inserted slowly
enough to prevent transitions to other excited states, but fast
enough to prevent a lapse to the original ground state at the end,
a possibility that exists in finite systems with disorder.} This
condition and analyticity point to the ansatz \beq \Psi_{qh}=
\prod_i (z_i-z_0) \Psi_{2s+1} \eeq This is a {\em quasihole}.
There is a more complicated state with a quasiparticle.

The prefactor is a {\em vortex} at $z_0$. Since vortices play a
significant role in the FQHE let us digress to understand them
better. Consider first a real flux tube inserted into the sample at a
point $z_0$. Clearly every electron will "see" this tube at $z_0$,
i.e., $\psi (z_1)$, the wavefunction, seen as a function of any one
coordinate, chosen to be $z_1$, will see a $2\pi$ phase shift as $z_1$
goes around $z_0$. The flux tube clearly has a reality and location of
its own, independent of the locations of the electrons. The vortex is
an analytic LLL version of the flux tube: the $2\pi$ phase change is
accompanied by a zero, i.e., there is an analytic zero at $z_0$ for
{\em every coordinate}. The zeros associated with the vortex have the
key feature that their location does not depend on the location of the
particles, i.e., it too has an independent reality and location like
the flux tube.

There are also vortices in Laughlin's ground state wavefunction.  The
only difference is that instead of sitting at some point $z_0$, they
are anchored to the particles themselves. To see this, consider $\nu
={1\over 3}$, and focus on the part involving just $z_1$: \beq
\prod_{j>1}(z_j-z_1)^3 \eeq If we freeze $z_1$ and view it as a
parameter like $z_0$, we see that there is a triple vortex on particle
$1$, and by symmetry, each particle. Apart from the vortex mandated by
the Pauli principle, there are two vortices bound to each
electron. (The Pauli zero comes with the turf, for being a fermion,
and is not included in the count of vortices {\em attached} to enforce
correlations.) The term vortex is again appropriate here since the
location of the vortex is independent of any coordinate except, of
course, the electron to which it is attached.

Contrast the zeros that constitute a vortex to the zeros of a generic
antisymmetric analytic polynomial of the same degree as the Laughlin
wavefunction. Once again, in $\psi (z_1)$, there is one zero at $z_i$
($i> 1$) independent of all other $z_j$ by the Pauli principle. This
zero is part of a vortex, since all particles will "see" (in the
wavefunctions $\psi$) the Pauli zeros anchored on the other
particles. The non-Pauli zeros of $\psi (z_1)$, by contrast, will be
parametric functions of all other particle coordinates $i=2,..N$ (what
else can they be?). If we now consider $\psi (z_2)$ its non-Pauli zeros
will in general bear no relation to those of $\psi (z_1)$. Since the
location of the non-Pauli zeros move parametrically with all the
$z$'s, they do not belong to or form vortices. The reader must
remember these distinctions in order to follow the discussion on the
internal structure of CF's that follows shortly\footnote{If the
Laughlin function is perturbed slightly, $\psi (z_1)$ will have one
Pauli zero on every other electron and two others nearby. As long as
the perturbation is small we can say, at length scales bigger than the
excursion of the zeros, that electrons are bound to double
vortices. For stronger perturbations that take us far from the
Laughlin ansatz, there will be no simple description of correlations
in terms of vortices.}.

Returning now to the quasihole, another  important property  is
that it represents  a charge deficit of $1/(2s+1)$ in electronic
units. One way to show this is to employ the plasma analogy
(Laughlin 1983a,b,1987). Here is a more general way (Su and
Schrieffer, 1981, Su, 1984) that only depends on the state being
gapped, incompressible, and having a quantized Hall conductance.
As the flux quantum $\Phi_0$ is adiabatically inserted to create
the quasihole, the charge driven out to infinity is given by
integrating the radial current density $j$ produced by the Hall
response to the induced azimuthal $E$ field
\begin{eqnarray} Q&=& -\int j(r,t) 2\pi r dt = \sig  \int E\ 2\pi
r \ dt \nonumber \\ &=& -\sig \int {d\Phi \over dt} dt = -\Phi_0
\sig\\ &=& -{2\pi \hbar \over e}\ {e^2 \over 2\pi \hbar} \ \nu = -e\nu
. \label{su-argument}\end{eqnarray} For non-Laughlin fractions, the
charge driven out by inserting one flux quantum is that of $p$
quasiparticles, each of which has a charge equal to $1/(2ps+1)$ (Su,
1984). {\em Note that the fractional $\sig$ is the cause behind the
fractional charge.} The fractional charge of the quasiparticles has
been confirmed experimentally (Goldman and Su, 1995, Goldman,
1996). The quasiparticles also have fractional statistics, as was
pointed out by Halperin (1984), and explicitly confirmed in a
wavefunction calculation by Arovas, Schrieffer, Wilczek, and Zee
(1985)\footnote{Fractional statistics was shown to be an emergent
property of Composite Fermions by Goldhaber and Jain (1995).}.

The insensitivity of the Hall conductance to disorder in the FQHE can
be established by an extension of the Laughlin-Halperin gauge
arguments for the IQHE. To show this in toroidal geometry, the old
arguments have to be supplemented by the fact that at fractions of the
form $p/q$, there are $q$ ground states, and the assumption that upon
adding $q$ flux quanta one returns to the starting state.  There seems
to be a consensus that Hall conductance is a ground state property
that is impervious to a reasonable amount of disorder, very much the
way superfluidity is.

Laughlin (1983a,b,1987) explained the plateaus as we move off the
magic fraction as follows. Suppose we move slightly off ${1\over 3}$,
to say $\nu= {1\over 3} \pm .01$. The system has two choices.  It can
either go to a FQHE ground state at this fraction, or it can be at
${1\over 3}$ plus some number of quasiparticles (holes) of charge $\pm
{1\over 3}$.  It turns out the latter has a lower energy. The reason
$\s_{xy}$ is locked at ${1\over 3}$ is that the
quasiparticles/quasiholes, being sharply defined entities, get
localized by the disorder potential and do not contribute to the Hall
current. Thus the Hall conductance is expected\footnote{There is
currently no way of rigorously calculating transport coefficients in
the FQHE in the presence of disorder, hence the qualifier "expected".}
to be unaffected by disorder. As we move further off ${1\over 3}$, the
system switches to another ground state with its own quasiparticles,
and $\s_{xy}$ jumps to the neighboring plateau.  It is clear that if
the fate of the quasiparticles/quasiholes is to get localized, their
fractional charge cannot explain the fractional Hall conductance. As
seen earlier, the connection is the other way around.

Thus a fairly complete description of the FQHE emerged  from the
Laughlin wavefunctions for the ground and quasiparticle/quasihole
 states. However numerous other issues surfaced and were
tackled by subsequent work in the intervening two decades. We now
steer our attention to those bearing directly on this paper.

Sticking for a moment to the Laughlin fractions, one important
question that was raised was whether the system was really gapped,
i.e., could there be lighter excitations than Laughlin's
quasiparticles and quasiholes? Perhaps a quasiparticle and quasihole
could bind to form a gapless excitation. Haldane and Rezayi (1985)
verified by exact diagonalization of small systems that the $\nu
={1\over 3}$ system was gapped.  Girvin, MacDonald and Platzmann
(1986) explored the question of neutral particle-hole excitations
using an analogy with Feynman's work on superfluids and found a way to
calculate the dispersion relations of the magnetoexciton within the
LLL and showed that they were gapped in this sector for $\nu ={1\over
3}$.

\subsubsection{Jain's Composite Fermions}

Let us now proceed to the non-Laughlin fractions. Here the central
question is  this: Given that non-Laughlin fractions like
${2\over 5}$ are seen in experiment, what is their wavefunction?
Simply replacing the factors $(z_i-z_j)^{2s+1}$ by
$(z_i-z_j)^{5/2}$ in Laughlin's ansatz is not acceptable since
this does not produce the mandatory change of sign under particle
exchange. In the hierarchy approach (Haldane, 1983, Halperin,
1983, 1984) the quasiparticles of the Laughlin states condense
into their own FQHE states, whose quasiparticles in turn do the
same thing and so on. While this approach gives a natural way to
generate additional fractions, it does not give explicit wave
functions in terms of electrons.

Explicit trial states for fractions $\nu =\n$ were provided by
Jain (1989,1990,1994), who also explained why they were natural in
terms of objects called {\em Composite Fermions} (CF's).

Jain's scheme  hinges upon the seminal idea of flux attachment,
which plays such a central role in the FQHE that it merits a
little digression.  It was introduced first by Leinaas and
Myrrheim (1977) in terms of wavefunctions, and explored in the
language of Chern-Simons field theories by Wilczek (1982a,b). The
crux of the idea is that in $d=2$ one can have particles (dubbed
anyons by Wilczek) that suffer a phase change $e^{i\theta}$ upon
exchange, with $\theta =0 \ ,\pi \ \ (\mbox{ mod} \ 2\pi )$
corresponding to bosons and fermions respectively. To obtain one
of these particles one takes a fermion and drives through its
center a point flux tube, the amount of flux being decided by the
desired $\theta$. In particular, if this tube contains an even/odd
number of flux quanta, the composite particle one gets is a
fermion/boson.

 Jain (1989,1990) exploited flux attachment for $\nu^{-1}=2s+{1\over
p}$, as follows.  Suppose we trade our electrons for CF's carrying
$2s$ point flux quanta pointing opposite to the external $B$. On
average the CF's effectively see ${1\over p}$ flux quanta per
particle and fill up exactly $p$ LL's. At the mean field level,
this approach gives the following trial wavefunction:

  \beq
\Psi_{p/2ps+1} = \prod_{i<j} \left[{(z_i -z_j)\over
|z_i-z_j|}\right]^{2s}\cdot \chi_p (z,\bar{z})\eeq  where $\chi_p$
is the CF wavefunction with $p$-filled LL's  and the prefactor
takes the CF wavefunction back to  the electronic wavefunction.
(This will be made clearer shortly.)

Jain improved this ansatz in two ways and proposed: \beq
\psi_{p/2ps+1} = {\cal P} \prod_{i<j} (z_i -z_j)^{2s}\cdot \chi_p
(z,\bar{z})\label{jainwave}\eeq where \beq \prod_{i<j} (z_i
-z_j)^{2s} \eeq is called the {\em Jastrow factor}, and \beq {\cal
P} : \bar{z} \to 2l^2 {\partial \over \partial z}\eeq projects the
wavefunction to the LLL by eliminating all $\bar{z}$'s in
$\chi_p$ by turning them to $z$-derivatives on the rest of the
wavefunction, except for the gaussian.  (We shall
see\footnote{For a nice review of states, operators, and matrix
elements within the LLL see Girvin and Jach (1984)} that in the
LLL $\left[ z,\bar{z}\right] =-2l^2$.)

 In making the change $\prod_{i<j} \left[{(z_i -z_j)\over
     |z_i-z_j|}\right]^{2s} \to \prod_{i<j} (z_i -z_j)^{2s}$, Jain
 replaces {\em flux tubes} by {\em vortices}. Although in both cases
 each particle picks up an extra phase shift of $4\pi s$ on going
 around another, only the vortex factor has the desirable multiple
 zero that keeps the particles apart and lies in the LLL. Jain does
 not need to justify these modifications since in writing a trial
 wavefunction, he is free to resort to any changes that improve its
 energy. However, in the hamiltonian approach, the replacement of flux
 tubes by vortices and the projection to the LLL prove difficult but
 not insurmountable.

 At $p=1$, since $\chi (z, \bar{z}) =\prod_{i<j} (z_i-z_j) \cdot
  \mbox{Gaussian}$, we do not need ${\cal P}$ to get back Laughlin's
  answer. For $p>1$ we have concrete expressions for $\Psi$ in terms
  of electron coordinates, although the action of $\cal{P}$ can be
  quite involved, and has a big impact on the wavefunction as we will
  see soon.

 Thus while the degeneracy of the noninteracting problem is present
for any $\nu <1$, at the Jain fractions one can beat it by
thinking in terms of Composite Fermions. As we move off the Jain
fractions, the incremental CF's (particles or holes) get
localized, giving rise to the plateaus. This is the sense in which
the CF approach allows one to think of the FQHE in terms of the
IQHE of the CF's. Since the IQHE can be understood without
invoking interactions, it is sometimes suggested that CF's are
free. Later we will  see why this cannot be so.

The reader will have noted that both Laughlin and Jain
wavefunctions {\em make no reference to the inter-electron
potential}. This feature, which permits them to work for a whole
class of potentials also renders them  insensitive to specific
features, including even the range of the interaction.

\subsubsection{Vortices, zeros, and the dipole picture re-examined}
\label{dipole-picture}

Let us return to the vortices that are attached to electrons in
the Jastrow factor.  We have already discussed Halperin's
observation (Halperin 1983, 1984)  that electrons are bound to
vortices in the Laughlin wavefunction. This concept was taken up
and vigorously pursued by Read (1989). In particular Read made a
clear distinction between flux attachment and vortex attachment,
which were often loosely interchanged. Attaching electrons to flux
tubes is a mathematical trick; flux tubes are unphysical and
neutral. Vortices on the other hand are physical excitations of
the Laughlin ground states, and electrons would naturally bind to
them since the two are oppositely charged. The reason the CF sees
a weaker field is not due to any mysterious capture of flux tubes.
Indeed, the external field is uniform and not quantized into point
flux tubes. What really happens in the Laughlin case is that each
electron pairs with $2s$ vortices and when the composite object
goes around on a loop, it sees a phase change of $-2\pi (2s+1)$
per enclosed particle due to the external $B$ and $2\pi (2s)$ per
enclosed CF due to  the vortices attached to each one.

Next, since the $2s$-fold vortex has a charge $-2s/(2s+1)$ as per
the flux insertion argument (Su, 1984), Eqn. (\ref{su-argument}),
the vortices reduce $e$ down to the quasiparticle or CF charge

\beq e^*=1-{2s \over 2s+1}={1 \over 2s+1}. \eeq

In a Fermi liquid an added electron is also screened by a
correlation hole; here the difference is that the possible
response of the Laughlin liquid to an extra electron is limited to
an integer number of vortices of quantized charge.

Read (1994, 1996) next applied the notion of electrons binding to
vortices to the case   $\nu={1\over 2}$ and derived what is called
the {\em dipole picture}. We now turn to a critical review of this
concept. Since the dipole picture comes from the projection
involved in the $\nu ={1\over 2}$ wavefunction, it  is instructive
to start with understanding the impact of   ${\cal P}$ in a
general Jain fraction  where  higher  CF-LL's ($p>1$) are filled.

Consider $N$ electrons at $\nu ={2\over 5}$. Prior to projection, each
electron in the Jain wavefunction has a double vortex sitting on it
due to the Jastrow factor. As we explained earlier, the term vortex is
merited since every other particle sees it at the same place.  At this
stage it is correct to view the CF as an electron plus a double
vortex\footnote{There is also another zero, generally non-analytic, in
$\chi_p$ which is antisymmetric.  This is not part of the vortex
count.}.  All this changes after projection to the LLL. Now there can
only be $5N/2$ zeros of the wavefunction as a function of, say, $z_1$;
$N$ of them must lie on other electrons by the Pauli principle
(forming single vortices anchored to electrons), with $3N/2$ left
over. Clearly they cannot all be on electrons, or be associated with
them uniquely (since we have $3/2$ zeros per electron). If the zeros
are not on electrons, their location must depend parametrically on the
locations of {\it all} the electrons and they cannot organize
themselves into vortices.

 Thus the wavefunction
analysis leaves us with the following quandry. The CF cannot be
viewed as an electron-vortex complex in the projected non-Laughlin
states, but the quasiparticle charge, $e^*=1-{2ps\over 2ps+1}$ is
robust under ${\cal P}$ since $e^*$ is tied to $\sigma_{xy}$ which
is presumed robust under projection. {\em What, if any, is the
entity that binds to the electron to bring $e$ down to $e^*$?} How
does this entity enter the theory? What makes it bind to the
electron? We will  see that the EHT answers such questions.

Let us now return to $\nu ={1\over 2}$. In the unprojected
wavefunction the double-vortex charge fully cancels the electron
charge so that $e^*=0$. Read (1994) has argued that  the neutral
CF of momentum $k$ has a dipole moment $d^*=kl^2$, based  the wave
function at $\nu ={1\over 2}$, also called the Rezayi-Read (Rezayi
and Read, 1994) wavefunction, in which $\chi_p$ in Eqn.
(\ref{jainwave}) is replaced by the filled Fermi sea:

\beq \Psi_{\half} = {\cal P} \prod_{i<j} (z_i-z_j)^2 Det \left|
e^{i{\bf k}_i\cdot {\bf r}_j}\right| . \eeq

He considers

\beq e^{i{\bf k\cdot r}} = \exp {i \over 2} ({k\bar{z}+
\bar{k}z})\ \ \ \ \ \  k= k_x+ik_y .\eeq Since under the action of
${\cal P}$,   $\bar{z}$ acts on the analytic part of the
wavefunction as $ 2l^2\partial/\partial z$, $e^{il^2k{\partial
\over \partial z}} $ causes the shift $z \to z+ikl^2 $ in the
Jastrow factor. This motion of the vortex off the electron
produces the dipole moment $d^*=kl^2$. The energy needed to
separate the vortex from the electron (the Coulomb attraction),
must begin with a term quadratic in the separation or momentum.
This then gives the LLL fermions a kinetic energy or an effective
mass $m^*$ resulting from interactions.

Attractive though this picture is, closer scrutiny reveals that
the preceding line of reasoning is incorrect in the following
ways:

\begin{itemize}
  \item Since {\em every} $z_i$ gets translated,  $(z_i-z_j)^{2s}\to
  (z_i-z_j+ik_i-ik_j)^{2s}$ and particle $i$ sees a
  multiple zero associated with $z_j$ at
  $z_j+i(k_j-k_i)$. We cannot call this a vortex since the location
  of the zero varies with the label $i$. Different particles see
  it in different places. This problem is actually moot because of
  the next, more serious  one.

  \item {\em Even this multiple zero is there for one particular assignment
  of $k$'s, or one term in the determinant. Upon antisymmetrization,
  we cannot relate the zeros of the sum over the $N!$ terms to the
  zeros of individual terms.} The situation would have been different
  if we had been talking about poles which can survive such a sum. In
  fact, one of the two zeros that moved off the particles (in each
  individual term) should return upon antisymmetrization to lie
  exactly on the particles by Pauli's principle,
  leaving another zero
  to depend parametrically on all other coordinates, and all the
  $k_i$. Thus there is no reason to expect any simple relation
  between the location of the electrons and the non-Pauli zeros in
  $\Psi_{\half}$ or to conclude that these zeros form vortices.
    We commend to the reader experimentation with
   the limited but nonetheless instructive case of
   just two particles, to see some of these ideas in concrete form.

  \item If one looks at $\nu =\half$ in isolation, the fact there is
  one non-Pauli zero per electron (in the thermodynamic limit) may
  tempt one to suggest that perhaps in this case they organize
  themselves into vortices.  Besides our analysis of the gapped
  fractions which shows that this is extremely unlikely, there is the
  general argument that zeros that do not lie on particles or external
  flux tubes (i.e., locations with an independent reality) must vary
  parametrically with all coordinates and therefore cannot be
  organized into vortices.

\end{itemize}

Thus at $\nu ={1\over 2}$, there seem to be  two choices: Either
use the unprojected wavefunction in which the Jastrow factor
explicitly has two vortices per electron, but the dipole moment is
zero (since the vortices are on the electrons), or go the LLL
where the vortices disintegrate into a smaller number of ordinary
zeros, not correlated with the electrons in any simple fashion. So
what happens to  the dipole?

Our EHT, which provides an operator realization of the CF,  will
show that the dipole picture  is more robust than the
wavefunction-based  arguments pointing to it. However one must
look for it  not in the wavefunction, but in operators and
correlation functions. Even then,  the value of $d^*$ will be
sensitive to the details of the wavefunction. In this respect, it
is quite unlike $\sig$ which is robust under changes in the ground
state wavefunctions, (including projection to the LLL), and
depends only on the filling factor.

\subsection{Hamiltonian approach }

Let us begin by reviewing the challenges facing any hamiltonian
theory. In the FQHE there is more than enough room in the LLL for
all the electrons, leading to a macroscopic number of degenerate
ground states in the absence of interactions. This preempts many
of the standard approximations. It is expected that when
interactions are turned on, there will emerge from this degenerate
manifold a unique ground state, separated by a gap from other
low-lying excited states. While the true ground state and
low-lying excitations will of course contain an admixture of
higher LL's, one expects that in the limit of bare mass $m \to 0$,
or $\omega_0 \to \infty$, there will emerge a low-energy sector
spanned by LLL states. {\em In other words, in the limit $m \to 0$
a nonsingular low-energy theory must emerge.} The $m$-independent
Laughlin wavefunctions for the ground state, quasiparticles and
quasiholes illustrate this point.

If all memory of $m$ is lost, the only energy scale is set by the
interactions. How is one to isolate the LLL physics, staring with
the full Hilbert space? How is one to battle the degeneracy of the
noninteracting ground state? How is one to get rid of the $1/m$
dependence of the kinetic energy? These are problems for the
hamiltonian approach.

Next, there is one vestige of   $m$-dependence even in the LLL,
discovered by  Simon, Stern and Halperin (1996): if  the external
field $B$ varies slowly in space, the zero point energy $eB/2m$,
can no longer be eliminated by redefinition of the zero. In fact
the particles behave as if  they have a magnetic moment
$\mu^*=e/2m$ coupled to $B$. The theory must reproduce this
moment. \footnote{This magnetic moment is actually a compact way
to describe an effect that is  orbital in origin.}

We also do not want to ban higher LL's too soon, since the Hall
conductance necessarily involves them\footnote{ The current operator
goes as $1/m$ while $\s_{xy}$ does not. The $1/m$ in the current is
cancelled by the energy denominator of order $\omega_0$ that comes
from virtual transitions to the higher LL's. For a more detailed
explanation, see Girvin, MacDonald, and Platzman (1986), and Sondhi
and Kivelson (1992).}.  Finally, Kohn's theorem (Kohn, 1961) assures us
that despite interactions, the cyclotron mode at $q=0$ must be at
$\omega_0 =eB/m$ and it must saturate certain sum rules at small
$q$. The theory must pass this test.

{\em The hamiltonian theory must succeed in keeping the
$m$-dependence where it belongs and exorcise it elsewhere.}

The next set of issues concern the quasiparticles, the CF's. A
heuristic picture that we saw arising from the unprojected
wavefunctions is that the quasiparticles are composites of
electrons and $2s$ vortices. How is this actually realized in the
hamiltonian approach? After all, this is not analogous to the
statement that mesons are made of quarks and antiquarks, for
unlike antiquarks,  which appear in the hamiltonian, vortices do
not. Vortices are not independent of electrons and to speak of
them and electrons at the same time surely paves the way for
overcounting. In the non-Laughlin cases, the issue is further
complicated by the fact that there are not enough zeros to form
$2s$ vortices per electron, and in any case the zeros are not
organized into vortices. Yet some object of the  same charge as
the $2s$-fold vortex seems to bind with the electron since in the
end  $e$ gets reduced to $e^*$. Somehow this object has to  enter
the theory and give rise to the  CF.

Other issues surround the CF. How strong are interactions between
CF's? Why do they sometimes appear to be free when we can give
persuasive arguments for why they cannot be? How is disorder to be
included?

Besides addressing these questions of principle, we need answers for
{\em quantitative questions}. For example, what is the gap at $\nu
={1\over 3}$? Now, very precise answers can be given for such a
question in the trial wavefunction approach\footnote{For a review of
this method with complete references, see Jain and Kamilla (1998).} or
by exact diagonalization\footnote{Due to limitations of space we
present only the earliest references: Girvin and Jach (1983), Haldane
and Rezayi (1985), Su (1985), Haldane (1985), Yoshioka (1986), Morf
and Halperin (1986, 1987), with the exception of Morf {\em et al}
(2002). }. While these results are founded on the microscopic
hamiltonian, the intermediate steps are computer-intensive. It would
be nice to have a theory which displays transparently the features of
the CF deduced from the study of wave functions, and furnishes
quantitative results, to say 10\% accuracy. The same goes for
polarizations in various states, and transitions between them.

A specially interesting set of questions appear at and near $\nu
=1/2$.  Can the notion of the CF survive in this region without a
robust gap? In particular, will they really behave like particles
of charge $e^*$ in this region? Since the average effective
magnetic field $B^*=0$ at $\nu =1/2$, we have fermions in zero
magnetic field, except for fluctuations. Will this Fermi system
form a Fermi liquid after including fluctuations? What are its
response functions? Will it be compressible? If so, how do
particles with $e^*=0$ manage to be compressible or even have a
nonzero Hall conductance? What are the collective modes and how
can they be detected? How is disorder to be handled? Exact
diagonalization and the wavefunction based approaches are not very
helpful for unequal-time correlations, and finite-temperature
physics. Hamiltonian theories should fill this void and answer all
the questions raised above.

\subsection{Organization of this review}

 Section II introduces the
reader to some notation and the physics of the LLL, including
Kohn's theorem which is often referred to in our work. Section III
describes  the Chern-Simons theory of flux attachment including
both composite bosons and fermions. Section IV deals with the
region at and near $\nu =1/2$ and is mainly about the work of
Halperin, Lee and Read (1993). Section V explains our extended
hamiltonian approach, EHT,  formulated in an enlarged Hilbert
space. It is shown that there are two approaches to solving our
final equations: the conserving approximation in which the
constraints are respected, but the CF physics is hidden, and a
shortcut in which the opposite is true. The compressibility
paradox at $\nu =1/2$ is discussed. Section VI illustrates the
conserving approximation in the EHT via the calculation of the
structure factor at small $q$ in the Jain series and the
magnetoexciton dispersion relations for $1/3$ and $2/3$. Section
VII is devoted to the computation of gaps and comparison to
numerical work and experiment. Sections VIII deals with magnetic
phenomena at $T=0$ for gapped and gapless systems. The question of
whether CF's are free or not is discussed and answered in the
negative, along with an explanation of why sometimes they appear
to be so. Section IX addresses physics at $T>0$, describing the
computation of polarization and relaxation at $\nu ={1\over 2}$
and polarization at ${1\over 3}$ and ${2\over 5}$. The results are
then compared to experiment. Section X deals with
 inhomogeneous states.   Section XI gives a critical evaluation of the
 EHT.  A summary and a discussion of open
problems follow in Section XII.

\section{Preliminaries and notation \label{prels}}
This section focuses on a single electron in two dimensions. The
hamiltonian is  \beq H_0={(\bp+e\bA)^2\over 2m}={\bPi^2 \over 2m}.
\eeq Here $\bA$ is the vector potential that leads to the external
magnetic field ${\bf B}=\nabla \times \bA = -B{\hat \bz}$, $\bp $
is the canonical momentum of the electron, $-e$ is its charge, and
$m$  is its bare or band mass.  Note that $\bB $ points along the
negative $z$-axis.

Although one needs to pick a gauge for $\bA$ in order to find the
wavefunctions, we can obtain the spectrum without making that
choice. Let us define a {\em cyclotron coordinate}
\beq
\etab=l^2\hz\times\bPi
\eeq
where $l=\sqrt{\hbar/eB}$ is the
magnetic length.  Despite the name, the two components of $\etab$
are not commuting but canonically conjugate  variables: \beq
\left[ \eta_{x} \ , \eta_{y} \right] = il^2. \eeq It follows
that  \beq H_0= {\etab^2 \over 2ml^4}\eeq describes a harmonic
oscillator with energies \beq E=(n+{1\over 2})\omega_0\eeq where
$n$ is the LL index.

The LL's are highly degenerate because another conjugate pair that
commutes with $\etab$, and called the {\em guiding center
coordinate} \beq \bR = \br -\etab \eeq is cyclic. The components
of $\bR$ obey  \beq \left[ R_{x}\ , R_{y} \right] =- il^2.\eeq
Thus $l^2$ plays the role of $\hbar$. Since in the LLL $\langle
\eta^2 \rangle =l^2$, $\bR$ roams all over the sample, whose area
$L^2$
 plays the role of phase space. The degeneracy of the
LLL (or any LL) is, from the Bohr-Sommerfeld quantization rule,
 \beq D = {L^2 \over "h"} = {L^2
\over 2 \pi l^2} = { eBL^2 \over 2\pi \hbar} ={\Phi \over \Phi_0}
\eeq where $\Phi_0$ is the flux quantum.This leads to the
following result worthy of committing to memory:

\beq  { \mbox{LLL states}\over \mbox{particles}}={D\over N}=  {
\mbox{flux quanta density}\over \mbox{particle density}} ={
eB\over 2\pi n\hbar } \eeq

At points where steps in  $\sigma_{xy}$ cross the straight line
dictated by Galilean invariance,  \beq \sigma_{xy }= {ne\over B} =
{e^2 \over 2\pi \hbar}\nu = {e^2 \over 2\pi \hbar} {p \over 2ps+1}
, \eeq and

\beq  { \mbox{LLL states}\over \mbox{particles}}={ \mbox{flux
quanta density}\over \mbox{particle density}} ={eB\over 2\pi \hbar
n}= 2s+{1\over p} = \nu^{-1}. \eeq
 Hereafter we set $\hbar =1$.

\subsection{Gauge choices}

There are two famous choices for gauge. In the {\em Landau gauge}
$\bA=- {\bf j} Bx$ the Hamiltonian is cyclic in $y$, and hence has
the eigenfunctions $e^{iky}$. Making the ansatz \beq
\psi(x,y)=e^{iky}\phi(x) \eeq we find that $\phi$ obeys the
equation for a displaced harmonic oscillator \beq -{1\over
2m}{d^2\phi\over dx^2} + \half m \omega_{0}^{2} (x-kl^2)^2
\phi=E\phi \label{lg}\eeq The LLL corresponds  to putting the
oscillator in its ground state. The degeneracy of any LL  can be
computed by considering a  sample with sides $L_x$ and $L_y$, with
periodic boundary conditions in the $y$ direction. This forces
$k={2\pi j\over L_y}$. Since the  wavefunction is centered around
$x=kl^2<L_x$, we must demand   $0\le j\le {L_xL_y\over 2\pi l^2}$.
Thus the  degeneracy $D$ of each LL, is (going back to a square
sample) $D={BL^2\over\Phi_0}$, in agreement with our prior
gauge-invariant counting of states. We shall use this gauge in the
computation of the magnetoexciton.

In the rest of this article we employ  the {\em symmetric gauge}
 in which
 \begin{eqnarray}
 \bA &=& {eB\over 2} ({\bf i}y -{\bf j}x)\\
\etab &=& {1 \over 2} {\bf r} +l^2 \hat{\bz}\times {\bf p}  \\
{\bf R}&=& {1 \over 2} {\bf r} -l^2 \hat{\bz}\times {\bf p}.
\end{eqnarray}

\subsection{Getting to know the LLL} As with any simple harmonic
oscillator, we can construct ladder operators from the canonically
conjugate pair $\eta_x$ and $\eta_y$ (with $l^2$ playing the role
of $\hbar$)
\begin{eqnarray}
a_{\eta}=&{1\over l\sqrt{2}}(\eta_x+i\eta_y)\\
a_{\eta}^{\dagger}=&{1\over l\sqrt{2}}(\eta_x-i\eta_y)
\end{eqnarray}

 The LLL
condition  $ a_{ \eta} |LLL\rangle =0$ gives \beq \psi =
e^{-|z|^2/4l^2} f(z) \eeq where $z=x+iy$. A basis for $\psi$  is
\beq     \psi_m(z) = z^m\ e^{-|z|^2/4l^2}\ \ \ \ \ \ \
m=0,1,\ldots \eeq The Gaussian is often suppressed. The state has
angular momentum $L_z=m$.

If  $\nu =1$, (one electron per LLL state) there is a unique
noninteracting ground state, which may then be perturbed by
standard means, \beq \chi_1 =\prod_{i<j} (z_i -z_j)\cdot Gaussian
= Det \left|
\begin{array}{cccc}
  z_{1}^{0} & z_{1}^{1} & z_{1}^{2} & .... \\
  z_{2}^{0} & z_{2}^{1} & z_{2}^{2} & ...\\
  .... & .... & .... &....
\end{array}
\right|\cdot Gaussian\eeq

For $\nu <1$, we often want to focus on the limit  $\omega_0 \to
\infty$ and  work entirely within the LLL. If in $H=T+V$ we set
$T$ equal to a constant ($eB/2m$ per particle), all the action is
in $V$. Why is this a problem if $V$ is a function of just
coordinates? After all,   \beq \rho (\br) =\sum_j \delta (\br
-\br_j)\eeq
 and
  \begin{eqnarray} V&=&{1 \over 2} \int
d^2r \int d^2 r' \rho (\br) v(\br -\br') \rho  (\br')\\ &=& {1
\over 2}\sum_{\bq }\rho (\bq )v(q) \rho (-\bq)\\ \rho (\bq) &=&
  \int d^2r \rho (\br)   e^{-i\bq \cdot \br}  =
\sum_j e^{-i\bq \cdot \br_j}.\end{eqnarray}

       The point is that if one sets $T= eB/2m$, the LLL value,
one must project the operator ${\bf r}$ to the LLL. The
coordinates $x$ and $y$, which commute in the full Hilbert space,
no longer commute in the LLL and $V$ now contains noncommuting
operators.

\subsection{     Projection to the LLL}

    Let ${\cal P}$ denote projection to the LLL\footnote{A very nice
introduction to LLL physics appears in Girvin and Jach (1984)}.
Then \beq {\cal P}: \ {\bf r}={\bf R} + \etab \Rightarrow {\bf R}
.  \eeq The projected components do not commute: \beq \left[
R_{x}\ , R_{y} \right] = - i{l^2}\ \ \ \mbox{or}\ \ \ \ \ \ \
\left[ {z}\ , \bar{z} \right] = -2{l^2} \ \ \ \mbox{in the LLL.}
\eeq As for the densities

\begin{eqnarray}
{\cal P}&:&\ e^{-i\bq\cdot \br}\Rightarrow \langle e^{-i{\bf q}
\cdot \etab} \rangle_{LLL}e^{-i\bq \cdot \bR}=
e^{-q^2l^2/4}e^{-i\bq \cdot \bR}
\end{eqnarray}

Thus the projected
problem is defined by
\begin{eqnarray}
\bar{\bar{H}}&=& {1\over2}\sum_{\bq}e^{-q^2l^2/2}\bar{\bar{
\rho}}(\bq)v(q)\bar{\bar {\rho}}(-\bq)\label{energy} \\ \barro
&=& \sum_j e^{-i{\bq \cdot \bR}_j}
\end{eqnarray}
The projection $\bar{\bar {\rho}}$ is the {\em Magnetic
Translation Operator}\footnote{These operators and the projective
group they form have a long history, and were first used to
describe symmetries of the noninteracting electron Hamiltonian in
a magnetic field. For references, see: Peterson (1960), Brown
(1964), Zak (1964a,b). To the best of our knowledge, Girvin,
MacDonald, and Platzman (1986) were the first to concentrate on
the {\it algebra} of this operator}, which differs from the
projected density by a factor $e^{-q^2l^2/4}$. We shall often
refer to it as the density, but take care to include the gaussian
in Eqn.(\ref{energy}). The commutation rules of $\bar{\bar
{\rho}}$ define the magnetic translation algebra: \beq \left[
\bar{\bar{\rho}}({\bf q}) , \bar{\bar{\rho}}({\bf q'}) \right] =2i
\sin \left[ { ({\bf q\times q'})\ l^2 \over 2}\right]
\bar{\bar{\rho} }({\bf q+q'}) \eeq which was thoroughly exploited
in the work of Girvin, MacDonald, and Platzman (1986).

There is no small parameter in $\bar{\bar{H}}$ and the overall
energy scale is set by $v(q)$. This is why the FQHE problem is
unique. As mentioned earlier the Hartree-Fock solution is also not
an option at fractional filling.

\subsection{Kohn's Theorem}
By Kohn's theorem we shall mean the following results. Consider,
in a translationally invariant system, the density-density
response function $K(\bq , \omega)$, the Fourier transform of

\beq K(\bq \ , t) = i\theta (t) \langle 0|\left[ \rho (\bq , t),
\rho (-\bq , 0)\right] |0\rangle . \label{k00}\eeq

In the frequency domain we define \beq K(\bq , \omega ) =
\int_{-\infty}^{\infty} K(\bq , t )e^{{i\omega t}} dt .\eeq

>From just the  canonical commutation rules it can be shown that
$K$ obeys the sum rule \beq \int_{0}^{\infty} Im \ K(\bq , \omega)
{d\omega \over \pi} = {q^2n\over 2m}.\eeq Kohn showed that $K$
must have a pole, the magnetoplasmon,  at the cyclotron frequency
$\omega_0 =eB/m$ with a residue that {\em saturates} the above
sum rule as $q\to0$.

{\em It follows that  $\barro$, the restriction of $\rho (\bq)$
to the LLL, cannot have  (transition) matrix elements that are linear
in $q$ for small $q$.}

 {\em The structure factor} $S(q,\omega )$   \beq
S(q,\omega )=\sum_n |\langle 0|\rho (q,0) |n\rangle |^2 \delta
(\omega -E_n)\eeq is related to $K(q,\omega )$ for $\omega >0$  by
\beq {1\over \pi} Im\ K (q,\omega )= S(q,\omega ).\eeq  \footnote{
To see this, introduce a complete  set of exact eigenstates
$|n\rangle$ of $H$ between the two factors of $\rho$ in Eqn.
(\ref{k00}), express the Heisenberg operator $\rho (t)$ in terms
of $\rho (0) $ and $H$,and do the time integral.} Kohn theorem
tells us that if we limit ourselves to the LLL, $S(q)\simeq q^4$
for small $q$.

\section{     Hamiltonian theory I\ - the CS approach}
  The hamiltonian for electrons in a vector potential $\bA$  is
\begin{eqnarray} H &=& \sum_j{({\bf p}_j + e{\bf A} (\br_j))^2 \over 2m}\ \ \
+V
\end{eqnarray}
where $V$ is the electron-electron interaction, say Coulombic.
Disorder is not included. As stated earlier, the field $\bA$ is
such that there are $2s+{1\over p}$ flux quanta or LLL states per
electron and the attendant degeneracy frustrates perturbative
analysis.

The first step in the CS approach is to deal with the degeneracy
by resorting to flux attachment. For the Laughlin fractions there
are actually {\em two} options involving either composite bosons
(CB) or composite fermions.

\subsection{Composite bosons}
Historically, the first treatment came from Zhang, Hansson and
Kivelson (1989) who considered Laughlin fractions $\nu =1/(2s+1)$.
They  traded electrons for CB's carrying $2s+1$ flux quanta in
opposition to the applied field so that at mean-field level the
bosons saw zero field and had a unique ground state\footnote{This
mean-field idea was first applied to anyon superconductivity by
Laughlin (1988). Many more works on this topic followed: Fetter,
Hanna, and Laughlin (1989), Hanna, Laughlin, Fetter (1989), Chen,
Halperin, Wilczek, and Witten (1989), Halperin, March-Russell, and
Wilczek (1989), Lee and Fisher (1989), Hanna, Laughlin, Fetter
(1991), Dai, Levy, Fetter, Hanna, and Laughlin (1992).}.  The
trading is done by introducing a Chern-Simons wave function
$\Psi_{CS}$ defined as follows:

\begin{eqnarray}
    \Psi_e &=&     \prod_{i<j} {(z_i - z_j) ^{2s+1}\over |z_i
-z_j|^{2s+1}}    \Psi_{\rm CS}\equiv   \exp
((2s+1)i\sum_{i<j}\phi_{ij})\ \Psi_{\rm CS}.\label{eq-phase}\\
H_{CS} &=& \sum_i {({\bf p}_{i} + e{\bf A} ({\bf r}_i)+     {\bf
a}_{cs}({\bf r}_i))^2 \over 2m} + V\label{26}
\end{eqnarray}
where $\phi_{ij}$ is the phase of the coordinate difference
$z_i-z_j$.
 Since $\Psi_{e}$ changes sign under particle exchange and the prefactor
 produces  $2s+1$  extra minus signs,
$\Psi_{CS}$ describes bosons.

The CS gauge field, ${\bf a}_{cs}$, comes from the action of ${\bf
p}$ on the prefactor (which is just the phase of the Jastrow
factor) that multiplies $\Psi_{CS}$:

\begin{eqnarray}
  {\bf a}_{cs}({\bf r}_i) &=&2s \nabla \sum_{j\ne i}\phi_{ij}\\
  \oint {\bf a}_{cs}({\bf r}_i)\cdot d{\bf
r}_i &=& 2s\oint \sum_{j\ne i}\nabla \phi_{ij}\cdot d{\bf r}_i
\\ &=& 2\pi (2s+1) \ \mbox{(number of particles enclosed)}\\
       \nabla \times {{\bf a}_{cs}} &=&
2\pi (2s+1) \rho \label{cscondition1}
\end{eqnarray}

 Eq.
(\ref{cscondition1}) shows explicitly that the flux quanta density
is $2s+1$ times the   particle density. This is what one means by
flux attachment. Eqs. (\ref{26}) and (\ref{cscondition1}) define a
Chern-Simons theory. \footnote{Note that these manipulations could
just as well be done in the path integral formulation. We have
chosen to use the first quantized operator version for all our
discussions, in the interest of uniformity.} The possibility that
the FQHE would be described by a CS theory was presaged by Girvin
(1987).

Since the idea of flux attachment is to cancel the applied field
on average, Zhang, Hansson, and Kivelson separate ${\bf a}_{cs}$
and $\rho$ into average and fluctuating parts:

\beq \nabla \times \langle {{\bf a}_{cs}}\rangle + \nabla \times
:{{\bf a}_{cs}}:= 2\pi (2s+1) n+ 2\pi (2s+1) :\rho : \eeq

This gives

\begin{eqnarray}
 H_{CS} &=& \sum_i {(   {\bf p} +e{\bf A}+\langle a_{cs}
\rangle+    :{\bf a}_{cs}:)_{i}^{2}      \over 2m} +  V \nonumber
\\ &=&\sum_i {(   \bp +    :{\bf a}_{cs}:)^{2}_{i}\over
2m}+V
\end{eqnarray}
upon using the fact that the flux due to $e\bA$ precisely cancels
that due to $\langle \ba_{cs} \rangle $. Bosons in zero field have
no degeneracy problem (assuming they have some repulsive
interactions) and allow one to describe much of the FQHE physics
in the familiar language of superfluids. For example, it has been
shown by computing response functions, that the superfluidity of
the bosons implies the FQHE for electrons (Zhang, Hansson, and
Kivelson , 1989), that the vortex in the superfluid is Laughlin's
quasi-hole (Lee and Zhang, 1991) The nature of the collective
modes has also been explored (Kane, Kivelson, Lee, and Zhang,
1991).

Neglecting $:{\bf a}_{cs}:$ (the mean-field approximation) and the
interaction, $\Psi_{CS}$ the wavefunction for bosons (in Eqn.
(\ref{eq-phase})) is just unity and that of electrons is \beq
\Psi_e = \prod_{i<j}\left({z_i-z_j \over |z_i-z_j|}\right)^{2s+1}
\cdot 1\eeq which is the phase of Laughlin's answer. Kane,
Kivelson, Lee, and Zhang (1991) showed that if long-wavelength
gaussian fluctuations are included, the full Laughlin wavefunction
is obtained at long distances. But the same fluctuations also
reduce the long-range order in the boson field down to the
power-law order found by Girvin and MacDonald (1987) who analyzed
a gauge-transformed version of the Laughlin
wavefunction.\footnote{This is to be expected given that the gauge
transformation of Girvin and MacDonald (1987) is the same CS
transformation of Eq. (\ref{eq-phase}).} Read (1989) then showed
that for Laughlin fractions, one could form an operator (which was
a composite of an electron and $2s+1$ vortices) which was neutral
and had true long-range order. The corresponding Landau-Ginzburg
theory for the order parameter was, however, very complicated.
Constraints of time and space prevent us from describing CB's any
further. We refer the reader to primary sources and excellent
reviews (Zhang, 1992, Karlhede, Kivelson, and Sondhi, 1993).

There are, however, some shortcomings in the CB approach. First, it is
restricted to Laughlin fractions. Next, since bosons have to be
interacting to be stable, a noninteracting starting point does not
exist. Finally, there is singular dependence on $m$ as $m\to 0$ and it
is hard to carry out quantitative computations.  Nonetheless, this
work has served as a paradigm for the hamiltonian approach.

\subsection{Composite fermions}

We turn now to composite fermions.  It was seen in the review of
Jain's work that by trading electrons for CF's which carry $2s$
flux quanta in opposition to the applied field, one could get a
fermionic system that on the average sees a field $\bA^*$ that is
just right to fill $p$ Landau levels.  Lopez and Fradkin
(1991,1992,1993,1998) were the first to accomplish this in the
hamiltonian approach. \footnote{ Note that we do not distinguish
between the functional integral formalism which they employed, and
the operator approach which we used in our work, and in the
interest of uniformity,  throughout this paper.} They traded the
electronic wavefunction $\Psi_{e}$ for $\Psi_{CS}$ defined as
follows:
\begin{eqnarray}
    \Psi_e &=&     \prod_{i<j} {(z_i - z_j) ^{2s}\over |z_i
-z_j|^{2s}}    \Psi_{\rm CS}\equiv   \exp
(2is\sum_{i<j}\phi_{ij})\ \Psi_{\rm CS}.\label{eq-phase2}\\ H_{CS}
&=& \sum_i {({\bf p}_{i} + e{\bf A} ({\bf r}_i)+     {\bf
a}_{cs}({\bf r}_i))^2 \over 2m} + V.
\end{eqnarray}

 Since $\Psi_{e}$ describes fermions, so does
$\Psi_{CS}$ since the phase factor is even under particle
interchange. The CS gauge field ${\bf a}_{cs}$ now obeys
\begin{eqnarray}
       \nabla \times {{\bf a}_{cs}} &=&
4\pi s \rho .\label{cscondition}
\end{eqnarray}
 Separating  ${\bf a}_{cs}$ and
$\rho$ into average and fluctuating
  parts
\begin{eqnarray}
 H_{CS} &=& \sum_i {(   {\bf p} +e{\bf A}+\langle \ba_{cs}
\rangle+    :{\bf a}_{cs}:)_{i}^{2}      \over 2m} +  V \nonumber
\\ &=&\sum_i {(   \Pib +    :{\bf a}_{cs}:)^{2}_{i}\over
2m}+V     \label{CSH}
\\      \Pib &=&      {\bf p} + e{\bf A}+\langle {\bf a}_{cs}\rangle \equiv {\bf p}+e{\bf A}^*
\label{pistar}\end{eqnarray}
\begin{eqnarray}
 \nabla \times
(e{\bf A}+\langle {\bf a}_{cs} \rangle) & =&      -eB+4\pi ns\\
 &=&-{eB\over 2ps+1}\equiv       -e
B^*      \ \ \ \ \ \  ({\bf A}^*= {{\bf A} \over 2ps+1})\\
l^*&=&{1\over \sqrt{eB^*}}=l\sqrt{2ps+1}
\end{eqnarray}

 The following important results emerged from the work of Lopez and
    Fradkin (1991,1992,1993):
\begin{itemize}
  \item If we ignore $:{\bf a}_{cs}:$ and $V$, the composite
fermions see ${1\over p}$ flux quanta each (since $2s+{1\over
p}\to {1\over p}$ under flux attachment) and have  a unique ground
state $\chi_p$ of $p$ filled LL's. Excitations are given by
pushing fermions into higher CF-LL's.

There is however a problem: If we excite a fermion from level $p$
to $p+1$, the energy
  cost (activation gap) of the particle-hole pair is
   $\Delta = eB^*/m$ plus corrections due to neglected terms.
   This divergent  dependence on $m$ flies in the face of the
   nonsingular $m \to 0$ limit we have
   argued must exist\footnote{Jain does
  not have this problem since  he does not use   $H_{CS}$ or $\chi_p$
  or its excitations directly.
 For him the CS picture is a step  towards  getting      electronic
  wavefunctions for the ground and excited states by attaching the
   Jastrow
  factor and projecting.
  The energy gap is computed  as the difference in
   $\langle V\rangle $
  between  the  ground and excited  electronic wavefunctions.}.
  We want $\Delta \simeq e^2/\varepsilon l$ in the Coulomb case.

\item  At the  mean-field level, the CF wavefunction $\chi_p$, transformed
   back to
  electrons is \beq
  \Psi_e =     \prod_{i<j}\left(  {z_i-z_j \over |z_i-z_j|}\right)^{2s}    \chi_p
  (z,\bar{z})\eeq

Lopez and Fradkin showed that fluctuations at one loop give the
square of the  wavefunction (at long distances) for Laughlin
fractions.
  (The factors
$|z_i-z_j|^{2s}$ in the denominator  of the mean-field wave
function
   are eliminated by fluctuations.)

  \item They calculated time-dependent density-density
  response functions in the
   Random Phase Approximation (RPA) (which will be explained in the next section).
  The cyclotron mode appears with the right position and residue.
  However, between the cyclotron mode ($eB/m$) and the LLL
  excitations,
   there are many spurious modes attributable to the ubiquitous
   presence of $m$ which prevents a clear separation of LLL
   and non-LLL energy scales. This is a problem common to all CS
   theories and was well appreciated by the authors.
\end{itemize}

The Lopez-Fradkin work paved the way for subsequent work to which
we now turn.

\section{Physics at and near    $\nu={1\over 2}$}

In the early days the CF, there was a widespread belief  that its
utility was confined to fractions with a robust gap: the
all-forgiving gap  allowed one to neglect, in a first
approximation,
 interactions, disorder and gauge field fluctuations.
 It was therefore quite a surprise to see that the CF
survived  even when $p \to \infty$ or $\nu \to 1/2s$, and the CF
cyclotron gap $eB^*/m = eB/(2ps+1)m$ approached zero.

\subsection{Physics at $\nu ={1\over 2}$}

Kalmeyer and Zhang (1992) were the first to discuss the case $\nu
=1/2$. Their work emphasized the following important point. One
may expect that the effect of disorder will be rather small since
the  electron donors lie at a respectable distance from the
electron gas itself. However any small charge inhomogeneity
induced by disorder is accompanied by a corresponding flux density
in a CS theory, and this can cause significant scattering.

 A landmark study of the region at and near $\nu =1/2$ was made by
 Halperin, Lee and Read (1993), (HLR) who seriously pursued the
 remarkable possibility that a Fermi liquid could be hiding deep in
 the FQH regime. HLR started with the following hamiltonian {\em at}
 $\nu =1/2:$ \beq H_{CS} = \sum_i {(\bp + :{\bf a}_{cs}:)^{2}_{i}\over
 2m}+V\eeq where $:\ba_{CS}:$ is related to charge {\em fluctuations}
 by the CS condition \beq \nabla \times :{\bf a}_{cs}:= 4\pi
 :\rho:.\eeq Note that CS theory dictates that $m$ must be the bare
 mass, since flux attachment by {\em minimal} coupling is a gauge
 transformation performed on the bare electron. If one wants to take
 the view that this is an effective low-energy theory with an
 effective mass $m^*$, one must also be prepared for the possibility
 that the coupling of the fermion to the gauge field is more
 complicated.

The cornerstone of HLR's work is the computation of the
electromagnetic response functions. Let us recall some general results
on this topic so we may better understand and appreciate their
work. In writing this description we were greatly aided by the reviews
of Halperin (in Das Sarma and Pinczuk, 1997) and Simon (in Heinonen,
1998).

If an external four-potential $eA_{\mu}^{ext}\bq , \omega)$ is
applied to the system, it will generate a four-current
$j_{\mu}(\bq , \omega)$ (whose components are the {\em number}
current and density) as per \beq j_{\mu}(\bq , \omega) = e K_{\mu
\nu}(\bq , \omega) A_{\nu}^{ext}(\bq , \omega).\label{JKA}\eeq

Linear response theory tells us that \beq K_{\mu \nu} =
\int_{-\infty}^{\infty} dt e^{i\omega t} i\theta (t) \langle
0|\left[ j_{\mu} (\bq , t), j_{\nu} (-\bq , 0)\right] |0\rangle
\eeq where $|0\rangle$ is the vacuum state.

For pedagogical purposes consider just $K_{00}$, the
density-density correlator.   In free-field theory, $K_{00}$ is
given by $K_{00}^{0}$, the particle-hole bubble, in which the
particle and hole created by one  $\rho$ are absorbed by the
other. The full theory of course requires us to include
interactions. Once again, for pedagogical purposes, let us begin
with the case with just the Coulomb interactions. The exact
$K_{00}$ is given by an infinite sum of Feynman graphs in which
the electrostatic propagator $v(q)=2\pi e^2/q$ appears in all
possible ways. It is possible to organize the sum as in Figure
(3). Each
 bubble $K_{00}^{v}$, called the {\em  Coulomb-irreducible
response}, has the property that it
 cannot be cut into two
disjoint pieces by snipping just one Coulomb propagator $v(q)$,
denoted by the  wiggly line  connecting the irreducible  bubbles.
Performing the geometric sum one finds \beq
K_{00}={K_{00}^{v}\over 1+v(q)K_{00}^{v}}={1\over \left[
K_{00}^{v}\right]^{-1}+v(q)} .\label{geom}\eeq

The response $K_{00}^{v}$ has the following significance. Consider
some conducting system with an applied potential $e\phi^{ext}$. An
electron inside the conductor feels in addition the potential
generated by the charges themselves, i.e., it feels the total
potential $e\phi^T= e\phi^{ext}-v(q)\rho (\omega , \bq)$, and
$K_{00}^{v}$ is the response to this total field. To verify this
let us write \beq \rho (\omega , \bq) = eK_{00}^{v} \left[
e\phi^{ext}-v(q)\rho (\omega , \bq)\right]\eeq solving which we
get \beq \rho (\omega , \bq) ={eK_{00}^{v}\over
1+v(q)K_{00}^{v}}\phi^{ext}.\eeq

Who cares about the total field? Answer: The voltmeter we use to
measure the drop in a wire. The voltmeter  responds to the total
electric field and not to the externally applied one that would
exist in the absence of the conductor. The longitudinal
conductivity $\sigma_{xx}$ can be related to ${K_{00}^{v}}$ as per
\beq \sigma_{xx}= {ej \over E^{T}}= e^2{\omega \over
iq^2}{K_{00}^{v}} \label{sigxx}\eeq where we have used the
continuity equation $qj=\omega \rho$ and $E=iq\phi$.\footnote{The
extra $e$ in front comes in because $\rho$ and $j$ refer to the
particle  density and not charge density.}

{\em In the RPA  $K_{00}^{v}$ is approximated by $ K_{00}^{0}$,
the free particle-hole bubble}. Thus RPA takes into account
Coulomb interaction via the internal field the particle themselves
generate, but ignores all  vertex and self-energy corrections
 coming from exchanging $v(q)$ inside the
 particle-hole bubble. Figure (3) shows the connection
between the irreducible and total responses.

\begin{figure}[t]
\begin{center}
\includegraphics[width=3in]{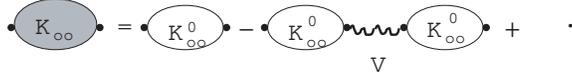}
\end{center}
\caption[]{\label{rpa} The diagrammatic relation between the full
 response function $K_{00}$ and the irreducible part $K_{00}^{v}$ for a problem with
 just Coulomb interactions, denoted by the wiggly line $V=v(q)$.
 In RPA $K_{00}^{v}$ is approximated by the particle-hole bubble
$K_{00}^{0}$. }
\end{figure}

The HLR work differs from this illustrative example in that  there
are now two types of gauge fields, the Coulombic $v(q)$ and the CS
gauge field, so that the wiggly line in Fig. (3) is described by a
$2\times 2$ matrix propagator. {\em The HLR version of RPA
consists of summing repeated bubbles irreducible with respect to
both propagators, and approximating the irreducible part by the
free-field response.} Thus the CS field and $v(q)$ are included
only to take into account the internal fields the induced charges
and currents produce.  In matrix notation it is still true that
\beq \left[ K_{ }\right]^{-1}=\left[K_{ }^{0}\right]^{-1}+U\eeq
where $U$ is a matrix propagator for the Coulombic $v(q)$ and the
CS gauge fields. Let us focus on just the $00$ element $K_{00}$,
which assumes the form
 \beq K_{00}= {1 \over v(q)+\left[
K_{00}^{0}\right]^{-1}+\left({2\pi \bar{\phi}\over
q}\right)^2K_{11}^{0} }\eeq where $K_{11}^{0}$ is the free-field
transverse-current response,  and $\bar{\phi}$ is the number of
flux quanta attached. Though $\bar{\phi}=2s$ in our notation, we
still use HLR's notation here to help readers who may wish to
consult that work for more details.

Let us now examine this expression in various regimes.

\subsubsection{Static compressibility}
The behavior of $K_{00}$ at $\omega =0, \ q\to 0$, gives the
static compressibility of the system. For $\omega =0$ and
$q<<k_F$, they find \beq K_{00}={1 \over v(q)+2\pi
(1+\bar{\phi}^2/6)/m} \eeq which shows that just as in a Fermi
liquid, the compressibility is nonzero if $v(q)$ is short ranged
and vanishes as $q$ if Coulombic. This vanishing  only means that
any applied external force  is unable to change the density
because the field inside the medium is strongly screened by the
medium.

\subsubsection{Cyclotron mode}

The poles of $K_{00}$ define the natural modes of oscillation of
the system, for at these poles we get a response with no applied
potential (see Eqn. (\ref{JKA})). If at any  $q$ there is a pole
at some $\omega (q)$, it means that there is a mode of  energy
$\omega (q)$. For example, in a 3D electron gas, a plasmon pole
appears at the plasma frequency as $q\to 0$ and then moves as a
function of $q$. At larger $q$, when decay into particle-hole
pairs is possible, the pole position acquires an imaginary part
denoting a finite lifetime.

 In the present case, as  $q\to 0$, at very high $\omega$, $K_{00}$
  has a pole at the
 cyclotron frequency $\omega_0 =4\pi n/m$ (where $m$ is the bare or
 band mass) with a residue in accordance with Kohn's theorem.

 \subsubsection{The overdamped mode.}
At very low $\omega$ and $q$,  one finds \beq
\left[K_{00}\right]^{-1}={2\pi \over m}(1+{\bar{\phi}^2\over 12})+
v(q) - i \left({2\pi \bar{\phi}\over q}\right)^2{2n\omega \over
qk_F}. \label{odm}\eeq

 The zero occurs at the {\em overdamped mode} with the dispersion
 relation \beq \omega \simeq i q^3v(q)\simeq q^2\eeq for the Coulomb
 case. If $v(q)$ is short ranged, the mode is sub-diffusive. The
 reason charge diffuses rather than moving ballistically (even in the
 clean system) is that the magnetic field makes the charge move
 perpendicular to the Coulomb force, which tries to even out the
 density gradient. The overdamped mode and the effects it produces lie
 at the heart of the HLR work.

 Note that if one first sends $m\to 0$, one would miss the
 over damped mode. Of course in a real system we can always find a $q$ such
 that $1/q$ dominates $1/m$. It is however a theoretical problem
 that we cannot send $m\to 0$ here. A way out  will
 be discussed later.

\subsubsection{Longitudinal conductivity}

Upon examining Eqn. (\ref{odm}), we can see that the
Coulomb-irreducible part (obtained by dropping $v$) is dominated
by the last term in the limit of small $q$ and $\omega$. Thus \beq
\sigma_{xx} = e^2{\omega \over iq^2}{K_{00}^{v}}={e^2 \over 8\pi }
{q\over k_F}. \label{sigmaxx} \eeq

The above result holds only for $q\gg 1/l_{m}$, where $l_{m}$ is the
mean free path for the CF. For  $q\ll 1/l_m$ \beq \sigma_{xx}={e^2
\over 4\pi k_Fl_{m}}.\label{sigxxdirty} \eeq  It is interesting to
see how one arrives at Eqn. (\ref{sigxxdirty}).

Let us define a CF conductivity by \beq e \bf{j} =
\sigma_{CF}\bf{E}^T\eeq where $\bf{E}^T$ is the total field, which is
the sum of the applied field and the internal field $\be$ generated by
the fermions themselves. To compute $\be$ imagine a particle current
$\bj$ and a unit length perpendicular to it. In one second $j$
particles cross it carrying with them $2j$ flux quanta of CS
flux. However the CF makes no distinction between real flux and CS
flux since they enter $H_{CS}$ only via their sum. It will therefore
sense a CS electric field $\be = {4\pi \over e}\hat{\bz}\times
\bj$. Thus \beq e \bj = \sigma_{CF}\left[ \bE + {4\pi \over
e}\hat{\bz}\times \bj\right].\eeq The electron resistivity tensor
$\rho$ (with components $\rho_{\alpha\beta}$ and not to be confused
with the density operator) defined by \beq E_{\alpha} = \rho_{\alpha\beta} \ (ej_{\beta})\eeq is then \beq
\rho = \rho_{CS}+\rho_{CF}\label{addingrhos}\eeq where \beq \rho_{CS}
= {4\pi \over e^2} \left[ \begin{array}{cc} 0 & 1 \\ -1 & 0 \
 \end{array}\right] \label{rhoadd}
 \eeq

At mean-field level  CF's do  not see a field of any kind. The
effect
 of disorder is to produce
 \beq
 \sigma_{CF, \ xx}=\sigma_{CF, \ yy}={ne^2\tau \over m}={ne^2
 l_{m}\over  k_F}={ k_{F}^{2}e^2l_{m}\over 4\pi k_F}={e^2
 k_Fl_{m}\over 4\pi }\eeq
 where $\tau$ is the elastic scattering time and $l_m$ is the mean
 free path.
 If we now use Eqn. (\ref{addingrhos}) and assume $k_Fl_{m}\gg 1$
we obtain Eqn. (\ref{sigxxdirty}).

{\em Note that the resistivity of the  electron is the sum of the
resistivities of the CF and the CS term.} If we move off to a
general Jain fraction and ignore disorder, we will have \beq
\rho_{CF}={2\pi \over e^2} \left[ \begin{array}{cc}0 & {1\over p}
\\
-{1\over p} & 0 \end{array} \right] \eeq since the CF's fill $p$
LL's of their own. When the resistivity matrices are added, one
obtains the correct Hall resistivity for electrons at $\nu =
p/(2p+1)$.

\subsubsection{Surface acoustic waves (SAW)}

 When a surface  acoustic
wave is coupled to the electronic system, it is predicted to
undergo a velocity shift and an attenuation described by \beq
{\delta v_s \over v_s}-{i\kappa \over q}={\alpha^2/2 \over
1+i\sigma_{xx}(q)/\sigma_m} \eeq where $ \alpha$ is a
piezoelectric constant, $ v_s$ is the sound velocity, $\kappa $
describes the attenuation, $\sigma_{xx}(q)=\sigma_{xx}(q,\omega =
qv_s)$ where $v_s$ is the sound velocity, and $\sigma_m ={
v_s\varepsilon \over 2\pi}$. Theory fits the experiments of
Willett {\em et al} (1990,1996) (for a review see Willett (1997))
with a $\sigma_m$ that is about five times larger than expected.
The reader is strongly urged to consider  HLR and the reviews for
more details.

\subsubsection{Mass divergences}

HLR (1993)  predicted a divergence in the
effective mass $m^*$ at the Fermi surface arising  from the
fermion
 self-energy diagram involving the emission and absorption of the
overdamped  mode. For the Coulomb case one has \beq m^*(\omega)
\simeq \ln \omega . \eeq (Shorter range interaction lead to more
violent divergences.)  Assuming this mass can be used near $\nu
=1/2$ one expects that the gaps will be given by \beq
E_{\nu}={eB^*\over m^*}\eeq where $m^*$ is self-consistently
defined by  $m^*(\omega = E_{\nu}).$ For $\nu =p/(2p+1)$, this
implies, as $p \to \infty$, that  $E_p\simeq 1/(p\ln p)$. The log
divergent effective mass will also imply a specific heat
$C(T)\simeq T \ln T.$ It has not been possible to confirm these
logarithms in numerical or experimental work.

Since  the nature of the mass divergence depends on the range of
the potential, it cannot be reproduced by trial wavefunctions,
which make no explicit reference to the potential, in particular
its range.

 It has been shown that the mass divergences do no affect bosonic (e.g.,
density-density) correlations\footnote{Kim, Furusaki, Wen, and Lee
(1994), Kim, Lee, Wen, and Stamp (1994), Kim, Lee, and Wen (1995),
Stern and Halperin (1995).}.

\subsection{Physics near $\nu =1/2$}

The HLR (1993)  predictions transcend $\nu =1/2$ and describe its
immediate vicinity. The key idea is that in this region, it is
useful to think of the CF as a particle seeing a weak effective
magnetic field $B^*=B/(2p+1)$. This means that it will describe a
cyclotron orbit of radius \beq R^*= {\hbar k_F \over eB^*} \eeq
with $k_F=\sqrt{4\pi n}$ (given that spin is fully polarized).
Note that this result is independent of the fermion mass, whose
treatment is quite tricky.

The SAW results of  Willett {\em et al} (1990,1996) found that
away from $\nu =1/2$, there was a resonance in the velocity shift
when the wavelength of the surface acoustic wave coincided with
$2R^*$.

Confirmation of the $R^*$  concept  was also found in the
experiments of Kang {\em et al} (1993), Goldman, Su, and Jain
(1994), and Smet {\em et al} (1996). To visualize  the Goldman
{\em et al} experiment, imagine a semi-infinite system in the
upper - half plane. If a current is introduced at the origin up
the $y$-axis, it should bend and return to the $x$-axis at
$x=2R^*$, after completing one semi-circle. It would then bounce
off and start the next semi-circle. It follows that if a return
path is provided on the $x$-axis, the maximum current will flow if
the drain is located an integral multiple of $2R^*$ from the
source.  Or if a return location is held fixed, and $B$ is varied,
a maximum is expected and (found)    whenever the spacing between
source and drain is a multiple of $2R^*$.

 Kang {\em et al} built an
array of antidots, where each antidot is a region where electrons
are absent.  It was found that the sheet conductance had peaks
when the antidot lattice constant equaled $2R^*$. While a detailed
formula for conductance is not known in this context, the
correlation is very suggestive and is what was seen with ordinary
electrons in a weak field.
 The fact  that CF's, which entered the theory as a mathematical
 devise,
 manifest themselves so clearly in transport is a stunning affirmation
  of the
 theoretical framework.

\subsection{ HLR - Room for improvement?}

Despite its many remarkable and experimentally confirmed
predictions, the HLR theory leaves room for improvement mainly
because of the CS approach on which it is based.

One such area has  to do with the dependence on bare mass $m$. We
have argued that at the starting point $H_{CS}$ must contain the
bare mass since flux attachment by minimal coupling the CS field
is an exact transformation done on electrons. This choice
certainly helps to get the cyclotron frequency in accord with
Kohn's theorem. However in our extraction of various low-energy
long wavelength quantities like the over-damped mode from Eqn.
(\ref{odm}), we had to assume that $1/q$ dominates over $1/m$.
While this is certainly valid for realistic and fixed values of
$m$, one would like to be able to extract what is evidently the
correct physics even if $m=0$ is imposed first.

 Simon and Halperin (1993) and Simon, Stern, and Halperin (1996),
proposed a way out. They assume, as in Landau theory, that $H_{CS}$ is
an effective theory with an effective $m^*$. Kohn's theorem and the
over damped mode can both be salvaged if a suitable Landau parameter
$F_1$ is introduced. This is reasonable since Kohn's theorem relies on
Galilean invariance and Landau's theory uses this principle to related
$m$, $m^*$ and $F_1$. However this leaves the origin of $m^*$ inside a
black box. \footnote{ Why do we view this inability to calculate $m^*$
as a weakness, when in Landau theory it is simply accepted as a fact
of life? As we shall see, in the FQHE we can go a long way towards
computing $m^*$ from the interactions by using the EHT.} It is also
not clear why in the effective theory the fermion the CS field should
be minimally coupled.

In the case of Coulomb interactions, one can also argue that the
bare mass is swamped by the renormalized  $m^* \simeq \ln \omega$
generated by the exchange of  the over-damped mode. Again it would
be nice to be able to follow in detail the separation of the low
energy physics controlled by an $m^*$ generated by interactions
and high-energy physics controlled by $m$.

Another shortcoming is that there is  no evidence of the neutral
fermion one expects at $\nu =1/2$ , and more generally a fermion
of charge $e^*=e/(2ps+1)$ at other Jain fractions. This was to be
expected since the CF in Halperin {\em et al} and in earlier CS
work of Lopez and Fradkin was an electron bound to two flux tubes
which carry no charge.

Also missing was the effective magnetic moment $\mu^*= e/2m$ that
Simon, Stern and Halperin (1996) would later argue must go with
each CF.

In short, the CF that appears in the CS theory does not have, in
obvious form, the right charge or energy scale of the ultimate
quasiparticle. While the CS transformation is exact and can yield
all these in principle,  they  are not manifest in the RPA.

Finally, Lee, Krotov, Gan, and Kivelson (1997, 1998) have raised the
following issue. Suppose we push all LL's up to infinity except the
LLL and assume particle-hole symmetric disorder at $\nu=\half$. Then
it can be verified that a nonzero $\sigma_{xx}$ for electrons implies
for CS-fermion a $\sigma_{xy}^{CS}=-e^2/(4\pi)$. However, both the
mean-field approximation and the RPA corrections to it, give
$\sigma_{xy}^{CS} =0$. We are not aware of a resolution of this issue,
arising from the fact that resistivities add.

\section{Hamiltonian theory II-  Extended  Hamiltonian Theory (EHT) \label{extended}}
We now turn to our extension of the CS formalism\footnote{Murthy and
Shankar (2002), Shankar and Murthy (1997), Murthy and Shankar
(1998a,b, 1999, 2002), Murthy (1999, 2000a,c, 2001b,c), Shankar (1999,
2000, 2001), Murthy, Park, Shankar, and Jain (1998).}.  In this
extension, a CF with all the known properties can be made manifest and
a variety of LLL quantities computed with no $1/m$ singularities.  But
the computation of certain low energy long wavelength quantities (like
the compressibility), straightforward in the CS approach, become
extremely delicate. Hopefully we will impart to the reader a sense of
which approach to use for what purpose.

We want to eschew the historical route and furnish an axiomatic
presentation of our work with the minimum of preamble, setting the
stage for exact or approximate calculations as rapidly as
possible. However, to establish the context, we will begin with a
brief sketch of our earlier work so as to take some of the mystery
out of the end product and provide some degree of motivation.

Let us recall the work of Bohm and Pines (Bohm and Pines, 1953) on
the electron gas in three dimensions. At small $q$ the spectrum
consists of particle-hole pairs at low energies and the plasmon at
high energies. The particle and hole are part of the original
hamiltonian while the plasmon comes from summing an infinite class
of diagrams in the density-density response. It is not an
independent entity, even though it is a sharply defined excitation
that can be experimentally produced and detected. Bohm and Pines
showed that by introducing extra canonical oscillator degrees of
freedom at small $q$, one could describe plasmons as independent
objects in an enlarged Hilbert space. In order to prevent double
counting, they imposed constraints on state vectors of the form
\beq \barchi |\mbox{physical state}\rangle =0\eeq with one
$\barchi $ for each $\bq$ at which a plasma mode was introduced as
an independent canonical oscillator. The plasmons, initially
coupled to the fermions, were approximately decoupled, leaving
behind fermions with a renormalized mass, and constraints that
essentially froze out any putative plasmons with small $q$. Such a
simple description of plasmons, or isolation of high-energy
physics, would have been impossible within the confines of the
electronic Hilbert space, wherein plasmons are complicated
collective excitations of the electrons themselves.

What we originally did for the FQHE (Shankar and Murthy, 1997, Murthy
and Shankar, 1998a) was similar in many ways. We enlarged the Hilbert
space to include at each $\bq$, a new set of independent canonically
conjugate variables - a transverse vector field $\ba (\bq)$ and a
longitudinal vector field $\bP (\bq)$. Using these it was possible, by
a unitary transformation, to get rid of the dependent CS vector
potential $\ba_{CS}$. While all the operators in our hamiltonian were
now independent, the physical sector was defined by the (CS) condition
\beq (\nabla \times \ba -4\pi s \rho)|\mbox{physical state}\rangle
=0. \eeq The conjugate variables $\ba$ and $\bP$ formed oscillators
near, but not exactly at, the cyclotron frequency. They were coupled
to fermions. We found a way to decouple the oscillators in the limit
$ql\to 0$ by a second unitary transformation. The decoupled
oscillators now complied with Kohn's theorem for pole position and
residue.  The formula for $\barro$, the electronic charge density
projected to the LLL, was derived to order $ql$, as was the constraint
$\barchi $ which now acted only within the particle sector. At this
point a guess was made (Shankar 1999) to extend the small $ql$ answer
to all orders by exponentiation of leading-order terms. The final
theory of the LLL sector took the following form:

\begin{eqnarray}
 \barH &=&
{1\over 2} \sum_{\bq }\bar{\bar{\rho}}(\bq) v(q) e^{-q^2l^2/2}
\bar{\bar{\rho}}(-\bq)\label{1}\\ \left[ \barH \ , \barchi
\right]&=&0\label{2}\\ 0&=& \barchi |\mbox{physical state }
\rangle \label{3}
\end{eqnarray}

The hamiltonian $\barH$ is just the potential energy $V$ projected
to the LLL, written in terms of the projected  electron charge
density $\barro$
\begin{eqnarray} \bar{\bar{\rho}}(\bq) &=&  \sum_j \exp (-i{\bf
q} \! \cdot \bR_{ej})\\ \bR_e  &=& {\bf r} -  {l^2\over
1+c}\hat{\bf z}\times {\bf \Pi} \label{re}
\end{eqnarray}
where we define a very important and frequently occurring
variable: \beq c^2 = {2ps\over 2ps+1} .\eeq

From the commutation relations   \beq
 \left[ \bR_{ex} \ ,
\ \bR_{ey} \right] ={-il^2}\eeq it is clear that $\bR_e$ is just
the electron guiding center coordinate {\em but now expressed in
terms of CF variables} $\br$ and $\bPi =\bp +e\bA^*$.

The physical states are annihilated by  the constraint
\begin{eqnarray} \bar{\bar{\chi}}(\bq) &=& \sum_j \exp (-i{\bf q}
\cdot \bR_{vj})\\ \bR_v &=&  {\bf r} + {l^2\over c(1+c)}\hat{\bf
z}\times {\bf \Pi}
\end{eqnarray}
The  {\em pseudovortex  coordinate} $\bR_v$ describes and object
of  charge $-c^2 =-2ps/(2ps+1)$:
\begin{eqnarray}  \left[ \bR_{vx} \ ,
\ \bR_{vy} \right] &=&{il^2\over c^2} \end{eqnarray} and commutes
with $\bR_e$: \beq \left[ \bR_e \ , \ \bR_v \right] =0 . \eeq

Thus the constraints commute with $\barH$, and    form an algebra:
\begin{eqnarray}
\left[ \bar{\bar{\chi}}(\bq) \ ,  \bar{\bar{\chi}}(\bq') \right]
&=& -2i \sin \left[ {l_{}^{2} ({\bf \bq\times \bq'}) \over
2c^2}\right] \bar{\bar{\chi}} (\bq +\bq').
 \end{eqnarray}
 The problem is just like Yang-Mills theory.

The expressions for $\bR_e$ and $\bR_v$ in terms of $\br$ and
$\Pib$ which were already encoded in the small-q theory, jumped
out upon exponentiation. They, together with the constraints,  lie
at the heart of our approach.

{\em We will now show how one can get to the final result,
Equations (\ref{1}- \ref{3}), if we simply make a certain
enlargement of the electronic Hilbert space and follow it with a
change of variables.} Any reader who wants to know more about what
makes us introduce these variables should
  consult our earlier work. Those who just want to use the
 results can go ahead.

\subsection{The axiomatic introduction to the EHT}
 Let
us   begin afresh with  the
 primordial hamiltonian in terms of electronic variables
(which carry the subscript $e$ to distinguish them from other
coordinates to be introduced): \beq H = \sum_j
{\etab_{ei}^{2}\over 2ml^4}+{1\over 2}\sum_{i,j,\bq }v(q)e^{ i\bq
\cdot (\br_{ei} - \br_{ej})}\equiv H_0 +V. \eeq

This hamiltonian contains complete information about the problem.
It can be used to study LL-mixing and the computation of Hall current
which requires higher LL's in an essential way. These topics will
be discussed in due course. We first focus on the main challenge
of hamiltonian theories:  extracting the $m$-independent physics
of the LLL. As discussed in Section (\ref{prels}), projecting  to
the LLL as such is no problem: one drops the first term and makes
the replacement $\br \to \bR$ in the density operator
 and $v(q) \to
            v(q) e^{-q^2l^2/2}$.
The catch is that the projected hamiltonian lives in  the highly
degenerate LLL, frustrating both  perturbation theory and the
Hartree-Fock approximation.

In the CS approach one resorts to flux attachment to beat the
degeneracy of the kinetic term, but that can be done only in the
full electronic Hilbert space. Consequently $m$ gets into
everything and the low (LLL) and high-energy sectors get
hopelessly entangled. What we really want to do is work within the
LLL and attach flux tubes, which in the LLL translates into
vortices, by analyticity. However when we say vortices, we do not
mean zeros of the wave function, for such a thing does not exist
as a degree of freedom within the hamiltonian and as we have seen,
there are not enough of them in the wavefunction to go around
anyway. {\em Instead we mean by a vortex some object which has the
charge of the $2s$-fold vortex and corresponds to an excitation
that can be created by inserting $2s$ flux quanta into the Hall
system.} Since such an object does not exist in the original
Hilbert space (as an independent entity) we enlarge it to make
room for this entity, which we call the {\em pseudovortex}, the
{\it vortex} to emphasize its similarity to the vortices in the
wavefunctions, and the {\it pseudo} to emphasize its differences.

The enlargement of Hilbert space can be explained  in terms of
just one electron, with cyclotron and guiding center coordinates $
\etab_e$ and $\bR_e$. Let us temporarily focus on just the LLL
physics and ignore $\etab_e$ which  does not participate in the
change of variables, and will be reinstated subsequently.

First we  introduce an extra guiding center coordinate $\bR_v$
(the pseudovortex), defining it algebraically by its commutation
relations which represent   a charge $-c^2$ \beq \left[ R_{vx} ,
R_{v y} \right]= {il^2\over c^2}.\eeq {\em Next we   combine
$\bR_e$ and $\bR_v$ to form the CF space}. (Note that it takes two
canonical pairs to make one regular fermion in two dimensions.)
This CF space can be defined in terms of either the position $\br$
and velocity  $\Pib$ of the CF, or in terms of its cyclotron  and
guiding center coordinates, $\etab$ and $\bR$. {\bf  Note that
henceforth  variables carrying not indentifying subscript will
 refer to the CF}. The CF coordinates  $\etab$ and $\bR$ obey the
  commutation rules of
 the cyclotron and guiding center coordinates of  an object of charge $e^* =
1-c^2 = 1/(2ps+1)$:

\begin{eqnarray}
\left[ \eta_x , \eta_y \right]&=& il^{*2} = {il^2\over 1-c^2}\\ \left[
R_x , R_y \right]&=& -il^{*2}.
\end{eqnarray}

The  rule for forming the CF from $\bR_e$ and $\bR_v$   is the
following:

\begin{eqnarray}
\bR &=& {\bR_e -c^2 \bR_v \over 1-c^2} \\ \etab &=& {c \over
1-c^2} (\bR_v-\bR_e)
\end{eqnarray}
The first equation says  that the  CF guiding center is
the weighted sum of its parts. The second can be found by
demanding that $\etab$ be linear in $\bR_e$ and $\bR_v$, commute
with $\bR$,  and have an overall scale that produces the right
commutator.

The inverse transformation is
\begin{eqnarray}
\bR_e &=& \bR + \etab c \\ \bR_v &=& \bR +\etab /c
\end{eqnarray}
In terms of $\br$ and $\bPi$, the CF coordinate and velocity
operators \begin{eqnarray}
 \bR_e &=&    {\bf r} -{l^2\over (1+c)}\hat{\bf z}\times {\bf \Pi
},    \\
 \bR_v &=&{\bf r} +{l^2\over c(1+c)}\hat{\bf z}\times {\bf \Pi
}
\end{eqnarray}
which are just  the expressions encountered in the brief
historical review.

 Ignoring the zero point energy, here is where we stand in the LLL
 sector:
\begin{eqnarray} \bar{\bar{H}}_{} &=&\half \sum_{i,j,\bq }
v(q)\ e^{-q^2l^2/2} \ e^{i\bq \cdot (\bR_{ei} - \bR_{ej})}\\ &=&
\half \sum_{i,j,\bq } v(q)\ e^{-q^2l^2/2} \exp \left({i\bq \cdot
\left[ (\bR_{i} - \bR_{j})+c(\etab_i-\etab_j)\right]}\right)
\end{eqnarray}

While it is true that we have managed to get rid of $m$ and
isolate the LLL cleanly, the reader may ask what we have gained,
since {\em algebraically} the problem is the same as in electronic
coordinates. {\it The answer is that now there is a natural
nondegenerate HF ground state in the extended space}. This is
because the HF hamiltonian is now  written in terms of CF
operators $\bR$ and $\etab$ and  the particle density is just
right to fill exactly  $p$-filled CF-LL's, i.e., $\chi_p$ is the
ground state.  The proof of its HF nature is found in Appendix
\ref{HFProof}. This  key step  opens up all the usual
approximation schemes.

Depending on what we want to compute, there are two distinct
 schemes. Both rely on the nondegenerate HF ground state,
 and both acknowledge a huge symmetry group of
$\barH$, which comes from the following fact: $\barH$, as embedded
in the CF space, does not depend on $\bR_v=\bR + \etab /c$, the
pseudovortex coordinate. Equivalently, it depends on $\bR$ and
$\etab$ only through the combination $\bR_e=\bR +c \etab $. The
other (commuting) combination $\bR_v=\bR+ \etab /c$ can be used to
define a {\em pseudovortex density } \beq \barchi = \sum_j
e^{-i\bq \cdot \bR_v}\eeq which commutes with $\barH$:
\begin{eqnarray} \left[ \barH \ , \barchi \right] &=&0.
\end{eqnarray}

Let us understand this symmetry.  If we view $\barH$ as a function
of $\etab $ and $ \bR $, or $\br$ and $\bPi$, its eigenfunctions
will depend on two coordinates, which can be chosen to be one
component each of $\etab$ and $\bR$ or just $x$ and $ y$. If we
view $H$ as a function of $\bR_e$ and $\bR_v$, it depends only on
$\bR_e$. The energy eigenfunctions will be of the form \beq \psi
(z_e,z_v) = \psi_e(z_e)\psi_v(z_v) \eeq {\em where $\psi_v(z_v)$
is arbitrary} since nothing in $\barH$ determines it.This
degeneracy is of the same type as that of the noninteracting
electron hamiltonian  which  depends on $\etab_e$ but not $\bR_e$.
However, since $\bR_v$ is unphysical, and  all physical
observables depend only on $\bR_e$, this is a gauge symmetry.

We deal with the gauge symmetry as usual,  by selecting a
representative from each orbit, which in this case means
eliminating the degeneracy due to the arbitrary function $
\psi(z_v)$ that tags along for the ride. Let us make a particular
choice $\psi_{v0}(z_v)$. {\em All we require is that it be
translationally invariant so that $\langle \barchi \rangle =0$ in
this state. } The gauge-fixed Hilbert space now consists of
functions of the form $\psi_e(z_e)\psi_{v0}(z_v)$. In this sector
$\barchi =0$ in the weak sense: any Green's function involving a
string of $\barchi$'s will vanish since (i) $\bar{\bar{\chi}}(\bq
, t)=\bar{\bar{\chi}}(\bq , 0)\equiv \barchi, $(ii) $\langle
\barchi \rangle =0$, \footnote{ An exception occurs if the string
contains only $\bar{\bar{\chi}}(0)=N$. This does not affect what
we plan to do.} in the one-dimensional space spanned by
$\psi_{v0}(z_v)$. (Imagine inserting the projector to this state
between  any two $\barchi$'s in the Green';s function.)

In the path integral language we can make $\barchi$ vanish {\em
weakly}, that is to say, vanish whenever it appears in a Green's
function, by a method similar to what is done in gauge theories.
Imagine writing a path integral in the full CF space and then
inserting a delta function imposing $\barchi =0$ for all $\bq$ at
any one time. \footnote{ Since $\barchi$ involves $\br$ and $\bp$,
this will have to be done in the phase space path integral.} Since
$\barchi$ does not change with time, this restriction will hold
automatically for all times and specify the fate of $\barchi$ the
way a gauge fixing terms does for the longitudinal degrees of
freedom.

The theory is thus  defined (in schematic form) by the equations
\begin{eqnarray}
\barH  &=&\half \sum_{\bq } v(q)\ e^{-q^2l^2/2} \barro
\bar{\bar{\rho}}(-\bq)\label{cons1}\\
\left[ \barH \ , \barchi \right] &=&0\label{cons2}\\
\barchi &\simeq & 0
\end{eqnarray}
where $\simeq 0$ means vanish weakly.

Now we turn to the two approximate ways of dealing with this
problem: the conserving approximation and the shortcut.

\subsection{ The conserving approximation}
Given a hamiltonian, a reasonable first approximation to try is
Hartree-Fock. It is shown in  Appendix \ref{HFProof} that CF-LL
states are Hartree-Fock states of $\barH$. How good is this
Hartree-Fock approximation likely to be? For example, how will we
fare  if we compute the activation or {\em transport gap} in a
fully polarized sample as per:
 \beq \Delta = \langle {\bf p} + PH | H|{\bf
p}+ PH \rangle -\langle {\bf p} | H| {\bf p} \rangle \eeq where
$|{\bf p}\rangle$ stands for the Hartree-Fock ground state with
$p$-filled LL's and  $PH$ stands for a widely separated
particle-hole pair?

There are at least two good reasons to expect that this naive HF
result will require fairly strong  corrections. First,  if we
compute the matrix element of the projected electron density
between any two HF states, the answer will be  linear in $q$,
whereas in the exact theory, and within the LLL, it must go as
$q^2$ as per Kohn's theorem. To see this note that \beq e^{-i\bq
\cdot \bR_e}= 1-i\bq \cdot (\bR + c \etab) + {\cal O}(q^2) .\eeq
While $\bR$ has no transition matrix elements between different CF
Landau levels, $\etab$ does. The second problem is that as $ql \to
0$, the projected electronic density (which reduces to $\sum_j
\exp (-i\bq \cdot \br_j)$) has unit contribution from each CF
while we would like it to be $e^*$. Evidently the HF result will
receive strong corrections that will renormalize these quantities
till they are in line with these expectations.

These shortcomings are to be expected since  the HF solution does
not obey the constraint, or equivalently, does not  factorize into
the form $\psi = \psi_e (z_e) \psi_v (z_v) $.

The conserving approximation (Anderson, 1958a,1958b, Rickayzen, 1958,
Nambu, 1960, Baym and Kadanoff, 1961) is a sophisticated procedure for
improving the HF state with additional diagrammatic corrections so
that $\barchi \simeq 0$ in Green's functions. For $\nu =1/2$ Read
(1998) showed that this procedure restores Kohn's theorem, and reveals
a dipolar structure for density-density correlations.  We shall say
more about this in connection with the compressibility paradox.

 Now we discuss the other approximation, the shortcut.

\subsection{The shortcut: The preferred charge and hamiltonian}
Consider the exact solution to the gauge fixed problem.
 Suppose, in the hamiltonian and elsewhere,   we  replace $\barro$ by the {\em
preferred combination} \beq  \barrop = \baro -c^2 \bar{\bar{\chi
}} .\eeq This makes no difference (to the computation of anything
physical) in an exact calculation, since $\bar{\bar{\chi }}$ is
essentially zero.

 However, in the HF
approximation it makes a big difference if we start with the
hamiltonian written in terms of   $\barrop$.
 To see why,
consider its expansion  in powers of $ql$: \beq \bar{\bar{\rho}}^p
= \sum_j e^{-i{\bf q \cdot r}_j}\!\! \left(\! {1 \over 2ps+1} \!
-     {i l_{}^{2} } \bq\times {\bf \Pi}_{j}  \! +{0} \cdot \left(
\bq \times {\bf \Pi}_{j} \right)^2 \! + \cdots \right)
.\label{rostar1} \eeq

\begin{itemize}
\item
The   transition matrix elements are of order $q^2$ between HF
states because  coefficient of $\bq$ is proportional to the CF
guiding center coordinate $ \br - l^{*2}\hat{\bz}\times \bPi$ with
no admixture of the CF cyclotron coordinate. This is more
transparent if we use $\bR$ and $\etab$ to write \begin{eqnarray}
\barrop &=& (1 -i \bq \cdot (\bR + c \etab) + ..)-c^2 (1 -i \bq
\cdot (\bR +  \etab /c +...)\\ &=& (1-c^2)(1 - \bq \cdot \bR +
{\cal O} \ (q^2).\end{eqnarray}
\item With no further fixing, we see that the electronic charge
 density
associated with $\barrop$ is now $1-c^2 =e^*$.
\item We see from Eqn. (\ref{rostar1}) that when $\nu =1/2$, the preferred density
couples to an external electric field like a dipole of size $l^2
\hat{\bz}\times \bp $. \end{itemize} Thus there is not {\em
apriori} need for strong corrections, at least in the long-wavelength limit.

The hamiltonian we  work with  \beq \barH^p = {1\over 2} \sum_{\bq
}\barrop v(q) e^{-q^2l^2/2} \bar{\bar{\rho}}^p(-\bq)\eeq subsumes
a lot of the right low-energy physics and CF properties. Unlike
$\barH$ which commutes with $\barchi $, $\barH^p$  is {\em weakly
gauge invariant}, that is \beq \left[ H(\bar{\bar{\rho}}^p) \ ,
\barchi \right] \simeq 0 \eeq where the $\simeq 0 $ symbol means
that it vanishes in the subspace obeying $\barchi =0$. Thus
neither $H(\bar{\bar{\rho}}^p)$ nor $\bar{\bar{\rho}}^p$ will mix
physical and unphysical states.

The      significance of $H(\bar{\bar{\rho}}^p)$   is the
following. If the constraint      $\bar{\bar{\chi}} =0$ is imposed
{\em exactly},       there are many equivalent hamiltonians
depending on how $\bar{\bar{\chi}}$ is insinuated into it.
However, in the HF {\em approximation}, these are not equivalent
and $H(\bar{\bar{\rho}}^p)$ best approximates, between HF states
and at long wavelengths, the true hamiltonian between true
eigenstates. In contrast      to a variational calculation where
one searches among states      for an optimal one, here the HF
states are the same for a      class of hamiltonians (where
$\bar{\bar{\chi}}$ is      introduced into $H$      in any
rotationally invariant form),  and we seek the best hamiltonian:
$H(\bar{\bar{\rho}}^p)$ encodes the fact that every electron is
accompanied by a correlation hole of some sort which leads to a
certain $e^*$, $d^*$ and obeys Kohn's theorem
 ($q^2$ matrix element for the LLL projected charge density).

Note that when we use the preferred charge and hamiltonian we will
make no further reference to constraints, and simply carry out the
Hartree-Fock approximation. This is based on the expectation that even
if we found some way to include the effect of constraints, it will
make no difference in the small $ql$ region. This is because the
leading renormalization of $e$ to $e^*$ and suppression of ${\cal} \
q$ matrix elements down to ${\cal} \ q^2$ that are achieved by the
conserving approximation by summing ladder diagrams following Read
(1998) or using the time-dependent HF (Murthy, 2001a, Section VI) are
built in here.

{\it It is in this approximate, operator sense, where we use
$\bar{\bar{\rho}}^p$ in place of $\bar{\bar{\rho}}$  that the
binding of electrons and pseudovortices to form CF's is realized
in the Hamiltonian theory}. Since the pseudovortices have
coordinates that are independent of the electrons, there is no
double-counting here. The problems that we encountered with
vortices in the wavefunction approach for non-Laughlin fractions
(with the limiting case being the dipole picture of $\nu=\half$)
are also absent since we are not talking about those vortices.
Since the pseudovortices {\em per electron} are independent of
electrons, their number does not change when $\nu$ changes (though
their charge, tied to $c^2=2s\nu$, does.) Whereas
antisymmetrization fragmented the vortices in the Jastrow factor
into ordinary parametric zeros (except for the Pauli zero), it
does nothing to the pseudovortices. Antisymmetrization is
accomplished by simply writing the operator $\barro$ in terms of
second quantized (composite) fermion operators.

 The reader will recall that any {\em simple} picture of
quasiparticles, whether it be in Landau's Fermi liquid theory, or
in BCS theory, is best captured by approximate and not exact
descriptions. The quasiparticles are all caricatures of some exact
reality and therein lies their utility. Similarly the CF in our
extended formalism appears only in the HF approximation to
$\barH^p$. Recall that we brought in the coordinate $\bR_v$ to
become the electron's partner in forming the CF.  However $\bR_v$
was cyclic in the exact hamiltonian $\barH$. {\em Thus the exact
dynamics never demanded that $\bR_v$ be bound to $\bR_e$ or even
be anywhere near $\bR_e$. } However, in the HF approximation,
since we wanted the right charge and transition matrix elements of
the density operator (Kohn's theorem) to be manifest, we needed to
replace $\barro$ by $\barrop$, and trade $\barH$ for $\barH^p$,
the preferred hamiltonian. In $\barH^p$, $\bR_v$ is coupled to
$\bR_e$. The HF approximation and this coupling go hand in hand.
The exact eigenfunctions of the original $\barH$ are factorized in
the analytic coordinates $z_e$ and $z_v$ and presumably reproduce
the electronic correlations of the FQHE states.  On the other
hand, in the HF approximation to $\barH^p$, the wavefunctions
(e.g., $p$-filled LL's) mix up $ z_e$ and $z_v$, and $\barH^p$,
the preferred hamiltonian, dynamically couples $\bR_e$ and
$\bR_v$. The net result is that, at least at long wavelengths,
these two wrongs make it right and mimic what happens in the exact
solution.

Another advantage of $\barH^p$ is that it gives an approximate
formula for $m^*$ originating entirely from interactions. This is
best seen at  $\nu =1/2$.  When we square $\bar{\bar{\rho}}^p$
(Eqn. (\ref{rostar1}), we get a double sum over particles whose
diagonal part is the one particle (free-field) term: \beq
H^{0}_{\nu ={1\over 2}}=2\sum_j \int {d^2q\over 4\pi^2} \sin^2
\left[{{\bf q \times k }_j l^2\over 2}\right] {v}(q)
e^{-q^2l^2/2}. \label{freeh} \eeq

This is not a hamiltonian of the form $k^2/2m^*$. However if the
potential is peaked at very small $q$, we can expand the sine and
read off an approximate $1/m^*$

\beq {1\over m^*}= \int {qdq d \theta \over 4\pi^2} \left[ (\sin^2
\theta)\ (ql)^2 \right] {v} (q)\ e^{-q^2l^2/2} \eeq which has its
origin in electron-electron interactions. However we can do more:
we have the full $H_0$ as well as the interactions. The point to
emphasize is that $H$ is not of the traditional form ($p^2/2m +V$
) and that there is no reason it had to be.

  The reader should verify  that if we use $H(\baro)$ instead, the
  one-particle piece will be a constant with no momentum
  dependence. The entire energy will be due to the Fock term,
  as in Read's (1998) conserving calculation.

For the benefit of readers who may be overwhelmed by seeing too
many approaches, we present   the key equations of the CS and EHT
approaches in Table \ref{basicequations}.

 To summarize, in any context where
$\bar{\bar{\rho}}^p$ can be reliably employed, we can say that the
CF appears to be the bound state of an electron and the
pseudovortex. We shall see that this includes calculations of the
gap and  magnetization properties at zero and nonzero
temperatures, even at $\nu =1/2$. We shall turn to these later on.
But first, we list some cases where $\barro^p$ and $\barH^p$ fail
to capture the right physics.

\begin{table}
\caption{Basic equations \label{basicequations}}
\begin{ruledtabular}
\begin{tabular}{|c|}
 {\bf CS Theory}\\ $H_{CS}=$    ${1\over 2m}\sum_i (   {\bf p} +e{\bf
A}+   {\bf a}_{cs})_{i}^{2}  +  V $\\
 $\nabla \times {{\bf a}_{cs}} =4\pi s \rho$\\
   {\bf Extended hamiltonian Theory (LLL only)}\\
 $ \barH =\half \sum_{\bq } v(q)\ e^{-q^2l^2/2} \barro
\bar{\bar{\rho}}(-\bq)$\\
 $\left[ \barH \ , \barchi \right]=0$ \\
 $ \barchi \simeq 0$\\
 $\bar{\bar{\rho}}(\bq)=\sum_j \exp (-i{\bf q} \! \cdot
\bR_{ej})$\\ $\bR_e  ={\bf r} -  {l^2\over 1+c}\hat{\bf z}\times
{\bf \Pi}=\bR + \etab c $\\ $\bar{\bar{\chi}}(\bq) = \sum_j \exp
(-i{\bf q} \cdot \bR_{vj})$ \\ $\bR_v={\bf r} + {l^2\over
c(1+c)}\hat{\bf z}\times {\bf \Pi}=\bR + \etab / c $\\
$\barH^p=\barH (\bar{\bar{\rho}}^p)$\\ $\bar{\bar{\rho}}^p(\bq
)=\bar{\bar{\rho}}(\bq)-c^2 \bar{\bar{\chi}}(\bq)$\\
\end{tabular}
\end{ruledtabular}
\end{table}

\subsection{The conserving approximation and compressibility paradox}

     HLR predicted that the $\nu=1/2$ system has a static
compressibility that is nonzero for short range forces and
vanishes as $q$ for the Coulomb interaction. This result appears
paradoxical in the present approach in which the dipolar nature of
the CF has been transparently exposed (using $\bar{\bar{\rho}}^p$
). Imagine coupling the system to an external potential $\Phi$.
The dipole will couple to the {\em gradient} of $\Phi$ and the
resulting response will be a dipolar density whose {\em
divergence} will give the induced charge. These two italicized
factors imply a $q^2$ in the response even for short range
interactions. Indeed, this is what we first obtained (Shankar and
Murthy, 1997, Murthy and Shankar, 1998a) upon doing a simple RPA
calculation.  How is this to be reconciled with the HLR (1993)
result, assuming their (compressible) answer is right?

Halperin and Stern (1998) first raised this question and provided
the key to its resolution. They first  established a matter of
principle, namely, that dipolar objects could be compressible, by
considering the following hamiltonian: \beq
H=\sum_i{\bp^{2}_{i}\over 2m} - {1 \over
2mn}\sum_{i,j,\bq}^{n,n,Q}\bp_i \cdot \bp_j e^{i\bq \cdot
(\br_i-\br_j)} .\eeq This hamiltonian arose in our earlier work
(Murthy and Shankar, 1998a) when we decoupled the magnetoplasmon
oscillators from the fermions. There we had chosen the upper
cut-off $Q=k_F$ so that the $i=j$ term from the double sum
contributes $-\bp^2/2m$ to each particle and cancels the first
sum, rendering $1/m^*=0$. Halperin and Stern considered the
limiting case $\bQ \to 0 $. In this limit $H$ takes the form \beq
H = \sum_{ij}{(\bp_i -\bp_j)^2\over 2m} = \sum_i {(\bp_i - {1\over
n}\sum_j \bp_j)^2\over 2m}\label{hq0} \eeq and is invariant under
the simultaneous shift of all momenta.  This symmetry called {\em
K-invariance} had also been pointed out by Haldane (1995) in
unpublished work and arose as part of a gauge symmetry in our
work. The symmetry implies that it costs no energy to move the
Fermi surface as a whole.  Consequently there are some very soft
modes that could lead to a singular density response which can
offset the $q^2$ from the dipolar factors, provided these soft
modes are not merely gauge artifacts that couple to nothing
physical. The detailed calculation of Stern, Halperin, von Oppen
and Simon (1999) (hereafter Stern {\it et al}) which we now
describe, demonstrated that gauge invariant soft modes do exist
and lead to nonzero compressibility.

The first order of business for them was to start with a
hamiltonian that had $K$-invariance for small $Q$ and not just
$Q=0$, since one needed to consider the response functions at
small but nonzero $q$. The hamiltonian in Eqn. (\ref{hq0}) had to
be augmented by more terms to assure this. These terms were
derived by Stern {\em et al} (1999) as follows. They go a step
back in our calculation and start with the following  hamiltonian
and constraints for the  coupled oscillators and particles:
\begin{eqnarray}
H&=& \sum_j{(\bp_j -\ba)^2\over 2m}+V\\ &=& \sum_j
{\bp_{j}^{2}\over 2m} -\sum_{\bq}^{Q}\bg (\bq) {1 \over
2mn_{CS}}\ba (-\bq) \nonumber
\\ &+& \int d^2r \left[ n_{CS}(\br) |\ba (\br)|^{2}\right] +V\\ \bg (\bq) &=& \sum_j
\bp_j e^{-i\bq \cdot \br_j}\\  0&=&\left(\nabla \times \ba -4\pi
n_{CS}\right) |\mbox{physical state}\rangle
\end{eqnarray}
where $n_{CS}$ is the fermion number density operator.  The only minor
difference from our work is that $a_x$ and $a_y$ (instead of the
longitudinal and transverse components $a$ and $P$) are canonically
conjugate. The cut-off on $q$ at $Q$ means the Chern-Simons fermions
now carry flux tubes, ones smeared over a distance $1/Q$ (sometimes
called ``fat'' flux tubes, see Halperin, 1992).

Next Stern {\em at al} (1999) approximate $n_{CS}$ by $ n$, the average
density and obtain \beq H=\sum_j{\bp_{j}^{2}\over 2m }- {1 \over
2mn} \sum_{\bq} \bg (\bq) \bg (-\bq) + \sum_{\bq} {n\over 2m}
\left|\ba (\bq) - {1 \over n}\bg (\bq)\right|^2 \eeq

Now they invoke  the unitary transformation we employed to
decouple the $\ba$ fields from the fermions in the small $ql$
limit: \beq U=\exp \left[ {i \over 4\pi n}\sum_{\bq}^{\bQ}\bg
(\bq)\times \ba (-\bq ) \right]  \eeq Since $a_x$ and $a_y$ are
conjugates, $U$   is just the  shift operator $\ba - \bg / n \to
\ba'$. Here comes the big difference. While we kept just the $H$
from Eqn. (\ref{hq0}) they augment $H$ with additional   terms to
ensure that the constraints \beq \rho (\bq) = -{i l^2 \over 2}\bq
\times \bg (\bq) \label{cons} \eeq commute with $H_{aug}$ to the
desired order. The reader should verify that this constraint is
just \beq \barchi =\sum_j e^{ -i\bq \cdot \bR_{vj}} = \sum_j \exp
(-i\bq \cdot (\br_j + {l^2\over 2} \hat{\bz}\times \bp_j))=0\eeq
expanded to order $ql$. \footnote{Since the (CS) constraint
commutes with $H$ prior to the action of $U$, it does so after the
action of $U$, but of course only to leading order in $ql$ since
$U$ was not implemented exactly.} To this order the electron
density is (dropping $\ba$ terms which do not matter at low
energies)
\begin{eqnarray} \rho^e (\bq) &=& \sum_j e^{ -i\bq \cdot
\bR_{ej}} = \sum_j \exp (-i\bq \cdot (\br_j - {l^2
\over2}\hat{\bz}\times \bp_j))\\ &\simeq & \sum_j \exp (-i\bq
\cdot \br_j)(1 - {il^2 \over 2} {\bq}\times \bp_j)\\ &=&\rho (\bq)
- {i l^2 \over 2}\bq \times \bg
(\bq).\label{echarge}\end{eqnarray}

 Recall the constraint generates gauge transformations.  To
zeroth order the constraint operator is just $\sum_j e^{-i\bq
\cdot \br_j}=0 $ and its action  is \beq {\bf r}_j \to {\bf r}_j \
\ \ \ \ \ \  {\bf p}_j \to {\bf p}_j + {\bf q} e^{-iqr_j}. \eeq
Thus respecting the constraint ensures $K$-invariance.

Stern {\em et al} (1999) now perform an RPA calculation of the
electronic density-density correlation. RPA  works because at a
given $\bq$, the gauge field (of that $\bq$) enters only in the
wiggles connecting irreducible bubbles in Figure  (3)  and nowhere
inside these bubbles: every internal exchange brings with it  a
sum over $q$ which introduces a small parameter $Q$. Since in the
limit $Q \to 0$, RPA diagrams are all we have, RPA respects the
symmetry of $H_{aug}$. We will now argue that if the constraint is
respected, compressibility follows. For the electron density
operator we have many choices starting with Eqn. (\ref{echarge})
and using the constraint Eqn. (\ref{cons}). In particular we can
write it as \beq \rho^e = 2 \rho \eeq
 or as \beq \rho^e =- 2 {i l^2 \over
2}\bq \times \bg (\bq).\eeq Written  the first way, a nonzero
compressibility is not surprising  since there are no powers of
$q$ in the operator. In the second way, we see the possible
paradox, since there is a $q$ up front in each of the two factors
of $\rho^e$. It is here that the overdamped mode appears in the
transverse sector, couples to $\bg (\bq)$  and saves the day with
a factor $q^2 v(q)$ in its static propagator, yielding results
that coincide with HLR.

The  details of this formidable calculation  are not shown here in
the interest of brevity. Suffice it to say that five different
operators  ($\rho$, and the two components each of $\bg$ and
$\bC$, where $\bC$ is another vector operator) get coupled and
$K^{-1}= K_{0}^{-1}+U$ is a relation among $5\times 5$ matrices.
It is remarkable that in the end all the physics of  HLR,
including the overdamped mode, finite compressibility and mass
divergences at the Fermi surface, all emerge from the more
physical quasiparticles, though after a lot of work.

However this is not quite the end since we have only discussed the
$Q\to 0$ limit, while the actual theory has no such limit on $Q$.
Stern {\em at al} (1999) finesse this question with a two-step
argument: (i) First they show that $K$-invariance guarantees that the
Landau parameter $F_1$ has to be $-1$. (ii) Then they show that this
condition generally produces very soft modes which restore
compressibility to the dipole gas. Let us elaborate a little. Suppose
we want define an effective low-energy theory for the composite
fermions. How do we ensure that it has $K$-invariance? We cannot
possibly find all the higher order correction in the $Q$
expansion. This is also the typical situation in Landau theory, that
although in principle the Landau parameters can be calculated given
the microscopic interaction, there is no way to do this reliably in
practice.  However, there are exceptions where a symmetry is
involved. For example, Galilean invariance can be used to relate the
bare mass, the physical mass, and the Landau parameter $F_1$. {\em It turns
out that here too, $K$-invariance can be assured for the actual
problem if we choose $F_1 =-1$.} The reason is that the energy cost of
boosting the Fermi sea is measured by $(1+F_1)$.  Choosing $F_1=-1$,
makes the boost cost-free, i.e., implies $K$-invariance. In other
words the dipolar fermion cannot have an arbitrary Fermi liquid
interaction: its $F_1$ must equal $-1$. At the level of diagrams, if
$F_1=-1$ is included as an interaction, the correlation function of
two dipolar densities will go as $q^0$ and not $q^2$.

The compressibility of the $\nu =1/2$ system was also established
by Read (1998) using a conserving approximation. He begins with
the problem of bosons at unit filling, a problem first studied by
Pasquier and Haldane (1998), though not in a conserving
approximation. The bosons  can be traded for fermions in zero
average field by attaching a flux quantum. The role of
fluctuations is the same as in the HLR problem except for the
strength of the gauge field-fermion coupling.  The hamiltonian is
just the electronic interaction written in the CF basis with \beq
\bR_e = \br - {l^2 \over 2} \bar{\bz}\times \bp \eeq and the
constraint is the density corresponding to \beq \bR_v = \br + {l^2
\over 2} \bar{\bz}\times \bp .\eeq

The advantage of this starting point is that the constraint
exactly commutes with $\barH$. Thus one can look for approximation
schemes in which the constraint is respected at the level of
Green's functions, i.e., Green's functions with any number of
$\barchi $'s in them vanish. Read's (1998) calculation begins with the
filled Fermi sea (which does not respect the constraint) and
embellishes it with diagrammatic corrections. An infinite sum of
ladders (whose legs contain fermions in zero field) leads to the
collective mode.  The density-density correlator  has the
appearance of dipolar objects that exchange a transverse gauge
field.  The transverse propagator (which is just the overdamped
mode)
 introduces  a $v(q)q^2$ in the denominator that  offsets
the $q^2$ upstairs. It is found that correlators containing
$\barchi$'s vanish and Ward identities are satisfied.

It is instructive to look at Read's (1998)  density-density
correlator which takes the form:

\begin{eqnarray}
K_{irr}(q,\omega)  &=& \int d^2kd^2k'(e^{il^2 \bk \times
\bq}-1)M(\bk , \bk', \bq, \omega)(e^{-i l^2\bk' \times \bq}-1)
\label{readcomp}
\end{eqnarray} where the factors at each end reduce to dipoles at
small $ql$ and $M =M_0+M_T $ is a sum of two terms, one from the
free Fermi sea and one from the exchange of the transverse
collective mode which peaks at $i\omega \simeq q^3\ v(q)$. At high
frequencies, $M_T$ can be ignored and the dipoles emerge as free
objects, while at low frequencies, they are coupled by the
overdamped mode and do not behave like classical dipoles.  We
begin to see how the dipoles appear in the correlators  (because
 we can tune the frequency and wavelength in $K$  to
expose them)  but not the wavefunctions, which describe equal-time
correlations  and thus  involve an integral over all frequencies.

Arguments for the compressibility of  the dipolar system were also
given by D.H. Lee (1998).

The resolution of the compressibility paradox did much to assure
the community that various descriptions of the quasiparticle, each
tailor-made for a different occasion,  were mutually compatible
and consistent.

\subsection{Higher Landau Levels}
The advantage of the  extended hamiltonian  is that it keeps track
of the electronic cyclotron coordinate and does not go to the LLL
prematurely. This has many benefits. If we want to compute the
Hall conductance, we can couple the system to an external
potential and find the response. As stated previously, the
response involves higher LL's so that the presence of $\etab_e$ is
crucial. The details are shown in Appendix \ref{hallext}. We can
also study the effects of LL mixing. Instead of using just
$H_{LLL}$ we can try to get an effective theory within the LLL
which subsumes the effects of virtual transitions to higher LL's.

To extract the leading correction due to higher LL's we write the
Schr\"{o}dinger equation in schematic form as as \beq \left[
\begin{array}{cc}
  H_{00} & H_{0n'} \\
  H_{n'0} & H_{n'n'}
\end{array} \right]   \left[ \begin{array}{c} \phi \\ \xi
\end{array} \right] = E \left[ \begin{array}{c} \phi \\ \xi
\end{array} \right]
\eeq where $\phi$ is  restricted to the space spanned by Fock
states composed  of just the LLL states and $\xi$ stands for
everything else. Likewise the subscripts $0$ and $n'$ stand for
collective labels in the LLL and above the LLL respectively.  The
exact equation obeyed by $\phi$ is \beq \left(H_{00} + H_{0n'}{1
\over E-H_{n'n'}}H_{n'0}\right) \phi = E \phi \eeq which is not an
eigenvalue problem since $E$ appears on both sides. However we may
approximate as follows: \beq {1 \over E-H_{n'n'}}=-{1 \over
H_{n'n'}} + {\cal O} (v/\omega_0)\eeq
 since the eigenvalue $E$ we are interested in is of order $v$ and
 the eigenvalues of $H_{n'n'}$ are of order $\omega_0$. To the
 same accuracy
 in $\kappa = v/\omega_0$ we can also replace
 \beq
 H_{n'n'}\simeq  H_{n'n'}^{0} = \sum_{\alpha}
 a^{\dag}_{\alpha}a_{\alpha} n^{'}_{\alpha} \omega_0
 \eeq
 which leads to \beq
 H^{eff}_{00}=H_{00} -\sum_{n'}H_{0n'}{1\over n'\omega_0}H_{n'0} \eeq

Once we have the effective theory in the LLL, we can switch to the
CF formalism: Introduce $\bR_v$, exchange $\bR_e$ and $\bR_v$ for
$\bR$ and $\etab$ and proceed as before by a Hartree-Fock
calculation. The results are in accord with earlier works that
showed that LL-mixing reduces the transport gap, but that finite
thickness reduces this effect (Yoshioka, 1986, Price, Platzman,
and He, 1993, Melik-Alaverdian and Bonesteel, 1995, Price and Das
Sarma, 1996, Melik-Alaverdian, Bonesteel, and Ortiz, 1997). Full
details can be found in Murthy and Shankar (2002).

\section{Correlation functions in the conserving
approximation\label{conserving}}

This section illustrates how one does the conserving calculation
within the hamiltonian approach by deriving density-density
correlation functions (Murthy, 2001a). This calculation differs
from  Read (1998) in two ways. First, it is done for nonzero
effective field ($\nu \ne 1/2$), so that the CF's are in Landau
levels instead of
  plane wave states. Next, the calculation is done in the
operator approach (Anderson, 1958a,b, Rickayzen, 1958) using
equations of motion as compared to diagrams (Nambu, 1960). The
second difference is only cosmetic, and introduced  here to
promote harmony   with the rest of the paper\footnote{For an
illustration of calculations in the gapped fractions using the
diagrammatic approach, see Green (2001).}.

The Hamiltonian  and constraint to be solved
 are
\begin{eqnarray}
H &=& {1 \over 2} \sum_{\bq}     \bar{\bar{\rho}}(\bq)\
v(q)e^{-(ql)^2/2}\ \bar{\bar{\rho}} (-\bq) \\ \barchi &\simeq & 0.
\end{eqnarray}

  Let us define
a time-ordered pseudovortex-electron density-density Green's
function $G_{ve}(q,t)$ as follows (with $G_{ee}$ and $G_{vv}$
similarly defined):

\eqa G_{ve}(\bq,t-t')&=&-i<T\bachi(\bq,t)\baro(-\bq,t')>\nonumber\\
&=&-i\Theta(t-t')<\bachi(\bq,t)\baro(-\bq,t')>\nonumber\\
&-&i\Theta(t'-t)<\baro(-\bq,t')\bachi(\bq,t)> \eea which evolves as
per \eqa -i\ddt
G_{ve}(\bq,t-t')&=&-\delta(t-t')<[\bachi(\bq,t),\baro(-\bq,t')]>\nonumber\\
&-&i<T[H,\bachi(\bq,t)]\baro(-\bq,t')> \eea Since $\bachi$ commutes
with $H$, one immediately sees that $G_{ve}$ is a constant. If the
constraint is set to zero initially, then it remains zero, and all its
correlators also remain zero.

The above is true in an exact treatment of the theory. Of course, we
can only do approximate calculations. A calculation that respects
$G_{ve}=G_{vv}=0$ respects the symmetries of the theory at the level
of correlators, and is called conserving.  Let us see what a natural
approximation scheme might be. Consider

\eqa i\ddt
G_{ee}(\bq,t)&=&\delta(t)<[\baro(\bq,t),\baro(-\bq,0)]>\nonumber\\
&-&i<T[H,\baro(\bq,t)]\baro(-\bq,0)>
\eea

Since $\left[ \baro \ , \baro \right]\simeq \baro$,   a Green's
function involving three densities will arise. Additional time
derivatives will produce higher-order density correlators, leading
to a hierarchy of equations for more and more complicated Green's
functions. The natural way to truncate this hierarchy is to make a
mean-field approximation at some stage that reduces a product of
two densities to a single density. One of the simplest of such
approximations (Baym and Kadanoff, 1961) reduces $[H,\baro]$,
which is a product of four fermi operators, to a product of only
two, by using the averages

\eqa \dd_{\a_1}\dd_{\a_2 }d_{\b_2}d_{\b_1}&\to
& <\dd_{\a_1}d_{\b_1}>\dd_{\a_2}d_{\b_2}
+<\dd_{\a_2}d_{\b_2}>\dd_{\a_1}d_{\b_1} \nonumber\\
&-&<\dd_{\a_1}d_{\b_2}>\dd_{\a_2}d_{\b_1}
-<\dd_{\a_2}d_{\b_1}>\dd_{\a_1}d_{\b_2}
\eea

Here
$<\dd_{\a}d_{\b}>=\delta_{\a\b}N_F(\a)$, where $N_F(\a)$ is the
Fermi occupation of the single-particle state $\a$.

Using the HF states and their occupations in the above truncation is
known as the time-dependent HF (TDHF) approximation\footnote{The TDHF
approximation has been the method of choice in computing
magnetoexciton dispersions in the IQHE. Here are some of the early
references: Chiu and Quinn (1974), Horing and Yildiz (1976), Theis
(1980), Bychkov, Iordanskii, and Eliashberg (1981), Bychkov and Rashba
(1983), Kallin and Halperin (1984), MacDonald (1985), Hawrylak and
Quinn (1985), Marmorkos and Das Sarma (1992), Longo and Kallin (1993).
}.  We will use the operator approach to TDHF as expounded by Anderson
(1958a,b) and Rickayzen (1958). We will explicitly see below that it
is conserving for all principal fractions.

The physical picture underlying our calculation is the following.
When a bosonic operator such as $\barro$ or $\barchi$ acts on the
ground state, it creates a linear combination of particle-hole
pairs. In the Landau gauge, each pair is labeled by an index
$\nu={n_1,n_2}$ (not to be confused with the filling factor!) that
keeps track of the CF-LL indices of the particle and hole, and a
total pair momentum $\bq$ which is conserved because the particle
and hole have opposite charges and do not bend in the magnetic
field. It is clear that in order to calculate time-dependent
response functions we have to understand the time-evolution of
these pairs. In the exact theory, the Hamiltonian can scatter a
pair into two pairs, two pairs into four pairs, etc. This is what
leads to the hierarchy of equations. However, the great simplicity
of the TDHF approximation is that in this approximation a
particle-hole pair scatters only into another particle-hole pair.
At a given $\bq$ we thus have a matrix labeled by the indices
$(\nu ,\nu')$ of the incident and scattered particle-hole pairs.
The magnetoexciton spectrum comes from the eigenvalues of this
matrix, which, together with its eigenvectors, will be seen to
explicitly determine the Green's function.

Clearly, we must begin with an operator that creates an exciton in
a state of definite momentum $\bq$, starting with states labeled
by  CF-LL indices $\nu =(n_1\ n_2)$. The following operator does
the job:

\eq O_{n_1n_2}(\bq)=\sum_{X} e^{-iq_xX}
\dd_{n_1,X-{q_yl^{*2}\over2}} d_{n_2,X+{q_yl^{*2}\over2}} \ee
 Why
is this so? First note that in this gauge, (Eqn.(\ref{lg})), the
wavefunctions are plane waves in $y$ with momentum $k$ and
localized in $x$ at $X=kl^{*2}$. Thus
$\dd_{m_1,X-{q_yl^{*2}\over2}} d_{m_2,X+{q_yl^{*2}\over2}}$ which
creates a hole at $X-{q_yl^{*2}\over2}$ and a particle at
$X+{q_yl^{*2}\over2}$, creates an exciton in a state of momentum
$q_y$ in the $y$-direction, centered at $X$. Multiplying by
$e^{-iq_xX}$ and summing over $X$ creates a state of momentum
$q_x$ in the $x$-direction.

  Now we define \eq
G(\nu;\nu';\bq;t-t')=-i<TO_{\nu}(\bq,t)O_{\nu'}(-\bq,t')> .\ee
Taking the time derivative  we get

\eqa -i\ddt
G(\nu;\nu';\bq;t)&=&-i<T[H,O_{\nu}(\bq,t)]O_{\nu'}(-\bq,0)>\nonumber\\
& & -\delta(t)<[O_{\nu}(\bq,t),O_{\nu"}(-\bq,0)]>
\label{eqofmotion}\eea

The last piece is the standard inhomogeneous ``source'' term. The
dynamics is controlled by the commutator with the Hamiltonian, to
which we now turn

\eqa
\left[H,O_{\nu}(\bq)\right]&=&(\e(n_1)-\e(n_2))O_{\nu}(\bq)+\nonumber\\
&& (N_F(n_2)-N_F(n_1))\sum\limits_{\nu"} \bigg({v(q)\over
2\pi(l^*)^2} e^{-q^2l^2/2}\rho_{n_{1}^{"}n_{2}^{"}}(\bq)\times
\rho_{n_2n_1}(-\bq)\nonumber\\ &-&
 \ints v(s)
e^{-s^2l^2/2}\rho_{n_{1}^{"}n_1}(\bs) \rho_{n_2n_{2}^{"}}(-\bs)
e^{i (l^{*})^2\bs\times\bq}\bigg) O_{\nu'}(\bq)\label{tdhfeq} \eea
where $\varepsilon (\nu)$ is the Fock energy in a state of CF-LL
index $\nu$, and the TDHF approximation has been made.

{\em Note that the action of commuting with $H$ on $O_{\nu}$ in
the TDHF approximation can be represented as the
right-multiplication by a matrix $\cH(\nu;\nu";\bq)$.} It follows
that if a we form a linear combination of operators
$O_{\Psi}=\sum_{\nu}\Psi(\nu;\bq)O_{\nu}$, the column vector
$\Psi$ will transform linearly under the action of $\cH$.
Diagonalizing $\cH$ will also enable us to solve for the Green's
function, for the eigenvectors represent linear combinations of
particle-hole states that are normal modes of $\cH$. Assume that
one has found the right and left eigenvectors and corresponding
eigenvalues $E_{\alpha}(q)$ , labeled by $\a$

\eqa
\cH(\nu;\nu";\bq)\Psi_{\a}^{R}(\nu";\bq)&=&E_{\a}(q)\Psi_{\a}^{R}(\nu;\bq)
\label{psir}\\
\Psi_{\a}^{L}(\nu;\bq)\cH(\nu,\nu";\bq)&=&E_{\a}(q)\Psi_{\a}^{L}(\nu";\bq)
\label{psileft} \eea

where sums over repeated indices are implicit.

Assuming  the matrix $\cH$ has a complete set of eigenvectors
 the Green's function can be written as  \eqa
G (\nu;\nu';\bq;\o)&=&{L^2\over2\pi(l^*)^2}\sum\limits_{\a}
\Psi_{\a}^R(n_1n_2;\bq){1\over\o-E_{\a}}\Psi_{\a}^L(n_2'n_1';\bq)(N_F(n_1')-N_F(n_2'))
\label{specrep}\eea where the factor $(N_F(n_1')-N_F(n_2'))$ comes
from the source term (the last term of Eq. (\ref{eqofmotion})).

An  important property of $\cH$ from the point of view
 of the conserving approximation is that it always has one left
 eigenvector with zero eigenvalue for every $\bq$, namely
\eq \Psi_{0}^{L}(n_1n_2;\bq)=\tchi_{n_1n_2}(\bq) \ee where
 \beq
\tchi_{n_1n_2}(\bq)=\langle n_1|e^{-i\bq \cdot (\etab
/c)}|n_2\rangle . \eeq

The existence of this eigenvector with zero eigenvalue (for any
$v(q)$) is shown explicitly in Murthy (2001a). This zero
eigenvalue is related to the constraint and is one of the
conditions for the TDHF approximation to be conserving. To see
this let us go back to the exact theory. There the Hamiltonian
commutes with $\barchi$. This means that $\barchi$ acting on the
ground state should produce a {\it zero energy} state at every
$\bq$. In the TDHF approximation the action of the Hamiltonian on
this state is approximated by the action of the matrix $\cH$ on
the left vector $\Psi_{0}^{L}(n_1n_2;\bq)$, corresponding to the
operator $O_{\barchi}=\sum\tchi_{n_1n_2}(\bq) O_{n_1n_2}(\bq)$.
The fact that $\cH$ admits $\Psi_{0}^{L}$ as a left eigenvector
with {\it zero eigenvalue} means that the zero energy state which
had to be present in the exact theory is also present in the TDHF
approximation.

To put it in slightly different words, the TDHF approximation
represents a truncation of the Hilbert space of neutral
excitations to states having only a single particle-hole pair
above the ground state. {\it A priori} this truncation need not
have respected the constraint, but it does.

The above condition is necessary for TDHF to be conserving, but
not sufficient. The other condition is that the physical sector
(excitations created by $\barro$) does not couple to the zero
energy sector of $\cH$. This can also be verified in a
straightforward way (Murthy 2001a).

\subsection{Small-$q$ structure factor}
We will  now put the conserving approximation to another test.
Recall that the density-density correlator has two parts; one
coming from the cyclotron pole, with a residue of $q^2$, which is
unrenormalized as per  Kohn's theorem and saturates the sum rule,
and another coming from the dynamics in the LLL. Hence the
LLL-projected structure factor $\bareS (q)$ has to vanish faster
than $q^2$, and if it is analytic in $q^2$, it has to go like
$q^4$. We will see if this is true. Luckily, to obtain the leading
behavior of $\bareS (q)$ in the TDHF approximation one needs to
keep only a finite-dimensional submatrix of the infinite TDHF
matrix.

The results are as follows. For $\nu=\third$ we have, upon
diagonalizing a $4 \times 4$ matrix,  \eq \bareS
(q)={(ql)^4\over8}+\cdots \ee

 We can extend this result to all the Laughlin fractions
${1\over2s+1}$ to obtain \eq \bareS (q)={1\over 8} (ql)^4 +\cdots \ee
The coefficient ${1 \over 8}$, independent of $s$, differs from the
result ${s \over 4}$ obtained from Laughlin's wavefunction (Girvin,
MacDonald, and Platzman, 1986). However no general principle requires
that we regain this coefficient.

One can carry out a very similar, but much more tedious,
calculation for all the principal fractions (Murthy, 2001a).  The
result, upon diagonalizing a $6\times 6$ matrix, is  \eq \bareS
(q)={(ql)^4\over2}{p^4-3p^3+{5\over4}p^2+3p+{7\over4}\over p^2-1}
\label{sfp3}\ee This expression, with its divergence as
$p\to\infty$ or $\nu\to\half$ is consistent with the result (Read,
1998) that for a problem equivalent to the $\nu=\half$ problem,
$\bareS (q)\simeq q^3\log(q)$. As $p\to\infty$, the radius of
convergence of the power series expansion of $\bareS (q)$ must go
to zero (else the structure factor would diverge for a range of
$q$ according to Eqn. (\ref{sfp3})), and the $p\to\infty$ limit
does not commute with the $q\to0$ limit. The formula also does not
hold for $p=1$, the Laughlin fractions, for which we must use the
results quoted earlier.

Surprisingly, the results are independent of the form of the
potential, which does enter the intermediate stages of the
calculation.

\subsection{Magnetoexciton dispersions for $\third$ and $\twof$\label{sec6}}

We saw that the small-$q$ behavior of $\bareS (q)$ can be
satisfactorily addressed using the TDHF approximation. It turns
out that this approximation also works well for computing the
dispersion of magnetoexcitons,  by which  we mean the lowest
energy physical eigenstate of $\cH$ at each $\bq$.

As shown in the previous section, for very small $q$, the naive
magnetoexcitons (bare particle -hole states created by $O_{\nu}$)
do not mix with others and become the true eigenstates of $\cH$.
As $q$ increases they become increasingly coupled, and both level
repulsion (between positive energy states) and level attraction
(between positive and negative energy states) manifest themselves,
giving rise to complicated magnetoexciton dispersions. The matrix
$\cH$ is infinite-dimensional, but in any numerical calculation,
only a finite matrix can be diagonalized. We saw that the lowest
nontrivial result for $\bareS (q)$ could be obtained by keeping at
most a $6\times6$ matrix. As $q$ increases more and more CF-LLs
have to be kept to obtain accurate results. The accuracy of the
truncation  was checked by two different methods. Firstly, the
number of CF-LLs kept was increased until the energy of the
magnetoexciton was stable. Secondly, we knew that at every $q$
there should be two zero eigenvalues corresponding to the
unphysical sector. The number of CF-LLs kept in the calculation
was increased until these null eigenvalues were at least four
orders of magnitude smaller than the smallest physical eigenvalue.

\begin{figure}[t]
\begin{center}
\includegraphics[width=3in]{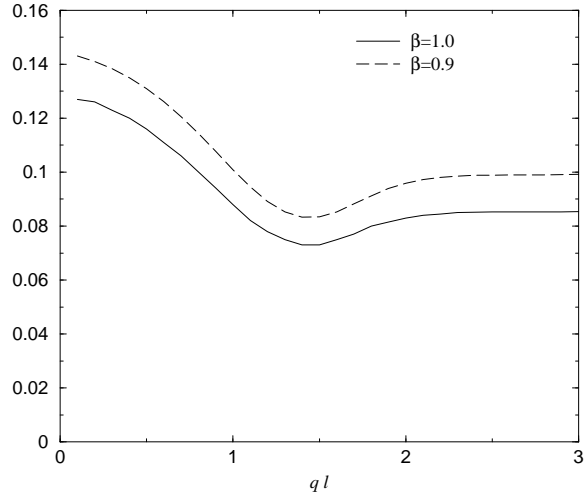}
\end{center}
 \caption[]{The lowest-energy spin-polarized magnetoexciton for $\nu=1/3$
for two values of $\beta$. Note the "magnetoroton" at $ql=1.4$.
\label{fig1-cons}}
\end{figure}

Figures 4 and 5 are for the potential $v(q)={2\pi e^2\over \e
q}e^{-q^2\beta^2/2}$ for various values of $\beta$. It is worth
noting that though the depths of the minima are $\beta$-dependent,
and different from what are found in numerical diagonalizations or
from CF-wavefunctions, the positions are correct\footnote{Haldane
and Rezayi (1985), Su and Wu (1987), Morf and Halperin (1987),
d'Ambrumenil and Morf (1989), He, Simon, and Halperin (1994), He
and Platzman (1996), Kamilla, Wu, and Jain (1996a,b), Jain and
Kamilla (1998).}. This indicates that the TDHF approximation to
the EHT does capture the important physics of the electronic
problem even at fairly large $q$.

In the case of ${1\over 3}$, the magntoexciton energy has
stabilized at large $ql$ enabling us to read off the {\em
activation gap}, $\Delta_a$, defined as the minimum energy needed
to produce a widely separated particle-hole pair. However, the
situation deteriorates rapidly as $p$ increases. It becomes
prohibitively hard, even at $p=2$, to get the large $ql$ limit of
the magnetoexciton spectrum, as is evident from Figure 5. This is
because the decoupling of the naive magnetoexcitons is only
asymptotic for large $q$. However, as $p$ increases, the {\em
shortcut} gives very good answers  (Figures 6 and 7) with hardly
any additional work. We now turn to it.

\begin{figure}[t]
\begin{center}
\includegraphics[width=3in]{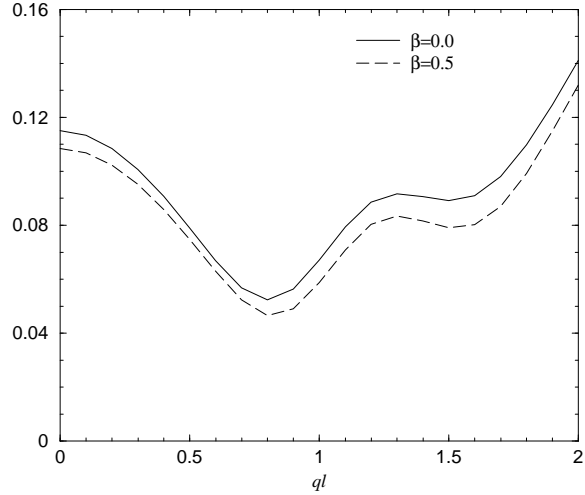}
\end{center}
 \caption[]{\label{fig2-cons}The lowest-energy spin-polarized magnetoexciton for $\nu=2/5$
for two values $\beta$. Note the minima at $ql=0.8$ and $ql=1.5$.
}
\end{figure}

\section{Gaps}

Having developed the EHT at length and having established several
qualitative results and matters of principle, we now turn to
mainly quantitative issues. As mentioned at the outset, one can
compute just about anything within this approach, some more
accurately than others. In the next few sections we will discuss a
variety of such quantities. Rather than run through an endless
list we will focus on a few that give the flavor of the method and
expose its strengths and shortcomings, and point out references
containing more details or examples. An invaluable benchmark will
be provided by results from trial wavefunctions and exact
diagonalization. Comparison with experiment will be done with the
clear understanding that no systematic attempt is made to
incorporate disorder.

In comparing to the results of exact diagonalization and trial
wavefunctions one must bear in mind their limitations. Exact
diagonalization (ignoring machine errors) suffers from the fact
that the systems are necessarily small. In comparing these results
to ours (valid for infinite systems) we must examine the approach
to the thermodynamic limit and the scaling of gaps. We refer the
reader to  Morf, d'Ambrumenil, and Das Sarma (2002) for the most
recent study of this kind.

While trial wavefunctions can be written down for any number of
particles, the evaluation of gaps requires once again that the
number of particles be small (though not as small as in exact
diagonalization). The details may be found in Park, Meskini and
Jain (1999) to whom we compare our numbers. It must also be borne
in mind that the correct wavefunctions may not be of the form
being tried, in which case even a flawless evaluation of gaps is
irrelevant.  So far there is good reason to believe that for $1/3<
\nu \le 1/2$ the Jain-like functions are good. At very low
densities the correct state is believed to be a Wigner crystal and
when higher LL's are involved, more complicated functions like
Pfaffians (the Moore-Read, 1991, state) need to be explored. With
these caveats in mind we proceed with the  comparison of our
theory with numerics.

{\em In the remaining  sections, we will use the preferred density
and pay no further regard to constraints.} As stated earlier, this
is the most efficient way to work on problems that do not depend
crucially on the deep infrared region ($\omega \simeq q^3\ v(q)$).
In the cases discussed this is ensured by either a gap, a nonzero
temperature, or both.

Let us begin with the activation or {\em transport gap} in a fully
polarized sample, defined as the minimum energy needed to produce
a widely separated particle hole pair:
 \beq \Delta = \langle {\bf p} + PH | H|{\bf
p}+ PH \rangle -\langle {\bf p} | H| {\bf p} \rangle \eeq where
$|{\bf p}\rangle$ stands for the Hartree-Fock ground state with
$p$-filled LL's and  $PH$ stands for a widely separated
particle-hole pair. We will use a  boldface symbol such as ${\bf
p}$  to label a Slater determinant with $p$  occupied Landau
Levels. Non-boldface symbols will label  single-particle states.
Note also that the highest occupied CF-LL index $n$ for the state labeled by
$p$ is $n=p-1$ since the CF-LLL has index $n=0$. As shown in Appendix
\ref{HFProof}, the  particle-hole excitations of $|{\bf p}\rangle$
are HF states of our $H$.

The expression for the gap written above is formally the same in
the wavefunction based approach of Park, Meskini and Jain (1999)
to whose results we will compare ours.  However the notation hides
a big difference.  They work in the {\em electronic basis} where
$\rho ({\bf q}) = \sum_j \exp (-i{\bf q\cdot r}_j)$, and the
states are the simple wavefunctions $\chi_p$, multiplied by the
Jastrow factor and then projected to the LLL.  (Projection leads
to a very complicated expression for the wavefunctions.) In the
present approach we have tried to incorporate these effects by
going in the reverse direction, from electrons to CF's, and
obtaining complicated expressions for the charge and other
operators, but with simple expressions for the wavefunctions.
While these operator expressions are unusual in form, they are
simple to evaluate within the HF calculation.

In all our calculations we shall use the Zhang-Das Sarma
potential (Zhang and Das Sarma, 1986)

       \beq
       v_{ZDS} = {2\pi e^2 e^{-ql\lambda}\over q}\label{zds}
       \eeq
       which is a crude model for the electron-electron interaction in a
sample of finite thickness. We simply take it to be a one
parameter family of potentials. In terms of the Haldane
pseudopotentials $V_m$ (which gives the interaction in a state of
relative angular momentum $m$, see Haldane, 1987) we know that just
one (typically $V_1$) dominates. We can think of $\lambda$ as
controlling the operative pseudopotential.

Rather than work with a widely separated particle-hole pair, we find
the energy in a state with just the particle and add to it the energy
of a state with just the hole and subtract double the ground state
energy. Relegating the details to Appendix \ref{gaps}, we present the
central idea.

We begin with the second quantized expression for the
preferred charge operator $\bar{\bar { \rho}}^p ({\bf q})$:

   \beq
 \bar{\bar{\rho}}^p({\bf q}) =
 \sum_{m_2n_2;m_1n_1}d^{\dag}_{m_2n_2}d_{m_1n_1}\rho_{m_2n_2;m_1n_1}(\bq
)
 \eeq
where $d^{\dag}_{mn}$ creates a particle in the state $|m\
n\rangle$ where $m$ is the angular momentum and $n$ is the LL
index of CF in the weakened field $A^*=A/(2ps+1)$ with a magnetic
length

\beq l^*=l\sqrt{2ps+1}.
\eeq

The key ingredient in the HF calculation is the matrix element
$\rho_{m_2n_2;m_1n_1}$ which factorizes (as shown in Appendix
\ref{matrixelement}):

 \beq
 \rho_{m_2n_2;m_1n_1}=\rho_{m_2m_1}^{m}\otimes \rho_{n_2n_1}^{n}
 \eeq

The gaps depend only on $\rho_{n_2n_1}^{n}$, the superscript on which
will be generally dropped.  Often we will use the {\em dimensionless
activation gap} $\delta_a $ defined by
\beq
\Delta_a = {e^2\over
\varepsilon l}\ \delta_a .
\label{reduced}
\eeq

\begin{figure}
\includegraphics[width=3in]{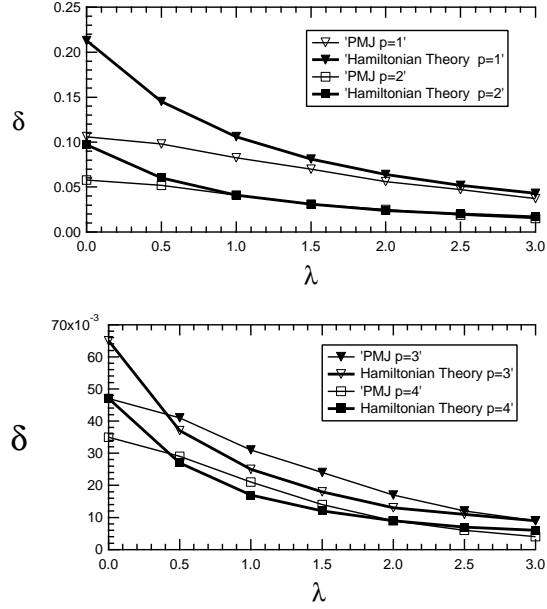}
\caption{\label{gapso} Comparison of dimensionless  activation
gaps $\delta_a$ to the work of Park, Meskini and Jain (1999) for
the fractions ${1\over 3},{2\over 5},{3\over 7}, {4\over 9}$
($p=1,2,3,4$, \mbox{and} \ s=1) as a function of $\lambda$,
  the thickness parameter in the ZDS potential.}
\end{figure}

Figure 6  shows the gaps computed for ${1\over 3},{2\over
5},{3\over 7}$ and $  {4\over 9}$ for the ZDS potential
 and compared to the work of Park, Meskini, and Jain (1999)
  in the region $0\le \lambda \le 3$.
The following features are noteworthy.
\begin{itemize}
\item At $\lambda =0$, the Coulomb case, the gaps are finite in contrast to
the small-$q$ theory (Murthy and Shankar, 1999). This is due
to the gaussian factor $e^{-q^2l^2/2}$ in $H(\baro)$ ((Eqn.
(\ref{energy})) which was absent in the {\em small-q} theory. The
slope of the graphs in the present theory is nonzero at this
point. It is readily
 verified that $d\Delta /d\lambda $ at $\lambda =0$ is the gap due to a
  delta-function potential, and  should vanish for spinless fermions.
The description of CF in terms of $\baro^p$ does not give good
answers for potentials as short ranged
 as the delta function, as anticipated earlier. Indeed even the Coulomb interaction is too
 singular, and
  the theory begins to work well
 only beyond $\lambda \simeq 1$.
\item Beyond $\lambda \simeq 1$ the agreement is quite fair in
 general and best for ${2\over 5}$. However we cannot go to
 too large a $\lambda$ since in this range the system may not be an
 FQHE state.
\item The gaps which do not vanish for any fraction and any finite
$\lambda$, exceed the Park, Meskini, and Jain (1999) values for
${1\over 3}$ and ${2\over 5}$ and lie below them for ${3\over 7}$ and
$ {4\over 9}$.  This result is at odds with the general belief that HF
always overestimates the gaps by neglecting fluctuations.
\end{itemize}

The situation is different when we compare with  the exact
diagonalization results of Morf, d'Ambrumenil, and Das Sarma
(2002) who used a potential  \beq v(q) = {2 \pi e^2\over q} {
e^{(qlb)^2)}\ \mbox{Erfc}\ (qlb)} \eeq where $b$ is the analog of
$\lambda$ \footnote{This potential is not very different in form
from the ZDS potential except at very short distances.}.  Our
numbers are compared in Figure 7. The calculated gaps always lie
above the exact diagonalization results for the two fractions
shown (as well as for the ${1\over 3}$ case, not shown).

 The general disagreement with our theory is worse for this potential than for
the ZDS case because at large $q$ this potential goes as $1/q$
while the ZDS potential falls exponentially.

\begin{figure}
\includegraphics[width=3in]{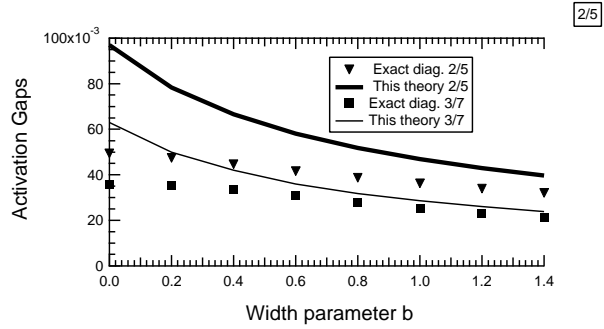}
\caption{\label{Morf}Comparison of hamiltonian theory to the exact
diagonalization results of Morf {\em et al} for $p=2 \ \mbox{and}\
3$ .}
\end{figure}

\subsection{Activation gaps}
We have computed gaps for many other fractions, including for
 $s=2$, when four vortices are attached to form CF's.
Rather than show more plots, we  will now analyze the theory in
terms of  the {\em activation mass} $m_a$ defined by

 \beq
 \Delta_a = {eB^*\over m_a}={eB\over(2ps+1)  m_a}.\label{acdef}
 \eeq

Comparison to Eqn. (\ref{reduced}) shows that

\beq {1 \over m_a} = {e^2 l\over \varepsilon}\delta_a (2ps+1)
\equiv {e^2 l\over \varepsilon} C_a . \eeq Thus  \beq C_a =
\delta_a (2ps+1). \eeq

  Based on the exact diagonalization results of d'Ambrumenil and Morf
(1989) Halperin, Lee, and Read (1993) pointed out that $C_a$
approaches a limit as we approach $\nu ={1\over 2}$ or $p\to \infty$.
The Halperin, Lee, Read theory expects $C_a$ to be modified by
logarithms. We too expect these logarithms once we approach
$\nu={1\over 2}$ and are forced to include the overdamped
mode. However, we find as did Halperin, Lee, and Read (1993), a good fit to
the calculated gaps without invoking the logarithms, which are
operative in a very tiny region near $\nu={1\over 2}$ (Morf,
d'Ambrumenil, and Das Sarma, 2002). The only difference here is that
$C_a$ approaches a limit that depends on $\lambda$, a parameter they
set equal to zero (Coulomb case).

We next turn to
  Pan {\em et
 al} (2000) whose experiments  detected
  that the   {\em normalized  mass } defined by
 \beq
 m^{nor}_{a} = {m_{a}\over m_e\sqrt{B(T)}}
 \eeq
where $m_e$ is the electron mass and $B(T)$ is the field in Tesla,
is nearly the same for  $s=1$ and $s=2$, i.e., two and four flux
tubes.
  The theory predicts that
   $m^{nor}$ are  comparable for
  $s=1$ and $s=2$, but makes it clear that no fundamental significance can be
  attached to this result  since it depends on  $\lambda$,
   or more generally, the potential. Further details may be found
   in Shankar (2001).

\begin{table}
\caption{\label{cscale} Activation gaps as a function of $\lambda$
for $1\le \lambda \le 2$ according to the hamiltonian theory. Note
the convergence of $C_{a}^{(2)}$ as $p \to \infty$. The
superscript (2) refers the number of flux tubes attached.}
\begin{ruledtabular}
\begin{tabular}{|c|c|c|c|}
p  & ${\Delta_{a}^{(2)} / k_B}=50 \sqrt{B(T)}\delta_{a}^{(2)}$ &
$\delta_{a}^{(2)}$ &$
 C_{a}^{(2)}$
 \\ \hline
  1 & $ {5.31 \sqrt{B(T)}/ \lambda}$ & ${.106/ \lambda}$ & ${.32/ \lambda}$ \\
  2 & ${2.08 \sqrt{B(T)}/ \lambda} $& ${.042/ \lambda}$ &${ .21/ \lambda}$ \\
  3 &$ {1.23 \sqrt{B(T)}/ \lambda} $& ${.025/ \lambda}$ & ${.17/ \lambda}$ \\
  4 & ${0.87 \sqrt{B(T)}/ \lambda}$ & ${.017/\lambda}$ &$ {.16/ \lambda} $\\
\end{tabular}
\end{ruledtabular}
\end{table}

\begin{table}
\caption{\label {numbers}
  Approximate numbers used in this
 paper,
with  $k_B$ the  Boltzmann's constant, $m_e$ the electron mass,
and  $B(T)$ the field in Tesla. }
\begin{ruledtabular}
\begin{tabular}{|c|c|c|}
${e  B\over m_ek_B}= 1.34 B(T){}^oK$ &
 $ {e^2 \over \varepsilon
lk_B}= 50 \sqrt{B(T)} {}^{o}K$  &
 $ {\varepsilon \over e^2l}=.026\ m_e \sqrt{B(T)}$\\
\end{tabular}
\end{ruledtabular}
\end{table}

Consider next the experiments of Du {\it et al} (1993), who have
extensive data on activation gaps. We  will limit ourselves  to
$\nu \le {1\over 2}$, to which states with $1\ge\nu\ge\half$ are
related by particle-hole symmetry if full polarization is assumed.
    Given that the experiments, unlike Park, Meskini, and Jain (1999),
   have an unknown contribution from   LL mixing and impurities,
    it is not clear how to
   apply the theory. There is no {\em ab initio } calculation that
   includes these effects. (There is however reliable evidence that LL mixing is a very small
   effect at the values of $\lambda$ under consideration.
   Recall also our results from Section \ref{extended}.)

    We will compute gaps using
 the ZDS potential with  $\lambda  $ as a free parameter,
     and ask what $\lambda$ fits the data, just to get a feel for its
     size.
   The results are summarized in
Table \ref{gapexp}.

\begin{table}
\caption{Comparison of activation masses to Du {\em et al}, sample
A, which has a density $n=1.12\cdot 10^{11} cm^{-2}$. The last
column gives the best fit to $\lambda$.\label{gapexp}}
\begin{ruledtabular}
\begin{tabular}{|c|c|c|c|c|}
$ \nu \ \ \  $  &$B(T))$& $\Delta_{a}^{exp} ({}^oK)$  &
$\Delta_{a}^{theo} ({}^oK)$ & $\lambda$\\ \hline
 $ {1\over 3}$  &13.9 & 8.2 &$5.3 \sqrt{B(T)}/\lambda$ & 2.4  \\
 ${2\over 5} $& 11.6&3 & $2.08 \sqrt{B(T)}/\lambda$ &2.4  \\
${3\over 7}$  &10.8 &2 &$1.23 \sqrt{B(T)}/\lambda$ &2.0 \\
\end{tabular}
\end{ruledtabular}
\end{table}

Comparing the above values of $\lambda$ extracted from data to the
Local Density Approximation (LDA) (Price and Das Sarma, 1996, and
references therein) and exact diagonalization calculations (Park,
Meskini, and Jain, 1999, Morf, d'Ambrumenil, and Das Sarma, 2002),
which suggest $\lambda \simeq 1$, we see that disorder has a
substantial effect on activation gaps.

It is possible to compute  the charge density in a state with a
widely separated particle-hole pair in some gapped fractions. The
details, and a comparison to the unpublished work of Park and Jain
 may be found in Shankar (2001).

\subsection{Other potentials}

\begin{figure}
\includegraphics[width=3in]{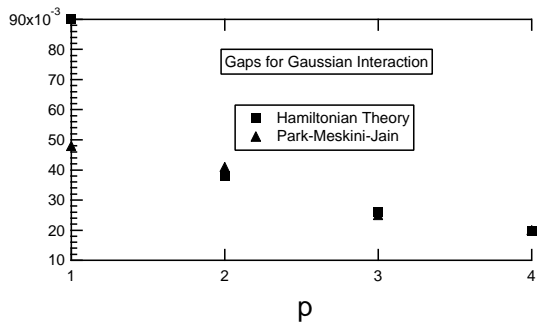}
\caption{\label{gaussian} Comparison of gaps for the gaussian
potential. For $p>1$ it is hard to separate the results from the
Hamiltonian theory and the wavefunction results of Park {\em et
al}. }
\end{figure}

Figure 8 shows a comparison to the Park, Meskini, and Jain (1999)
results for a gaussian potential \beq v(q) = {2\pi e^2 l} \
e^{-q^2l^2/2} \eeq Note that except for $\nu = {1\over 3}$ the
agreement is exceptional. This is the kind of ultraviolet-soft
potential for which the present theory works best, though
unfortunately we do not presently know of any system where it is
operative. At present, its study serves to instruct us on the
domain of validity of our approach. Not surprisingly, work on the
potential, \beq v(r) = {e^{-\kappa r}\over r} \eeq shows an
agreement that is worse than for the Coulomb case since this
potential is just as bad as $r\to 0$ and does not give the large
$r$ values a chance.  Likewise $1/r^2$ fares worse than $1/r$.

Note that if $\lambda \to \infty$ we may be in trouble since the
potential may not lead to  FQHE states.

{\em In summary, it appears as if our approach works well in any
problem which is not sensitive to distances smaller than about a
magnetic length.} Since the extended formalism  is mathematically
equivalent to the original electronic problem in the LLL; it is
the twin approximations -- the use of $\baro^p$ to deal with
constraints and Hartee-Fock -- that are responsible for deviations
from our benchmarks.

While our approach can reproduce the numbers of the wavefunction and
exact diagonalization approaches to within about $10 \%$, (if $\lambda
>1$), with a lot less work (a few seconds on a PC), without these
benchmarks, we would have known neither its range of validity nor
its degree of accuracy. Without such feedback, we would not have
applied the method with the such confidence to other phenomena not
treated by wavefunctions or exact diagonalization such as
relaxation rates and  polarization at $T>0$.

\section{Magnetic transitions at $T=0$.}

Now we turn to the behavior of the spin of the system, which so far
was assumed to be frozen along the applied field. The reader might
wonder why we bother, since the magnetic fields are of order at least
a few Tesla. The reason (originally pointed out by Halperin (1983)) is
that the Zeeman energy $E_Z=g^* ({e\over 2m})B$ is the smallest energy
scale in the problem. This is because of a conspiracy of two factors:
First, due to band structure effects, the band mass of the electron in
$GaAs$ is $0.068m_e$, with $m_e$ the electron's mass in vaccuum. This
makes the cyclotron frequency about 14 times what it would have been
in empty space. Next, due to spin-orbit coupling, the effective
$g$-factor of the electron in $GaAs$ is $g=0.44$ instead of the empty
space value of $2$. Thus, the Zeeman energy is about 64 times smaller
than the cyclotron energy $\omega_0$.  Typically the interaction
energy is of the same order as $\omega_0$, leading to $E_Z$ being the
smallest energy. Since the original realization by Halperin (1983),
spin transitions have been seen in experiment\footnote{Clark, Haynes,
Suckling, Mallett, Wright, Harris, and Foxon (1989), Eisenstein,
St\"ormer, Pfeiffer, and West (1989), Furneaux, Syphers, and Swanson
(1989), Buckthought, Boulet, Sachrajda, Wasilewski, Zawadski, and
Guillon (1991), Du, Yeh, St\"ormer, Tsui, Pfeiffer, and West,
(1995,1997). }, and been explored by exact diagonalization
\footnote{Chakraborty and Zhang (1984a,b), Rasolt, Perrot, and
MacDonald (1985), Chakraborty, Pietil\"ainen, and Zhang (1986).
}. We will compare our results with the recent work of
Park and Jain (1998, 1999) in the CF-wavefunction approach.

 The coupling of electron spin to the applied field is given by
the  Zeeman term
 \beq H_Z
=-g \left({e \over 2m_e}\right){S\over 2} B \eeq where $g=.44$,
$m_e$ is the electron mass in free space, $S$ is
  given by \beq
S=n\ P \eeq where $n$ is the density and $P$ is the polarization,
{\em to which each electron contributes $\pm 1$.}

Since the uniform external field couples to  the $q=0$ component
of the spin density which  is unaffected by the canonical
transformations, $H_Z$ will have the same form in the final CF
representation.

\subsection{Magnetic transitions in gapped fractions}

When $H_Z$ is large, we expect the system to be fully polarized
($P=1$). As we lower $H_Z$, we expect $P$ to drop. If
translationally invariant CF states are considered for the gapped
fractions, there is a discrete set of allowed values of $P$. At
$\nu =p/(2ps+1)$, these correspond to states of the form $|{\bf
p-r,r}\rangle$ in which $p-r$ CF-LL's are occupied by up spins and
$r$ CF-LL's by down spins.\footnote{In Jain's approach, the actual
wavefunction will be such a state times the Jastrow factor,
followed by projection to the LLL. In the present approach, $|{\bf
p-r,r}\rangle$ is literally the state, but the operators for
charge and spin are obtained by transformations to the CF basis.
For the interested reader we mention that these transformations
are spin-independent.} Thus the allowed values of polarization are
given  by
 \beq
P= {p-2r\over p} . \eeq
 For example, when $p=4$, the allowed
values are $P=1,.5,\mbox{and}\  0$ corresponding to $|{\bf
4,0}\rangle \ $,$|{\bf 3,1}\rangle \ $ and $|\bf{2,2}\rangle$.

Our goal is to calculate the critical fields at which the system
will jump from one value of $r$ to the next as $H_Z$ is varied.
Let
 \beq E (p-r,r)   = \langle {\bf p-r,r}|H|{\bf p-r,r}\rangle  \eeq
 {\em where $H$ does not contain the energy due to $H_Z$}.
 This will be the case for the single-particle and ground state energies,
 with one exception which will be clearly pointed out. Since $H_Z$  is diagonal in
 the
HF states  which   have definite spin, its effects can be
 trivially incorporated.

The HF calculation of $E(p-r,r)$ is detailed in the Appendix
\ref{bc}.

 The critical field $B^c$ for the transition from  $r$ to $r+1$ is
given by : \beq E(p-r,r)-E(p-r-1,r+1) = g{ e  B^c\over
2m_e}{n\over p} \label{criticalB}\eeq where the right hand side
denotes the Zeeman cost of flipping  the $n/p$  spins in  the  LL
that switched its spin. This discussion assumes that $B$ is
perpendicular to the sample. If there is a tilt $\theta$,  we
write
 \beq E(p-r,r)-E(p-r-1,r+1) = g{ e  B^{c}_{\perp}\over
2m_e \cos \theta }{n\over p} \label{criticalB2}\eeq

When these energy differences  are calculated, the same
remarkable regularity first noted by Park and Jain (1998, 1999)
emerges: they can be fit by a theory of free fermions of
mass $m_p$  {\em the polarization mass},  that  occupy LL's with a
gap $\Delta_p ={e B^*/ m_p}$. By this we mean the following. In a
free theory of gap $\Delta_p$, we would have

\beq E(p\! -\! r,r)-E(p\! -\! r\! -\!1,r\! +\! 1)= {n(p-2r-1
)\over p} \Delta_p \label{gapdefine} \eeq since    $(n/p)$ spin-up
fermions of energy $(p-r-1+{1\over 2})\Delta_p$ drop to the
spin-down level with energy  $ (r+{1\over 2}) \Delta_p$. Suppose
we evaluate the left-hand-side of Eqn. (\ref{gapdefine}) in the
HF approximation and {\em define}

\beq \Delta_{p}(r)^{def} = {p\over n} { E(p-r,r)-E(p-r-1,r+1)\over
p-2r -1}.\eeq

Given that  $H$ is not free, there is no reason why
$\Delta_p(r)^{def}$ should be  $r$-independent. But  it is very
nearly so. For example at $p=6, \lambda =1$,

\beq \Delta_p(0,1,2)^{def} = {e^2\over \varepsilon l}(0.00660,
0.00649,  0.00641) \eeq which describe the transitions $|{\bf
6,0}\rangle \to |{\bf 5,1}\rangle$, $|{\bf 5,1}\rangle \to |{\bf
4,2}\rangle$, and $|{\bf 4,2}\rangle \to |{\bf 3,3}\rangle$. This
$r$-independence of the gaps was true for every fraction and every
value of $\lambda$ we looked at. We will soon demystify this
apparent free-field behavior, which has a counterpart in the
gapless case as well.

The transition $|{\bf p-r,r}\rangle \to  |{\bf p-r-1,r+1}\rangle $
occurs when the spin-flip energy equals the energy difference of
the two competing ground states: \beq
 g{e\over 2m_e} {B_{\perp}^{c}  \over \cos \theta}
 =(p-2r-1)\Delta_p .\label{tran}
             \eeq

\subsection{Magnetic properties of gapless fractions}

Let us now turn to the gapless fractions ${1\over 2}$ and ${1\over4}$.
The discrete labels $p-r$ and $r$ of the HF states  that count the
spin-up and down LL's are now replaced by continuous variables
$k_{\pm F}$ which label the Fermi momenta of the spin-up and down
seas. These momenta are such that the total number of particles
equals $n$:  \beq k_{+F}^{2}+k_{-F}^{2} = k_{F}^{2}=4\pi n \eeq
where $k_F$ denotes the Fermi momentum of a fully polarized sea.

In the gapped case there were several critical fields $B^c$, each
corresponding to one more CF-LL flipping its spin, each describing
one more jump in the allowed values of $P$. In the gapless case the
situation is different. For very large Zeeman energy, the sea will be
fully polarized. It will not be worth including even one fermion of
the opposite spin since the Zeeman energy cost alone will exceed the
Fermi energy of the polarized sea. As we lower the Zeeman term, we
will reach a critical field at which it will be worth introducing one
fermion of the other spin with zero (effective) kinetic energy. At
this point the energy of a particle on top of the spin-up sea obeys
\beq {\cal E}_+(k_{+ F}) =g{e\over 2m_e} {B_{\perp}^{c} \over \cos
\theta}. \eeq If we lower the Zeeman term further, the polarization
will fall continuously and be determined by ${\cal E}_{\pm}(k_{\pm
F})$, the energies of the particles on top of these two seas according
to \beq {\cal E}_+(k_{+F}) -{\cal E}_-(k_{-F}) = g{e\over 2m_e}
{B_{\perp} \over \cos \theta}.  \eeq This equation states that the
system is indifferent to the transfer of a particle from one sea to
another, i.e., has minimized its energy with respect to polarization.

Since  the effective magnetic field vanishes at the gapless
fractions, we deal with a very simple expression for $\barro^p$:
\beq \bar{\bar{\rho}}^p({\bf q})=\int {d^2k\over 4\pi^2}(-2i) \sin
({{\bf q \times k }\ l^2\over 2}) d^{\dag}_{{\bf k - q}}d_{{\bf
k}}. \eeq It is easy to do a HF calculation and obtain
\begin{eqnarray*}
\lefteqn{{\cal E}_{\pm}(k))=}\\ && 2\int{d^2q \over
4\pi^2}\check{v}(q)\sin^2 \left[{{\bf k \times q}\ l^2\over
2}\right] \\ && -4\int{d^2k' \over
4\pi^2}n^{F}_{\pm}(|k'|)\check{v}(|{\bf k -k'}|)\sin^2 \left[{{\bf
k' \times k}\ l^2\over 2}\right]\\ &&\equiv {\cal E}_0+{\cal E}_I
\end{eqnarray*}
where the Zeeman energy is not included,  $n^{F}_{\pm}$ is the
(step)  Fermi function for the two species, ${\cal E}_0$ and
${\cal E}_I$ represent single particle energy (due to what was
called $H_0$ earlier) and the energy of interaction of this
particle at the Fermi surface with those inside the sea,  and \beq
\check{v}(k)=v(k)e^{-k^2l^2/2}. \eeq

When this result is used to compute ${\cal E_+}(k_{+F}) -{\cal
E}_-(k_{-F})$, we  find once again  that the numbers fit a free
theory in the following sense. Imagine that CF were free and had a
mass $m_p$. We would then have \beq {\cal E}_+(k_{+F}) -{\cal
E}_{-}(k_{-F})= {k_{+F}^{2}-k_{-F}^{2}\over 2m_p} \eeq What we
find is that the HF number for ${\cal E}(k_{+F}) -{\cal
E}(k_{-F})$ may be fit very well to the above form  with an $m_p$
that is essentially constant as we vary $k_{\pm F}$ i.e., the
relative sizes of the up and down seas (which is analogous to an
$m_p$ that does not depend on the index $r$ in the gapped case).
This constant   $m_p$ defined by \beq {1\over m_p} =2{{\cal
E}_+(k_{+F}) -{\cal E}_-(k_{-F})\over k_{+F}^{2}-k_{-F}^{2}}\eeq
matches smoothly with that defined for the nearby gapped
fractions.

This free-field behavior is surprising because  we will shortly
see that there are many reasons to believe that  the CF's are not free. For
now, note that  the HF energies are  not even  quadratic in
momenta: for example at $\nu ={1\over 2}$ and $\lambda =1$ there
is a hefty quartic term:
 \beq
{{\cal E}(k_{\pm F}) \over (e^2/\varepsilon l)} = a \left({k_{\pm
F} \over k_F}\right)^2 +b \left({k_{\pm F} \over k_F}\right)^4
\label{dispersion}\eeq where $a=.075$,    $b=-.030$.

There is  no reason the CF kinetic energy should be quadratic in
momentum. These particles owe their kinetic energy to
electron-electron interactions, and given this fact, all we can
say is that their energy must be an even function of $k$, starting
out as $k^2$ at small $k$. What constitutes small $k$ is an open
question that is answered unambiguously here: our expression of
the energy has substantial $k^4$ terms for momenta of interest,
but not from higher powers. An analogous result holds for the
gapped fractions, where the CF-LL's are not equally spaced (Murthy
1999, Mandal and Jain, 2001a)

The proper interpretation of this free-field behaviour will be
taken up next.

\subsection{Composite Fermions: Free at last?}
The fact that magnetic phenomena at $T=0$ can be described (to
excellent accuracy) by free fermions of mass $m_p$ (Park and Jain,
1998, 1999, Shankar, 2000, 2001) needs to be properly understood
and interpreted. In particular, one must resist the thought that
perhaps by some further change of variables one could take the
present hamiltonian and convert it to a free one. This is because
if there were really an underlying free theory, it would have a
single mass $m_{CF}$ for both activation and polarization
phenomena, with the CF's forming LL's of spacing $eB^*/m_{CF}$.
But we know from the extended  hamiltonian theory, Jain's
approach, or experiment, that there are two masses $m_a$ and $m_p$
that differ by at least a factor of two. Furthermore the shape of
the magnetoexciton dispersions (Kamilla and Jain, 1998, and Figure
5 also points to sizeable CF interactions: as $ql^2$, the distance
between the particle and hole varies, the (binding) energy varies
by an amount comparable to their individual energies, whose sum is
given by the value at large $ql$.

It has been shown (Shankar, 2000, 2001)   that a single assumption
about the form of the ground state energy, an assumption that is
not equivalent to the free-field assumption or even to a quadratic
dispersion relation in the gapless cases, will explain this
behavior for gapped and gapless fractions. Consider $E(S)$, the
ground state energy as a function of $S=nP$, where $P$ is the
polarization. By rotational invariance it must have only even
powers of $S$ in its series. Assume the series is dominated by the
first two terms: \beq E(S) = E(0)+{\alpha \over 2}S^2\eeq where
$\alpha$ is the inverse linear static susceptibility.

Consider first  the gapless case.
  When  $dn$  particles go from  spin-down to spin-up,
 \begin{eqnarray}
 dE &=&{\alpha   } \ S\  dS = {\alpha   }\  S \ (2 \ dn)\\
 &=& \alpha {k_{+F}^{2}- k_{-F}^{2}\over 4\pi} (2 \   dn)
 \end{eqnarray}
using  the volumes of the Fermi seas.
  We see that $dE$  has precisely the form of the kinetic
 energy difference of particles of mass  $m_p$ given by
\beq
 {1 \over m_p} = {\alpha \over \pi}.
  \eeq
 {\em Thus $m_p$ is essentially  the static susceptibility,
  which happens to have dimensions of mass in
 $d=2$. } The statement that $m_p$ has no $r$-dependence in the gapped
  case or no spin dependence in the gapless case is the same
 as saying that the full nonlinear susceptibility does not depend on the
  spin
 $S$, which in turn means $E(S)$ is quadratic in $S$.

  Note that the free-field form of $dE$
  comes from  $E \simeq S^2$   {\em and}
   $d=2$: in $d=3$, we would have $dE/dn \simeq  S
 \simeq (k_{+F}^{3}- k_{-F}^{3}) $ which no one would interpret   as a
 difference of  kinetic energies.

Let us see how this general argument applies to the  specific
example we have been working on.

   Consider the HF energies
 quoted earlier

\beq {{\cal E}(k_{\pm F}) \over (e^2/\varepsilon l)} = a
\left({k_{\pm F} \over k_F}\right)^2 +b \left({k_{\pm F} \over
k_F}\right)^4 \label{dispersion2}\eeq

The quartic terms miraculously drop out in the energy cost of {\em
transferring} a particle from the top of the spin-down sea to the
top of the spin-up sea:

\begin{eqnarray}
{dE \over (e^2/\varepsilon l)}&=&a{k_{+F}^{2}-k_{-F}^{2}\over
k_{F}^{2}}+b{k_{+F}^{4}-k_{-F}^{4}\over k_{F}^{4}}\\ &=&
a{k_{+F}^{2}-k_{-F}^{2}\over
k_{F}^{2}}+b{(k_{+F}^{2}-k_{-F}^{2})(k_{+F}^{2}+k_{-F}^{2})\over
k_{F}^{4}}\\ &=&{(a+b)\over k_{F}^{2}}(k_{+F}^{2}-k_{-F}^{2})
\end{eqnarray}
using \beq k_{+F}^{2}+k_{-F}^{2}=k_{F}^{2}. \eeq Note how $d=2$
was essential to this argument: in $d=3$ we would have
$k_{+F}^{3}+k_{-F}^{3}=k_{F}^{3}$.

 Thus the $k^4$ terms in ${\cal E}(k_{\pm})$ are not the cause
of the
 $S^4$ term. However,    a  $ k^6$ term
 in ${\cal E}(k_{\pm})$, can be shown to  produce an $S^4$  term in
 $E(S)$.

Thus the apparent free-field behavior is tied to the smallness of
terms of order $k^6$ and higher. To understand why the $k^6$ term
is so small, we turn to Eqn. (\ref{freeh}) for $H_0$. Expanding
the $sine^2$ in a series, we find the $k^6$ term is down by a
 factor of at least 15 (50) relative to the $k^2$ term, at $\lambda
=0$  ($\lambda =1$), all the way up to $k=k_F$. Presumably this
feature
 (and its counterpart in the gapped case)
persists in the HF approximation to  $H$ and keeps $E(S)$
essentially quadratic, which in turn mimics free-field behavior.

The reader is referred the original work (Shankar,  2001) for a
 proof that  $E(S) = E(0)+ {\alpha \over
2}S^2$ implies that $\Delta (r)$ will be $r$-independent in the
gapped case as well.

Composite Fermions are not free fermions but are like
Landau quasiparticles in a Fermi liquid\footnote{This was already
suggested by Halperin, Lee, and Read (1993). Also in this context,
see the work of Mandal and Jain (2001a).}. These objects too are
labeled by free-particle quantum numbers and long-lived. They do
have fairly strong interactions: the dimensionless Landau
parameters that describe these interactions are not small and
produce effects like zero sound. They are adiabatically connected
to free fermions in zero field just as the CF's are adiabatically
connected to free fermions in a reduced field $B^*$.

\subsection{ Effective potentials for experimental systems with disorder}

In comparing to experimental results one cannot neglect disorder. Our
theory ignores disorder and our results are completely determined by
the electron-electron interaction.  Here we ask if it is possible that
a ZDS potential with some effective $\lambda$ can describe a dirty
system. First of all, we realize this cannot be true with respect to
all observables, if at all it is true for any. For example, if one
were considering conductance, one knows the electron in a disordered
potential will typically get localized whereas no ZDS interaction will
predict this. As for transport gaps, the present day samples, with a
disorder broadening of the same order as the gaps, again preclude this
possibility. Magnetic transitions, on the other hand, are controlled
by total energies and one may expect that disorder will have a rather
innocuous effect and can be represented in an average way by some
translationally invariant interaction. We raise this issue because in
several magnetic phenomena to be described shortly, it appears that a
single $\lambda$ characterizes a sample. Specifically, $\lambda$
extracted from one data point can be used to explain the rest of the
data from that sample. If the other data points differ only in the
temperature $T$, the same $\lambda$ is used. If it differs in $B$ or
$n$ or $\nu$, the following scaling argument applies (Ando, Fowler,
and Stern, 1982): In a heterojunction, the donors of density $n$
produce a confining linear potential of slope that goes as $n$. If one
considers a variational wavefunction of the Fang-Howard (1966) form
$\psi (z) = A(w) z \exp (-z/w)$ in the transverse direction, then the
optimal $\bar{w}$ (to which $\Lambda$, the well width must be
proportional), varies as $\bar{w} \simeq n^{-{1\over 3}}$.
Consequently the dimensionless width, $\lambda = \Lambda / l$ varies
as \beq \lambda \simeq n^{-{1\over 3}}B^{{1\over 2}}\simeq
B^{1/6}\nu^{-{1\over 3}}\simeq n^{1/6}\nu^{-{1\over
2}}.\label{lamscale}\eeq

Arguments can be given (Shankar 2001) for why in certain limiting
conditions, not realized in today's experiments,  an effective
potential exists.  The question of why this works in realistic
situations far from this limit remains unanswered.

With this preamble, let us turn comparison with experimental
results. Kukushkin, von Klitzing, and Eberl (1999) vary both $n$ and
$B$ and drive the system through various transitions at $T=0$ (by
extrapolation). The field $B$ is always perpendicular to the
sample. We will compare the hamiltonian theory to these experiments by
calculating the critical fields at which the $\nu ={1\over 2}$ and
$\nu ={1\over4}$ systems saturate ($P=1$) and the gapped fractions undergo
transitions from one quantized value of $P$ to the next.

Let us recall that as far as these transitions go, the systems
behave like free femions of  mass $m_p$ which is independent of
the index $r$ which labels how many LL's have reversed their spins
in the gapped case or  the size of the up and down Fermi circles
in the gapless cases.

We consider $B^c$'s at which the systems at $1/4,{2\over
5},{3\over 7}, {4\over 9}, $ and ${1\over 2}$ lose full
polarization ($r=0$ for gapped cases, saturation for the gapless
cases) and, for $ {4\over 9}$, also  the $r=1$ transition,
$|\bf{3,1}\rangle \to |\bf{2,2}\rangle$.

An experimental complication needs to be addressed first. Each of
these transitions seems to take place via a narrow intermediate step
(Kukushkin, von Klitzing, and Eberl, 1999) with a polarization
half-way between the ones allowed by CF theory based on spatially
homogeneous states. We use the center of these narrow steps as the
transition points for comparison to the present theory. The physics of
these intermediate steps will be addressed in Section \ref{inhomo}.

In accordance with our strategy we find $\lambda$ by
fitting the theory to the experiment for a particular transition.
We will use this value (or for samples with changing field and
density, Eqn. (\ref{lamscale})) to predict $B^c$ for other
transitions using Eqn. (\ref{tran}). We obtain $\lambda_{{3\over
7}}=1.42$ from the transition $|\bf{3,0}\rangle \to
|\bf{2,1}\rangle$ at $B^c=4.5 T$.

 For the gapless cases, there are two equivalent
approaches. First, at the critical field  the Fermi energy of the
up spins equals the Zeeman energy of the down spins:
  \begin{eqnarray}
g\left[ {e  B^c \over 2m_e}\right]&=&{k_{F}^{2}\over 2m_p} = {2\pi
n\over m_p}={eB \nu \over m_p}
\end{eqnarray}

Equivalently we can write for the total ground state energy
density $E^Z(S)$, (where the superscript indicates that the Zeeman
energy is included),
 \beq
 E^Z(S)={\alpha \over 2 } S^2 -g {e\over 2m_e} {B_{\perp} S \over \cos \theta}
\eeq where  $\alpha  ={\pi / m_p}$. This expression  is minimized
(for$P\le 1$) to give
 $P$.
Setting $P=1$ gives the critical fields.

The comparison to experiment is made in Table (\ref{critical}.
Note that in rows above (below) ${3\over 7}$, where we fit
$\lambda$, the predicted $B^c$'s are lower (higher) than the
observed values, i.e.,  the actual $\lambda$'s are less (more)
than what Eqn. (\ref{lamscale}) gives.
  This is consistent with the
expectation that
 interactions
   will increase
 the effective thickness with increased  density.
If we  fit  to the ${2\over 5}$ point, we  obtain similar numbers,
with the agreement worsening as we move  off in density from
${2\over 5}$. Thus ${3\over 7}$ was chosen as the fitting point
since its density was somewhere in the middle of all the densities
considered.

\begin{table}
\caption{Critical fields based on a fit at ${3\over 7}$.The rows
are ordered by the last column which measures
density.\label{critical} }
\begin{ruledtabular}
\begin{tabular}{|c|c|c|c|c|}
 $ \nu \ \ \  $ & comment &  \  $B^{c}$ (exp) &\
$B^{c}$ (theo) &\ \ $\nu B^c$ (exp)
\\ \hline
  ${4\over 9}$ & $(3,1)\to (2,2)$ & \ \ 2.7 T & \ \ 1.6 T & \ \  1.2\\
 ${2\over 5}$ & $(2,0)\to (1,1)$ &\ \ 3 T & \ \   2.65 T & \ \  1.2\\
${1 \over 4}$ & saturation &\ \   5.2 T &\ \  4.4 T &\ \  1.3\\
${3\over 7}$ & $(3,0)\to (2,1)$ &\ \ 4.5 T&\ \   4.5 T & \ \
1.93\\ ${4\over 9}$ & $(4,0)\to (3,1)$ & \ \ 5.9 T & \ \  5.9 T &
\ \ 2.62\\ ${1\over 2}$ & saturation&\ \  9.3 T&\ \ 11.8 T&\ \
4.65
\\
\end{tabular}
\end{ruledtabular}
\end{table}

\section{Physics at nonzero temperatures $T>0$.}

So far we have seen that  the EHT may be used to compute
quantities such as gaps, particle-hole profiles, critical fields
for magnetic transitions and so on to 10-20\% accuracy. All such
quantities have been readily computed using trial wavefunctions,
giving numbers that are superior to ours. Our main emphasis has
been to expose the underlying physics as transparently as possible
and to resolve questions such as why CF's behave like free
particles on some occasions.

We turn to physics at finite $T$ (Murthy, 2000c, Shankar, 2000,
2001) where the hamiltonian method has few rivals. Exact
diagonalization (Chakraborty and Pietil\"ainen, 1996, Chakraborty,
Niemel\"a, and Pietil\"ainen, 1998) is limited to very small
systems and trial wavefunctions typically cover the ground state
and very low-energy  excitations. The hamiltonian approach is able to
yield the temperature dependence of polarization $P$ and the
relaxation rate $1/T_1$ for the gapless states in the
thermodynamic limit.  If $\lambda$ is treated as before (fit to
one data point per sample) we will see it is possible to give a
very satisfactory account of experiments in gapless systems up to
about $1{}^oK$, which is of the order of the Fermi energy
(Shankar, 2000, 2001).

We then address finite-temperature polarization in gapped states,
which is complicated by a nonzero spontaneous polarization and the
attendant spin waves. It turns out to be essential to take the
finite-$T$ behavior of these spin-waves into account. Once this is
done, the theoretical predictions are in excellent agreement with the
experiment up to several Kelvin (Murthy, 2000c).

We start with the gapless case since it is simpler.

\subsection{Polarization and relaxation  in gapless  states}

The polarization $P$ is computed as follows. First we compute the HF
energy of a particle {\em including the Zeeman energy}, which is the
self-consistent solution to
\begin{eqnarray*}
\lefteqn{ {\cal E}_{\pm}^{Z}(k)=}\\ &&\mp {1 \over 2}g \left[ { e
B\over 2m}\right] + 2\int{d^2q \over 4\pi^2}\check{v}(q)\sin^2
\left[{{\bf k \times q}l^2\over 2}\right] \\ && -4\int{d^2k' \over
4\pi^2}n^{F}_{\pm}(|k'|)\check{v}(|{\bf k -k'}|)\sin^2 \left[{{\bf
k' \times k}l^2\over 2}\right]
\end{eqnarray*}
where the superscript on ${\cal E}_{\pm}^{Z}$ reminds us it is the
total energy including the Zeeman part, the Fermi functions \beq
n^{F}_{\pm}(|k|)={1 \over \exp \left[ ({\cal E}_{\pm}^{Z}(k)-\mu
)/kT\right] + 1 }\eeq depend on the energies ${\cal
E}_{\pm}^{Z}(k)$ and the chemical potential $\mu$. Fig. 9 shows
some typical results. At each $T$, one must choose a $\mu$, solve
for ${\cal E}_{\pm}^{Z}(k)$ till a self-consistent answer with the
right total particle density $n$ is obtained. From this one may
obtain the polarization by taking the difference of up and down
densities. As usual we use the ZDS potential for which \beq
\check{v}(q) = e^{-q^2l^2/2}{2\pi e^2 e^{-ql\lambda} \over q}
.\eeq

\begin{figure}
\includegraphics[width=3in]{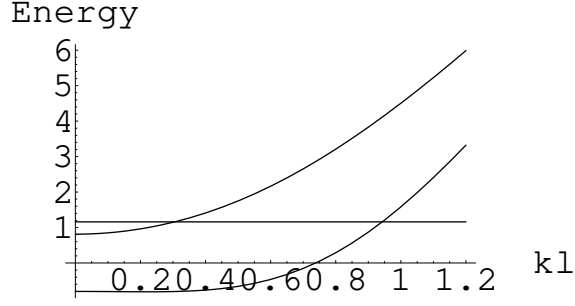}
\caption{\label{HFE}Hartree Fock energies at $\nu ={1\over 2}$ for
up and down spins (upper and lower curves) at $T=.3{}^oK$ at
$B=5.52 \ $ and zero tilt. Note that they are not simply quadratic
in momenta, and that at the chemical potential, indicated by the
horizontal line, the two graphs have very different slopes, i.e.,
density of states. }
\end{figure}

The computation of the longitudinal nuclear relaxation rate $1/T_1$ is
more involved (Shankar, 2000, 2001).  The question we ask is the
following. The fermions are in a quantum well, with their density
varying across the width. So the nuclear relaxation rate will be a
function of position.  Consider a nucleus at the center of the quantum
well, (as well as the $x-y$ plane) where the density is the
largest. Let us call this point the origin and let $1/T_1$ be the
relaxation rate here. The theory predicts
\begin{eqnarray}
1 \over T_{1} &=& 4\pi k_BT\left({K^{max}_{\nu}\over
n}\right)^2\nonumber \\ &&\! \! \! \! \! \! \! \times
 \int_{E_0}^{\infty} dE \left(
{dn^F(E)\over dE}\right)
\rho_{+}(E)\rho_{-}(E)F(k_+,k_-)\label{oneovert}\\
F&=&e^{-(k_{+}^{2}+k_{-}^{2})l^2/2}I_0(k_+k_-l^2)\\
 \rho_{\pm}(E) &=&\int{kdk\over
2\pi}\delta (E-{\cal E}_{\pm}^{Z}(k))\label{oneovert1}
\end{eqnarray}
where $k_{\pm}$  are solutions to ${\cal E}_{\pm}^{Z}(k_{\pm})=E$,
$I_0$ is the  Bessel function,  $E_0$ is the lowest possible
energy for up spin fermions, and $K^{max}_{\nu}$ is the measured
maximum Knight shift (at the center of the sample)  for the
fraction $\nu ={1\over 2}\ \mbox{or}
 \  {1\over 4}$.

Here is a rough description of the derivation, the details of
which may be found in   Shankar (2000). Suppose for a moment we
were dealing with electrons and not CF's. The Knight shift at the
chosen point, the origin, will be determined by the spin density
there. The same parameter enters the $1/T_1$ calculation
quadratically. This is why $K^{max}_{\nu}$ enters the answer. Thus
 $K^{max}_{\nu}$ is not calculated {\em ab initio} but
taken from the same experiment.  The density of states and Fermi
factor are standard. The only new feature here is the presence of
$F(k_+,k_-)$ which  reflects the fact that the spin density has to
be projected into the LLL when going to the CF basis. The effect
of this factor (which is none other than the $e^{-q^2l^2/2}$ which
appeared on the projected charge density) is to suppress processes
with momenta much larger than $1/l$, as these have no place within
the LLL.

\begin{figure}
\includegraphics[width=3in]{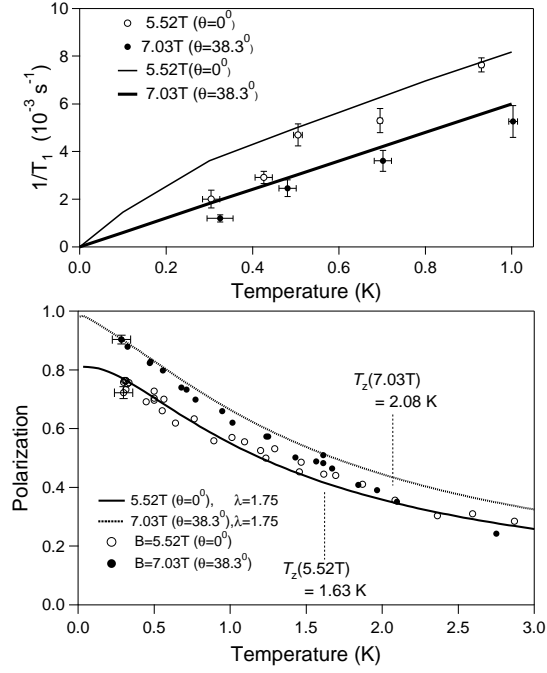}
\caption{\label{fig1}Comparison to the work of Dementyev {\em et
al}.
  The value of $\lambda$ is fit to $P$ at $300 \ mK,\  B_{\perp}=5.52\ T$
   and the  rest follows from the theory.
Notice the correlation between the curvature of
 $1/T_1$ and the limit of $P$ as
$T\to 0^{0}K$.  }
\end{figure}

We now compare to some experiments at $\nu={1\over 2}$ and $T>0$.
 Consider first Dementyev
{\em et al} (1999).  From their data point  $P=.75$ for
$B=B_{\perp}= 5.52 T$ at $300 \ mK$ We  deduce \beq \lambda =1.75.
\eeq  We have once again  chosen  to see to what extent a {\em
sole} parameter $\lambda$, can
 describe  $P$ and $1/T_1$ for the
given sample at a given $B_{\perp}$, but various temperatures and
tilts.

 Since there does not exist a model, including disorder,  that describes
   how $\lambda$
should vary with tilt we  include no such variation.

 Dementyev {\em et al}  (1999) find
$K_{{1\over 3}}^{max}= 4.856 \cdot 10^{-7}{}^{o}K$, which is
believed to describe a saturated system  at $P=1$. They estimate
that $K_{{1\over 2}}^{max}=.953 K_{{1\over 3}}^{max}$, which is
what we need here.  Given this information, $1/T_1$ follows.

 The top and  bottom halves of  Figure 10  compare the   HF calculation of
 $1/T_1$ and  $P$ respectively,  to the
  data. (The graphs for $1/T_1$ differ slightly from those in
   Shankar (2000)  since the present calculation
    treats the spin of the CF more carefully. The $1/T_1$ graph
    at $5.52 \ T$
      appears a little jagged since it was computed at just six points
      which were then connected.This is not apparent in the tilted
      case since the points lie on a straight line.

Dementyev {\em et al} (1999) had pointed out that a two parameter fit
(using a mass $m$ and interaction $J$), led to disjoint sets of values
(in the $m-J$ plane) for these four curves. Given that $H$ is neither
free nor of the standard form ($p^2/2m +V (x) $) this is to be
expected. {\em By contrast, a single $\lambda$ is able to describe the
data here rather well since $H$ has the right functional form.} Given
how the theory fits the polarization data up to the Fermi energy of
$\simeq 1^o K$, it is clear that changing the data point used to fix
$\lambda$ will be inconsequential.

   The present work establishes a phenomenological, nontrivial, and
nonobvious fact that a single $\lambda$ parameter, determined from one
data point, can describe both $P$ and $1/T_1$ for the given sample
under a variety of conditions.  That the fitted $\lambda$ is larger
than the Local Density Approximation value makes sense, as both
disorder and LL mixing will lower the gap and raise $\lambda$.

 Consider next sample M280 of Melinte {\em et al} (2000) which had
$P=.76 $ at $.06^{o}K$ and $B=B_{\perp} = 7.1 T $, from which we
deduced $\lambda =1.6$. Figure 11 compares the theoretical $T$
-dependence of $P(T)$ with data.  There is a factor of 2 between
theory and experiment for $1/T_1$, not shown.
\begin{figure}
\includegraphics[width=3in]{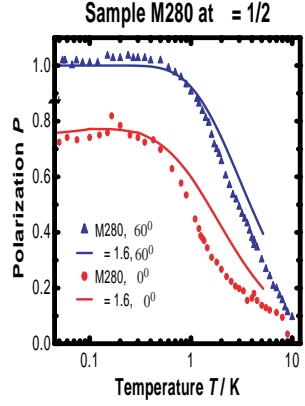}
\caption{\label{fig2}Comparison to  Melinte {\em et al} at $\theta
=0 \ and\  60{}^0$ with $\lambda =1.6$ fit to $P$ at $60 \ mK,\
B_{\perp}=7.1\ T$. }
\end{figure}

The fair  agreement for the tilted cases is unexpected in both the
Dementyev {\em et al} (1999) and Melinte {\em et al} (2000) data . First of all,
orbital effects have to be considered due to the tilt. The
thickness parameter $\Lambda$ can be affected by it. As pointed
out by Jungwirth in a private communication, once there is an
in-plane component of $B$, the problem is no longer rotationally
invariant. This means that our states are no longer HF states and
get scattered into each other by the potential. No attempt is made
here to take into account all the effects of the tilt. Instead we
included just the increased Zeeman coupling and hoped for the
best.

 For the benefit of others who measure $1/T_1$ at $\nu ={1\over 2}$ in
the future on similar samples, we give some  very approximate
formulae (to be used for zero or small tilts). From Figure 10, we
note that in general, the graphs of $1/T_1$ become linear and
parallel for temperatures above $.3\ {}^oK$. In this region we can
write \beq {d(1/T_1)\over dT}\simeq 3 \left[ {\bar{K}\over
\bar{n}}\right]^2 \cdot 10^{-3} s^{-1}\left[ {}^o K\right]^{-1} \
\ \ \  \mbox{for}\  T>.3 {}^o K\eeq with ${\bar K}$ the Knight
shift in KHz and \beq \bar{n} = {n \over 10^{10}/cm^2}\eeq

(In this approximate formula, we ignore the $\lambda$ dependence
of
 Eqns.
  (\ref{oneovert}-\ref{oneovert1}), and the distinction between the average and
maximum Knight shift.)

The graphs do not generally obey the Korringa-like law because as
$T\to 0$  they are sublinear (superlinear)for saturated
(unsaturated) cases. Only the critical case with $P(0)\to 1$ as
$T\to 0$ is linear. For $T>.3 \ {}^o \ K$, (which in general must
be replaced by either the energy gap or energy overlap between the
up and down Fermi energies) we have the approximate result
\begin{eqnarray} {1\over T_1} &\simeq & \left[ 3 \  T({}^oK) + C\right]
\left[ {\bar{K}\over \bar{n}}\right]^2 \cdot 10^{-3} s^{-1} \\
C&=& 0 \ \ \  \ \ \mbox{(critical)}\\
 &=& >0\ \  \ \  \mbox{(unsaturated)}\\
  &=& <0 \ \ \ \ \mbox{(saturated)}
  \end{eqnarray}

 For the critical case (only),  we have a  Korringa law
  \beq
{1\over T_1 \ T({}^oK)} \simeq \left[ 3 \   \right] \left[
{\bar{K}\over \bar{n}}\right]^2 \cdot 10^{-3} s^{-1} ({}^o\
K)^{-1} \eeq

For Dementyev {\em et al} (1999) $\ C\simeq 1$. This value may be used
as a first approximation.  For more accurate results one must solve
Eqns.  (\ref{oneovert}-\ref{oneovert1}).

\subsection{Polarization in gapped states}

Let us now turn to the theoretical description of the
finite-temperature polarization of the $\nu=\third$ state,
detailed measurements of which have recently been carried out
(Khandelwal {\it et al}, 1998, Melinte {\it et al}, 2000),
Freytag (2001). Theoretical details can be found in Murthy
(2000c).

While the CF-HF approximation provided an adequate description of
$\nu=\half$ it is {\it qualitatively} incorrect for a spontaneously
polarized state like $\nu =\third$. The reason is that it
underestimates the effects of the excitations that destroy order. For
example, it predicts a nonzero spontaneous polarization at $E_Z=g\mu
B_{tot} =0 $ and $T>0$, in violation of the Hohenberg-Mermin-Wagner
theorem (Mermin and Wagner, 1966, Hohenberg, 1967) that forbids
spontaneous breaking of a continuous symmetry in two dimensions,
except at $T=0$.  The $T=0$ spontaneous polarization is driven by the
fact that fermions of the same spin will avoid each other due to the
Pauli principle and thus have a lower interaction energy, just as in
the $\nu=1$ spontaneous quantum ferromagnet\footnote{The classic study
of nontrivial interaction effects (beyond ferromagnetism) at $\nu=1$
is Sondhi, Karlhede, Kivelson, and Rezayi (1993).}, which has been
extensively studied theoretically\footnote{Kopietz and Chakravarty
(1989), Read and Sachdev (1996), Kasner and MacDonald (1996),
Haussmann (1997), Timm, Girvin, Henelius, and Sandvik (1998), Kasner,
Palacios, and MacDonald (2000),} and experimentally\footnote{Tycko,
Barrett, Dabbagh, Pfeiffer, and West, (1995), Barrett, Dabbagh,
Pfeiffer, West, and Tycko, (1995), Aifer, Goldberg, and Broido (1996),
Manfra, Aifer, Goldberg, Broido, Pfeiffer, and West (1996). }.

Thus we must begin by understanding the disordering mechanism and
seeking a proper description of it. Consider a fully polarized $\nu
=\third$ state at $T=0$ with all the fermions in the CF-LLL state. As
the system is heated, some fermions will go to the $n=1$ CF-LL with
spin up (which does not change the polarization) and some will go to
spin down $n=0$ CF-LL, which reduces the polarization and costs an
energy per spin-flip of $\Delta_{SR}$, the {\em spin-reversed gap}.

This description completely misses the spin-waves which are
related to the particle-hole excitations as follows. Just as in
the case of the magnetoexciton (Figure 4) where no spin was
flipped, a spin-flip particle-hole excitation has an energy that
varies with $q$\ \footnote{For an example of this calculation see
Murthy (1999)}. As $q\to \infty$, the dispersion settles down at
$\Delta_{SR}$, the energy to create a widely separated pair.
However as $ q \to 0$ the pair energy vanishes, for $E_Z=0$, by
Goldstone's theorem. For nonzero $E_Z$ the long wavelength limit
for the spin-wave energy is $E_Z$ by Larmor's theorem.  These are
the modes to reckon with for they are very low in energy and
plentiful at low temperatures.

The simplest way to describe these spin-waves (including their
self-interaction) is the continuum quantum ferromagnet (CQFM)(Read
and Sachdev, 1996).  This model assumes all high-energy modes (at
the electronic and CF cyclotronic scales) have been integrated
out. The only modes left are slow and long-wavelength fluctuations
of the spin polarization, which have the action

 \eqa S=&\int d^dx\int\limits_{0}^{1/T} d\tau (iM_0 \bA
(\bn)\cdot\nabla_{\tau}\bn + {\rho_s\over2}(\nabla_{x}\bn)^2
\nonumber\\ &- M_0 {\bf {H}\cdot n}+\cdots) \eea where $M_0$ is the
magnetization density, $\bn$ is a local vector of unit length pointing
in the direction of the magnetization, $\bA(\bn)$ is the field that
implements the Berry's phase needed to obtain the correct quantum
commutation relations between the spin components, $\rho_s$ is the
spin stiffness, and $\bH =g^*\mu_B \bB $ is the Zeeman field
($|\bH|=E_Z$).

This model is still nontrivial (because of the condition $|\bn
|^2=1$). However it can be solved (Read and Sachdev, 1996) in the
limit when $N \to \infty$, where $N$ is the number of components
of $\bn$.  The limit $N \to \infty$  appears to describe the
actual case of $N=3$ in the case $\nu=1$.

Our strategy then will be deduce reasonable values for $M_0$,  the
magnetization density, and $\rho_s$, the spin stiffness, and then
plug them into the known results from the large-N limit for the
magnetization as a function of $T$ is given by \beq
P(T)=M_0\Phi_M(r,h)\label{em}\eeq where $r=\rho_s/T$ and $h=E_Z/T$
are scaling variables, and $\Phi_M$ is a known scaling function
(Read and Sachdev, 1996)\footnote{There are actually two different
large-$N$ approximations corresponding to the fact that the
symmetry group can be viewed as an example of $O(N)$ with $N=3$,
or as an example of $SU(N)$ with $N=2$. Both are considered by
Read and Sachdev (1996) and scaling functions given. }.

To find the values of the parameters $M_0$ and $\rho_s$
corresponding to the $\nu={1\over 3}$ state, we will use some
Hartree-Fock results. Since the underlying fermionic theory
responds to temperature by self-consistently modifying occupations
and energies, we expect to obtain temperature-dependent
values\footnote{This should be contrasted with the case $\nu=1$
where due to the huge exchange gap, particle-hole excitations are
frozen at all temperatures of interest, and the parameters of the
CQFM are $T$-independent.} $M_0(T)$ and $\rho_s(T)$.

First consider the spin stiffness $\rho_s(T)$. At a given temperature
$T$ the self-consistent occupations $N_{F,GS}(\s,n)$ and energies
$\e(\s,n)$ in the ground state are computed using the procedure
described in the gapless case. Now one creates a twisted spin state
and computes the HF energy of the twisted ground state (Murthy,
2000c), and thence the excess energy to order $q^2$. Comparing to the
energy cost of a twist in the CQFM, which is $(\rho_s/2) L^2q^2$, one
finds the spin stiffness \eqa
\rho_s=&{1\over16\pi}\int{d^2s\over(2\pi)^2} v(s)
\sum\limits_{n_1,n_2} |\rho_{n_1n_2}(s)|^2\times\nonumber\\ &
(N_F(\ua,n_1)-N_F(\da,n_1))(N_F(\ua,n_2)-N_F(\da,n_2))
\label{rhos}\eea where $L^2$ is the area of the system, and
$\rhot_{n_1n_2}$ is the  matrix element of equation
(\ref{rhomat}).  The above should be regarded as an estimate for
the twist rather than a rigorous calculation (even in HF), since
ideally one should compute the free energy cost of a twist,
rather than just the internal energy cost, as we have done.

To find $M_0(T)$ we need a more devious approach (Murthy, 2000c),
based on Eqn.  (\ref{em}). We already know how to compute the CF-HF
magnetization $P_{HF}(T)$. We set

\eq
P_{HF}(T)=M_0(T)\Phi_M(r=0,h=\Delta_{SR}/T) \label{m0}
\ee

and justify it as follows. In the CF-HF theory the particles and
holes are treated as independent, or noninteracting, with a gap equal
to $\Delta_{SR}$ independent of the distance between them.  This
corresponds to a collective mode dispersion that is completely flat,
$\omega(q)=\Delta_{SR}$. The CQFM description that corresponds most
closely to the CF-HF is the one that has {\it the same spin-flip
excitation spectrum}, namely one with no spin stiffness ($r=0$), and
an effective Zeeman field $E_Z^{eff}=\Delta_{SR}$.

\begin{figure}
\includegraphics[width=3in]{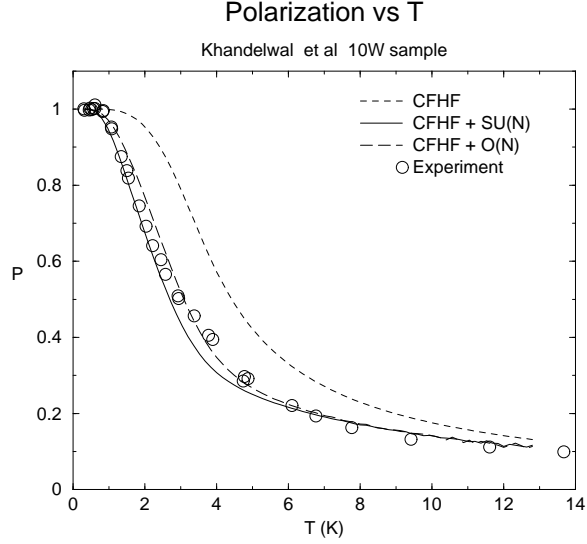}
\caption{\label{pol} Polarization $P(T)$ as a function of $T$ for
${1\over 3}$ for the data of Khandelwal {\it et al}'s (1998) 10W
sample. The thickness parameter has been set to $\lambda=1.5$.
Note the important role played by the spin-waves in bringing the
HF value of the polarization down to the nearly experimental
value.}
\end{figure}

Armed with $M_0(T)$ and $\rho_s (T)$, we evaluate the large-$N$
scaling functions and hence $P(T)$.  The predictions are shown in
Figure 12 for parameters corresponding to the experiment of
Khandelwal {\it et al} (1998). As can be seen, CF-HF overestimates
the spin magnetization considerably at low temperatures, but the
inclusion of interacting spin-waves gives a prediction in almost
perfect agreement with the data. The results are quite insensitive
to $\lambda$ and a typical value of $\lambda =1.5$ was used. The
results are also insensitive to which of the two large-$N$
approximations is used. Similar good agreement over a wide range
of temperatures is found in the comparison to the data of Melinte
{\it et al}'s (2000) M242 sample, shown in Figure 13.  Here we
have plotted the Knight shift vs. the temperature. The Knight
shift at very low temperature shows considerable scatter, and an
appropriate intermediate value has been used to fit to the theory.

\begin{figure}
\includegraphics[width=3in]{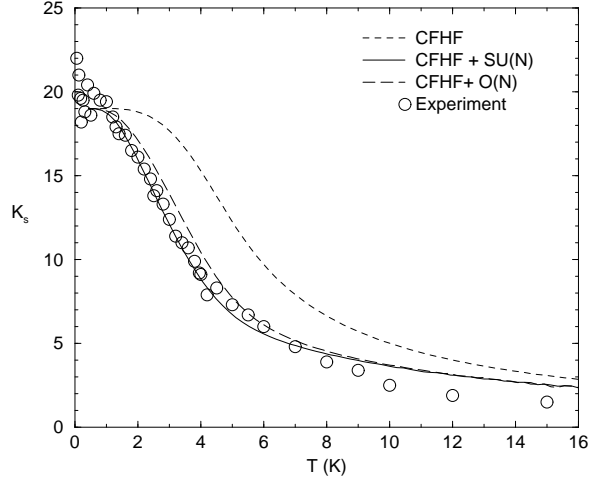}
\caption{\label{pol-melinte} Polarization $P(T)$ as a function of
$T$ for ${1\over 3}$ for the data of Melinte {\it et al}'s (2000)
M242 sample. The thickness parameter has again been set to
$\lambda=1.5$. Once again spin fluctuations are seen to be crucial
in reproducing the experimental polarization. }
\end{figure}

It is somewhat surprising that the theory agrees so well with the data
up to temperatures much higher than those for which agreement was
found for the gapless fractions. We expect the theory to work only
when CF's are well-defined, which should be true up to temperatures of
the order of the ${1\over3}$ gap. In fact, the agreement persists to
about 10$K$, which is higher than a typical activation gap for
${1\over 3}$ in such samples. For more details see Murthy (2000c).

The same considerations can be applied to the $\nu={2\over 5}$
state. Since it is unpolarized for $E_Z=0$ we can stop at the
CF-HF stage. Figure 14 shows the $P(T)$ curves for $\nu={2\over
5}$ for $\lambda=1.5l$ for a range of Zeeman couplings.  The rise
and fall of $P(T)$ at small $E_Z$ can be understood as follows. At
$T=0$, the system is in a singlet state with the  up and down
spins occupying the CF-LLL. As the system is heated, some
particles go the next LL. Here the ones with spin parallel to the
external field are preferred due to the Zeeman term and  hence a
nonzero polarization develops.  At higher $T$, the two spin
species start to get occupied with nearly equal probabilities and
$P$ starts to decline.  There is a transition to the fully
polarized state around $E_Z=0.01E_C$.

One may compute $P(T)$ for arbitrary fractions by exact
diagonalization (keeping all the excited states) and subsequent
calculation of thermodynamic quantities (Chakraborty and
Pietil\"ainen, 1996). Due to computational limitations, this
method is restricted to fairly small systems. For example, the
largest system studied by Chakraborty and Pietil\"ainen (1996) for
$\nu={1\over 3}$ has 5 electrons, and for $\nu={2\over 5}$ has 4
electrons. Allowing for this, the results seem fairly consistent
with ours. Currently experimental data are not available for
comparison.

Finally one use the CS theory to do finite $T$ calculations, using
$m^*$ as a free parameter instead of $\lambda$.

\begin{figure}
\includegraphics[width=3in]{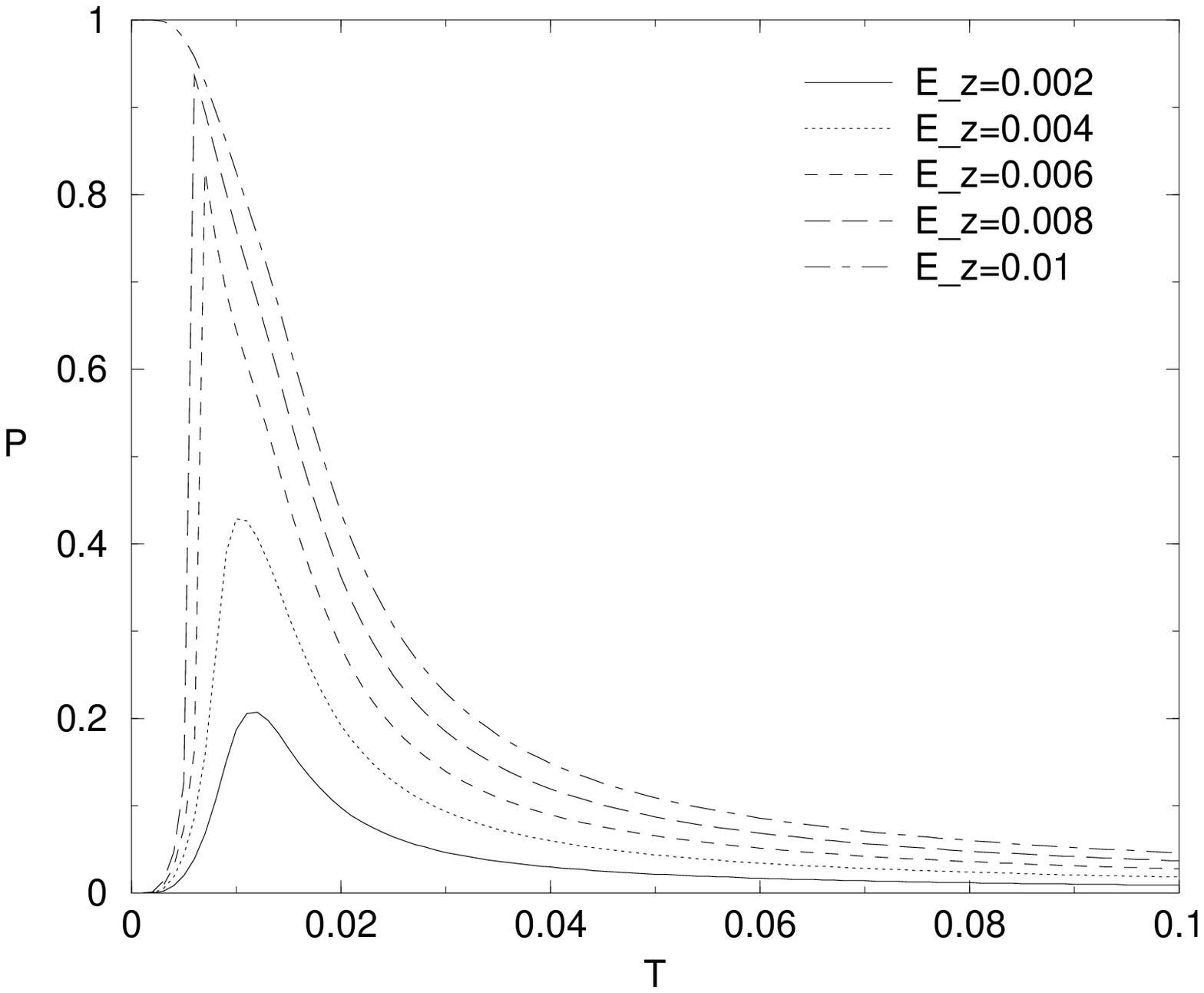}
\caption{\label{pol25}  Polarization as a function of $T$ for
${2\over 5}$. }
\end{figure}

\section{Inhomogeneous states of CF's\label{inhomo}}

We have repeatedly emphasized  that CF's are interacting particles.
The uniform liquid states studied so far were not contingent on CF
interactions-- such states could exist even for noninteracting CF's.
We now turn to some interesting inhomogeneous states, which would
not exist if the CF's were free, such as the high-field Wigner
Crystal, and possibly the partial polarized states  seen in the
gapped fractions by Kukushkin, von Klitzing, and Eberl (1999).

\subsection{The high-field Wigner Crystal}

The clearest instance of an inhomogeneous state in the fractional
quantum Hall regime is the high-field Wigner crystal (WC)
(Yoshioka and Fukuyama, 1979). If we start with a free electron gas
and slowly crank up the interactions, or lower the density, the
emphasis goes from the kinetic to potential energy. In the Wigner
crystal  the electrons seek an arrangement  aimed at minimizing
the potential energy. In the Hall case, due the quenching of
kinetic energy, the WC  seems even more likely  at low
densities.  It was initially proposed (Fukuyama and Platzman, 1982)
as a possible explanation of the fractional quantum Hall effect.
However, it was soon realized (Yoshioka and Lee, 1982) that this
state is very different from the fractional quantum Hall states,
because both its longitudinal and Hall conductances vanish at zero
temperature and when the crystal is pinned by disorder (Yoshioka
1983, MacDonald 1984), leading to an insulating state. Further,
the WC state is not tied to any particular commensurate filling.

There are now experimental observations\footnote{ Jiang, Willett,
St\"ormer, Tsui, Pfeiffer and West (1990), Jiang, St\"ormer, Tsui,
Pfeiffer and West (1991), Engel, Sajoto, Li, Tsui and Shayegan (1992),
Goldman, Wang, Su, and Shayegan (1993), Goldys, Brown, Dunford,
Davies, Newbury, Clark, Simmonds, Harris and Foxon (1992). For a
recent review of the experimental situation, see Shayegan (1997)}
which support the existence of the WC near $\nu={1\over 5}$. In fact,
experiments see a {\it re-entrant} transition (Jiang {\it et al},
1990, Jiang {\it et al}, 1991) in which there is a putative WC both
above and below ${1\over 5}$. To see how this might happen, recall
that the incompressibility of the Laughlin state results in a downward
cusp in the ground state energy as a function of filling factor in the
neighborhood of (say) ${1\over 5}$. Thus, one can easily imagine that
in a small neighborhood of ${1\over 5}$ the Laughlin liquid has a
lower energy than the WC, leading to the re-entrant WC near
${1\over5}$.

Theoretical work on the WC\footnote{Maki and Zotos (1983), Lam and
Girvin (1984), Levesque, Weiss and MacDonald (1984), Price, Platzman,
and He (1993). For a recent review of the theory,  see Fertig (1997). }, based on a study of trial wavefunctions and
collective excitations has established that the Laughlin state becomes
unstable around $\nu\approx {1\over 6}$ to a density wave even in the
absence of disorder. In the wavefunction approach, Hartree-Fock and
(weakly) correlated wavefunctions have also been written down, and
their energy evaluated. By studying the excitonic instabilities of the
Laughlin liquid, Jain and Kamilla (1998) showed that it becomes
unstable to crystallization around $\nu ={1\over 9}$.

Let us begin with   one of the simplest HF wavefunctions for the
crystal,  that of Maki and Zotos (1983):

\begin{equation}
\label{HF-trial-wf}
\Psi_{HF}(\{{\mathbf{r}}_{i}\})={\mathcal{A}}\prod _{i}\phi
_{{\mathbf{R}}_{i}}({\mathbf{r}}_{i}).
\end{equation}
where $ {\mathcal{A}} $ is the antisymmetrization operator, and $
\phi _{{\mathbf{R}}_{i}} $ is a single-particle wavefunction that
is localized at $ {\mathbf{R}}_{i} $ (lattice site) and belongs to
the LLL. It is given by
\begin{equation}
\label{single-particle} \phi
_{{\mathbf{R}}_{i}}({\mathbf{r}})=e^{-\left|
{\mathbf{r}}-{\mathbf{R}}_{i}\right|
^{2}/4l_{0}^{2}-i{\mathbf{r}}\times {\mathbf{R}}_{i}\cdot
\hat{z}/2l_{0}^{2}}.
\end{equation}
  The wavefunction (\ref{HF-trial-wf}) can be improved by adding a
correlation factor  corresponding to including small
fluctuations around the HF state. The energy of this state becomes
lower than that of the liquid state at about the experimentally right
filling fraction ($ \nu \approx \frac{1}{7} $) (Maki and Zotos, 1983,
Lam and Girvin, 1984, Levesque, Weiss, and MacDonald, 1984).

Experiments (Jiang {\it et al}, 1990, Jiang {\it et al}, 1991) show
transport gaps two orders of magnitude smaller than the theoretical
estimate as calculated using the Hartree-Fock approximation (Fukuyama,
Platzman, and Anderson, 1979, Yoshioka and Fukuyama, 1979). The
measurements of the Hall resistivity $ \rho _{xy} $ are surprising as
well (Goldman, Wang, Su, and Shayegan, 1993, Goldys {\it et al}, 1992)
. As mentioned above, the WC is expected to have a vanishing Hall
conductance $\sigma_{xy}=0$ (when pinned), which implies a vanishing
Hall resistance $\rho_{xy}=0$. However, the experiments see ``Hall
insulating'' behavior (Zhang, Kivelson, and Lee 1992), that is,
$\rho_{xy}\approx {2\pi \hbar \over \nu e^2}$. These problems led Yi
and Fertig (1998) to consider crystalline states with correlation
zeroes that keep electrons apart. Each electron is combined with $ 2s
$ vortices to obtain the trial wavefunction
\begin{equation}
\label{trial wf} \Psi (\{{\mathbf{r}}_{i}\})={\mathcal{A}}\prod
_{i\neq j}(z_{i}-z_{j})^{2s}\prod _{i}\phi
_{{\mathbf{R}}_{i}}({\mathbf{r}}_{i}).
\end{equation}
They also find that near $1/5$ the best energies are obtained by
attaching $4$ zeroes to each electron. In other words, they take
the Laughlin FQHE wavefunction at ${1\over 5}$ which is a product
of a Jastrow factor with quartic zeros and a single filled CF
Landau level $\chi_1$, and  replace $\chi_1$ with a crystal. The
Coulomb energy for this wavefunction is computed using Monte
Carlo methods. Yi and Fertig (1998) have shown that the ground
state energy of the correlated WC is lower than that of the usual
WC at experimentally relevant filling fractions.  Moreover by
introducing Laughlin-Jastrow correlations between the
interstitials and the lattice electrons the experimentally
observed $ \rho _{xy} $ (Zheng and Fertig, 1994a, 1994b) can be
explained. Unfortunately, the method becomes too computationally
demanding to allow one to calculate other quantities of interest,
such as the excitation spectrum.

This is a situation tailor-made for the EHT.  Before we
describe the theory and its results, let us note that there are two
features of the experiment which hint that a CF-WC is involved.  The
first is the nonmonotonic behavior of the gap near ${1\over5}$, which
shows that the ${1\over5}$ Laughlin liquid correlations are being felt
in the nearby WC state. Secondly, the threshold electric field beyond
which the WC becomes depinned and starts sliding {\it increases} near
${1\over5}$. Below we will see its connection with the structure and
properties of the CF-WC.

Now let us see how to set up the EHT. Since attaching the zeros to
electrons converts them into CF's, we are naturally led to consider a
WC of CF's.  Here is how one proceeds (details can be found in
Narevich, Murthy, and Fertig, 2001). The crystal is characterized by
density wave order parameters at the reciprocal lattice vectors $\bG$.

\begin{equation}
\label{order parameter} \Delta _{nn'}({\bG})={2\pi(l^*)^2\over
L^2}\sum _{X}e^{-iG_{x}X}<d^{\dagger
}_{n,X-G_{y}l^{*2}/2}d_{n',X+G_{y}l^{*2}/2}>
\end{equation}

First, one has to assume a particular lattice structure (shape and
size). An important parameter of lattice is the number of quanta of
effective flux that penetrate each unit cell\footnote{Since the CF's
see effective flux rather than external flux this point is crucial.}.
Things become simple when this number is rational, of the form $p/q$,
where $p$ and $q$ have no common factors. In this case each CF-LL
breaks up into $p$ sub-bands with the total number of states in the
original CF-LL being equally divided among the sub-bands (Yoshioka and
Fukuyama, 1979). While the original CF-LL had a sharp energy, the
sub-bands have a nonzero energy dispersion, which can be found from
the HF hamiltonian. One finally closes the circle by demanding
self-consistency; the ground state formed by filling up the sub-bands
with the correct number of particles should reproduce the assumed form
of Eqn. (\ref{order parameter}). Once one has a self-consistent HF
solution, the transport gap is found as the energy difference between
the centers of the highest occupied sub-band and the lowest unoccupied
sub-band. The WC is characterized by one particle  per unit cell,
and depending on the filling, this translates to different $p$ and
$q$.

Here are the results: The gaps calculated from the Hamiltonian
theory are within a factor of two of the experimental results,
(Figure 15)  which is to be contrasted to the more than two orders
of magnitude disagreement with electronic HF calculations. The
theory also predicts that the gap below $1/5$ should be
substantially more than the gap between $1/5$ and $2/9$ (the next
quantum Hall liquid state). Narevich, Murthy, and Fertig (2001)
are not able to go very close to $1/5$, because then $p$ and $q$
become large, and the problem becomes computationally prohibitive.

Two other results emerge at a qualitative level. First, the shear
modulus becomes very small as ${1\over5}$ is approached, which means
that the crystal becomes very soft to deformations. The reason is
that as ${1\over 5}$ approaches the WC tends more and more to a
Laughlin liquid. The density wave order parameter decreases, and
the shape of the crystal matters less and less.  This is connected
to the depinning threshold of the WC to electric fields. This
connection can be understood as follows (Fukuyama and Lee, 1978,
Blatter {\it et al}, 1994); when the crystal is stiff it is not
able to take advantage of all the minima in the disorder
potential, and so is not strongly pinned. However, when the
crystal becomes soft, it does not cost a lot of energy to deform
and take advantage of local minima of the disorder, and the
crystal becomes more strongly pinned. Disorder may also have a
dominant role in the behavior of the gap near ${1\over5}$. Since the
crystal is soft, its actual shape is determined by the local
disorder. Different shapes give gaps varying by about a factor of
two.

The second qualitative result is that the density inhomogeniety in the
CF-WC state is quite small in absolute terms (about 20\% of the
background density) for all $\nu$, not just near ${1\over 5}$.  This
is at odds with the conventional view that electrons are localized in
a WC. The correlation zeroes seem to prefer a more homogeneous
state\footnote{It is easy to see that correlation zeroes become
meaningless if electrons are strictly localized; then they do not come
close enough for the zeros to be operative.}. The electronic WC has to
partially melt and become more homogeneous to accommodate
Laughlin-Jastrow correlations.  This has interesting similarities with
earlier ideas concerning cooperative ring-exchanges in a WC, and the
melting of the WC (Kivelson, Kallin, Arovas, and Schrieffer,  1986a,b).

\begin{figure}
\includegraphics[width=3.2in]{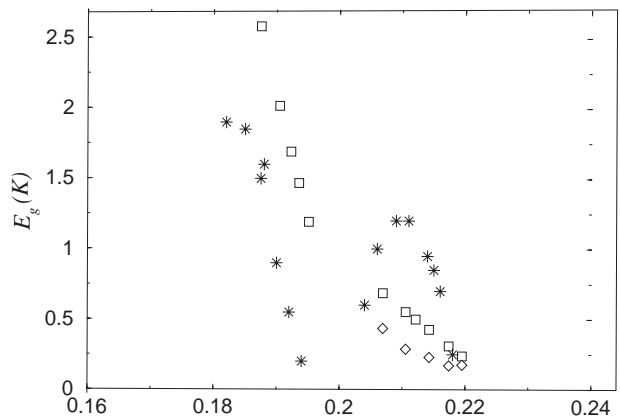}
\caption{\label{wignergap} Experimental (stars, from Jiang {\em et
al} 1991) and theoretical (squares and diamonds, from Narevich,
Murthy, and Fertig 2001) gaps versus $\nu$  for the Wigner
crystal. The squares are theoretical gaps for the triangular
lattice, while the diamonds represent gaps for an oblique lattice.
A change of shape of the lattice just above ${1\over 5}$ could
explain the nonmonotonicity of the experimental gap. }
\end{figure}

\subsection{Partially polarized Hall Crystal states}

Another class of candidates for an inhomogeneous state of CF's stems from
the observation of Kukushkin, von Klitzing and Eberl (1999), who
measured the spin polarization of various principal fractions as a
function of $E_Z$, the Zeeman energy and found polarizations not expected
in the CF theory with uniform states.  As an example, consider
$\nu={2\over 5}$. There are two filled CF-LL's, and at small $E_Z$
we expect that the state is unpolarized with the
$n=0,\uparrow$ and $n=0,\downarrow$ CF-LL's occupied and the rest
empty. At large $E_Z$ we expect a fully polarized state with
$n=0,\uparrow$ and $n=1,\uparrow$ occupied. If these are the only
states we are allowed to think about, then at some intermediate
$E_Z$ there should be a first order transition. Such transitions
between $P=0$ and $P=1$  were extensively discussed earlier.

However, Kukushkin, von Klitzing, and Eberl (1999) observe a
plateau at half the maximum polarization in a narrow region of
$E_Z$ for ${2\over5}$. Analogous plateaus are observed at other
principal fractions\footnote{Very recently, polarizations not
allowed by translationally invariant CF states have been seen at
${2\over 3}$ by NMR techniques as well: Freytag, Tokunaga,
Horvati\'c, Berthier, Shayegan, and L\'evy (2001).}.  Other
puzzling phenomena near this transition are that of hysteresis
over a range of fillings near $2/5$ (Cho {\it et al} 1998) and
very slow dynamics (Eom {\it et al} 2000). One of us (Murthy
2000a) has proposed that these observations might be explained by
considering partially polarized Hall Crystals of CFs. \footnote{
An alternate explanation is provided by Mariani {\em et al}
(2002).}

Let us first review some terminology. A state which shows the
co-existence of density wave and quantum Hall order (a quantized
$\sigma_{xy}$) is called a {\em Hall Crystal}. Such states were
implicit in the ring-exchange theory (Kivelson, Kallin, Arovas,
and Schrieffer, 1986a,b), but were first postulated explicitly by
Halperin, Tesanovic, and Axel (1986), and analyzed in great detail
with explicit examples by Tesanovic, Axel, and Halperin
(1989)\footnote{The topological numbers associated with these
states had been noticed before by a number of authors: Wannier
(1978), Thouless, Kohmoto, Nightingale, and den Nijs (1981),
Johnson and Moser (1982), Thouless (1983), MacDonald (1983), Dana,
Avron, and Zak (1985), Avron and Seiler (1985), Avron and Yaffe
(1986), Kunz (1986). However, Tesanovic, Axel, and Halperin (1989)
were the first to give a concrete example where nontrivial
topological integers were realized, and to make an important
connection between the topological integers and gapless collective
excitations. For an extension of the Hall Crystal concept to the
FQH regime, see Kol and Read (1993).}. A WC is an example of a
Hall Crystal, as is the lattice of quasiparticles formed at a
filling $\nu=\third+\delta$, provided the density wave is pinned.
The reason the partially polarized state at $2/5$ has to be a Hall
Crystal (if it is a crystal at all) is that no changes in the
quantized Hall conductance are observed as $E_Z$ is varied.

Now let us understand the nature of the proposed state (Murthy, 2000a). Imagine that
CF's were really free. Then as $E_Z$ is raised from $0$ the system
remains in a singlet state until the $n=0,\downarrow$ CF-LL (whose
energy is increasing) crosses the $n=1,\uparrow$ CF-LL (whose energy
is decreasing) at a special Zeeman energy $E_Z^*$.  This is when the
system makes the first-order transition. Now let us ``turn on''
interactions between CF's, which we know are present in the full
theory. At $E_Z^*$ there are twice the number of degenerate states
(from the $n=0,\downarrow$ and $n=1,\uparrow$ CF-LL's) as there are
CF's at the Fermi level. The true many-body ground state will
therefore be picked by the interactions. This state should be stable
in a small range of $E_Z$ near $E_Z^*$ determined by the strength of the CF
interactions.  One possibility is a charge/spin density-wave state.

As we saw in the previous subsection, when the system
self-consistently chooses a particular density-wave order with $p/q$
quanta of effective flux per unit cell, each CF-LL splits up into $p$
subbands of equal degeneracy (Yoshioka and Fukuyama, 1979). The
partially polarized ${2\over 5}$ state has two quanta of effective
flux penetrating every unit cell, which means $p=2$, $q=1$, and each
CF-LL is split into two subbands. The proposed state corresponds to
occupying 3 subbands of $\uparrow$-CFs and 1 subband of
$\downarrow$-CFs. Once again a CF-HF calculation allows us to
calculate the gap. The theoretical prediction for the range in $E_Z$
over which the state should be stable is consistent with the
experiment (Murthy 2000a).

This possibility can be investigated in more detail in the integer
quantum Hall analogue of $\nu=2/5$, which is $\nu=2$. There it can
be shown (Murthy 2000b) that there are indeed charge/spin density
wave states whose energy (in HF) is less than both the singlet and
fully polarized liquid states. It can also be shown that these
states are Hall Crystals with a nonzero quantized Hall conductance
(Murthy 2000b).

However, there are other possibilities for the partially polarized
state at $\nu=2/5$. For example, Apalkov, Chakraborty, Pietil\"ainen,
and Niemel\"a (2001), have proposed that the CF's belonging to the
$n=0,\downarrow$ and $n=1,\uparrow$ CF-LL's form a Halperin(111) state
(Halperin, 1983), which is a liquid. Their proposal is based on a
finite-size calculation in a spherical geometry where only the CF's
belonging to the two ``active'' levels were kept, and their
interaction was approximated as being a two-body interaction. While
their state remains a contender, Apalkov, Chakraborty, Pietil\"ainen,
and Niemel\"a (2001) wrongly conclude, by neglecting some terms in the
ground state energy (see Murthy 2001c), that the ground state energy
of the charge/spin density wave state would be higher than the
Halperin(111) analogue state.

At the moment it is fair to say that the situation is not
resolved. There are some key differences between the two proposals
which could be used to distinguish them experimentally. For example,
in the liquid state of Apalkov, Chakraborty, Pietil\"ainen, and
Niemel\"a (2001) one would expect no gapless excitations whatsoever,
and no spatial variation in the spin density, and thus the NMR Knight
shift. On the other hand, there are generically gapless density
excitations in the partially polarized Hall Crystal state (Tesanovic,
Axel, and Halperin 1989, Murthy 2000b), and one would expect (at very
low temperatures when the crystal is frozen) periodic spatial
variation in the Knight shift, with the maximum Knight shift being the
same as in the fully polarized state. A complete understanding of
these partially polarized states will hinge on future experiments.

\section{Critical review of EHT  and its approximate solutions}

The EHT was arrived at by starting with the electronic
hamiltonian, introducing a new (pseudovortex)  coordinate $\bR_v$,
combining it with the electronic guiding center $\bR_e$ to form
the CF variables $\etab$ and $\bR$. A (weak)  constraint $\barchi
\simeq 0$ completed the picture. For $\nu =1/2$ the formalism
coincides with that of Pasquier and Haldane (1998).

As an exact restatement of the original electronic problem, the
EHT is as good as any, with no loss of information, including that
pertaining to higher LL's. Its main claim to superiority over
other formulations is its amenability to Hartree-Fock and
attendant approximations, because the problem is now expressed in
terms of CF variables ($\etab$ and $\bR$) which have a natural HF
ground state that fills an integral number of LL's.

We discussed two approximation schemes, both of which rely on a
nondegenerate HF state, but which  differ in how the constraints
are handled. Since the separation of the LLL physics (and thus the
limit $m \to 0$) was straightforward in this formalism, we focused
on this limit and the LLL.

In a {\em conserving approximation}, one finds approximate Green's
functions that respect the constraints. The time-dependent
Hartree-Fock (TDHF) approximation is conserving. It was employed
(Murthy 2001a) to show that in the LLL, $S(q) \simeq q^4$ for
gapped states, in compliance with Kohn's theorem. For $\nu =1/2$
it was used by Read (1998) to show that the state is compressible,
and the fermion effective mass has a logarithmic divergence at the
Fermi surface, because the overdamped mode of HLR responsible for
all this is generated by summing ladder diagrams. Thus there is no
conflict between the HLR results based on CS fermions (that had
unit charge) and those based on Composite Fermions with charge
$e^*$ (which vanishes for $\nu =1/2$.)  However results that are
easily obtained in the CS approach are obtained only following a
careful, gauge-invariant calculation in the EHT.  It may well be
that without the CS results, the compressibility at $\nu =1/2$
would have been overlooked in a description that succeeded in
displaying dipolar expressions for the electron charge operator.
While the primary use of the conserving approximation thus far has
been to demonstrate that correct results in the infrared,
long-wavelength limit require a proper implementation of
constraints, it seems to numerically capture some of the
short-distance physics as well, as evidenced by the magnetoexciton
dispersions for $\third$ and ${2\over5}$ (Murthy, 2001a, section
VIB).

In the other approximation, ({\em the shortcut})  one  uses the
preferred charge $\barrop =\barro -c^2 \barchi$ in place of
$\barro$ on the grounds that in an exact calculation such a
substitution is permitted. This single substitution manages to
accomplish at tree level some of the work of the constraints: $e
\to e^*$,\ $d \to d^*, \mbox{and} \ \ S(q)\simeq q^2 \to
S(q)\simeq q^4$. The constraints are then ignored, so that their
effects at higher orders in $ql$ are not built in, and may or may
not be described well  by the simple HF calculation that follows.
Thus, although the EHT itself is exact, the small parameter $ql$
is in the shortcut needed because of the way we handled the
constraints. Note that it is $ql$ and not $ql^*$ that has to be
small, as evidenced by the success of the approach near and at
$\nu =1/2$ where $l^* \to \infty$.

Also missing in the shortcut is the overdamped mode and its
attendant consequences. However in problems with a gap or $T>0$
this does not seem to matter. On this basis one expects it to work
in problems where the potential favors small $ql$, and the problem
is still described by CF's\footnote{If one lets the "thickness"
parameter of the Zhang-Das Sarma potential become very large, CF's
will likely become unstable, and the state will not be describable
in the EHT.}.

These expectations were borne out in the computation of gaps,
magnetic transitions and $T$-dependence of polarization and
nuclear spin relaxation, which were compared to results from exact
diagonalization, trial wavefunctions and experiment. The very
special form the hamiltonian that comes out of this theory does
what {\em ad hoc} hamiltonians parametrized by standard kinetic
and interaction terms cannot: Explain a host of phenomena in any
given  sample (polarization, relaxation) with a single parameter
describing electro-electron interactions. We are not aware of any
other quantitative, analytical approximation scheme for computing
these quantities which is not plagued by singularities as  $m\to
0$.

The preferred charge description also answers many qualitative
issues: In what sense does an electron bind to its correlation
hole (represented here by pseudovortices) to form the CF? How can
CF's appear to be free in some magnetic phenomena when they are
surely not free?  What does the dipole picture really mean? How is
an effective mass generated from the interactions alone?

On the down side, the preferred charge as a way to implement
constraint is peculiar to the FQHE and  unprecedented. We do not
know {\em a priori} how well this procedure will work and how to
systematically improve it. We have no internal signal that the CF
description is failing as the interaction is varied (say by
sending $\lambda \to \infty$ in the ZDS interaction) or by mapping
$\nu
>1$  problems into effective LL problems with
modified interactions, where Pfaffian states and striped  states
might be the answer.

\section{Summary and open questions}

The goal of the hamiltonian theory was to start with the
hamiltonian for two-dimensional interacting electrons in a
magnetic field and arrive at a comprehensive description of
fractional quantum Hall phenomena by a sequence of transformations
to the final quasiparticles. The final theory was expected to
encode and display the wisdom inherited from the study of trial
wavefunctions, make the various pictures precise, and permit the
computation of all physical quantities (to some accuracy) by the
use of standard approximation techniques, such as HF.

Thus we first surveyed the wavefunction approach to see what it
could teach us and then described the two hamiltonian approaches:
the Chern-Simons (CS)  approach and our extension of it, the
Extended Hamiltonian Theory (EHT).

The wavefunctions taught us the following. The FQHE ground states
are described by incompressible fluids. The wavefunctions for the
ground states and quaiparticle (or hole) excitations are
independent of the mass $m$ (in the $m\to 0$ limit) and is built
entirely of LLL wavefunctions. In  the Laughlin fractions, the
elementary excitation is a vortex of charge $-1/(2s+1)$ in
electronic units. When an extra electron is introduced into the
system, it is screened by $2s$ such vortices, leading to a
composite fermion of charge $e^*=1/(2s+1)$. These vortices sit on
the electron. In the Jain fractions the wavefunctions are obtained
by projection to the LLL. Prior to projection one can see the CF
as made of an electron and $2s$ vortices sitting on them. After
projection, many zeros are annihilated, and typically move off the
electrons and are no longer organized into vortices. This holds
for $\nu =1/2$ as well, as is clear when $\nu =1/2$ is seen as a
limit of nearby Jain fractions. This means that the naive dipole
picture based on zeros of  the wavefunctions  is not tenable.
Remarkably, despite the dissolution of vortices on projection, the
CF still has the charge of an electron and $2s$ vortices. The
proper interpretation of this requires the hamiltonian approach,
in particular  the EHT, which in turn is an offshoot of  CS
theory,  which was described next.

{\em We emphasized  that our aim was not simply to get yet another
exact reformulation of the problem, but one that lends itself to
approximations.} For example the problem in terms of electrons:
\beq H = \sum_j {\etab_{ei}^{2}\over 2ml^4}+{1\over
2}\sum_{i,j,\bq }v(q)e^{ i\bq \cdot (\br_{ei} - \br_{ej})}\equiv
H_0 +V \label{primordial}\eeq contains all the answers in
principle, but is flawed in practice because of the  LLL
degeneracy of $H_0$ precludes the use of HF or perturbation
theory.

The degeneracy of the noninteracting problem is overcome by the
 flux attachment transformation of Chern-Simons
theory. Lack of space prevented us from discussing the very first
 application of this idea to the
 Laughlin fractions  by Zhang,
Hansson, and Kivelson (1989). They  converted  the problem  to one
of composite bosons in zero average field, which led to a detailed
analogy to superfluidity. This review focused on  CF's, turning to
the work of Lopez and Fradkin (1991,1992,1993) who implemented
Jain's idea in operator form, by  attached $2s$ flux quanta to
electrons
 converting them to CS fermions described by
\begin{eqnarray}
 H_{CS} &=& \sum_i {(   \Pib +    :{\bf a}_{cs}:)^{2}_{i}\over
2m}+V     \label{CSH2}
\\      \Pib &=&      {\bf p} + e{\bf A}^* \ \ \ \ ({\bf A}^*= {{\bf A} \over 2ps+1})\\
 \nabla \times
: {\bf a}_{cs} : & =&      4\pi s :\rho :
\end{eqnarray}

At mean-field one now has a unique vacuum of $p$-filled LL's of
CF's upon neglecting $V$ and $: {\bf a}_{cs} :$. Their effects
could be included perturbatively. In particular in an RPA
approximation one could include both these and obtain the
cyclotron mode at the right frequency, and in the Laughlin
fraction, the wavefunction in the long-distance limit (Lopez and
Fradkin, 1991,1992,1993).

However, two problems remained: A singular limit as $m$, the bare
mass of the electron, vanished, and a quasiparticle charge of
unity (at tree level) instead of $e^*$, owing to the fact that
flux tubes and not vortices were attached. These features  made it
hard to make quantitative predictions.

We mentioned Kalmeyer and Zhang (1992) who considered $\nu =1/2$
and pointed out that impurity scattering would be stronger than
naively expected since any charge inhomogeneity led to additional
gauge flux (by the CS condition)  which produced strong
scattering.

We then moved to the   Halperin, Lee, and Read (1993) treatment of
$\nu \simeq \half$, which was extensive and aimed at  confronting
theory with experiment at a quantitative level. The success of the
CF theory in the region of small or zero gap was quite unexpected.
They not only highlighted the fact that at $\nu =1/2$ the system
saw zero gauge field on average and therefore had a Fermi surface,
they also argued that the best way to think of the region near
$\nu=1/2$ was in terms of CF's seeing an effective field
$B^*=B-B_{1/2}$. This meant the particles would bend with a radius
$ R^*= {\hbar k_F \over eB^*} $ with $k_F=\sqrt{4\pi n}$ , a
result that was verified experimentally by Kang {\em et al}
(1993), Goldman, Su, and Jain (1994), and Smet {\em et al} (1996).

HLR computed the electromagnetic response within  the RPA. They
showed that the system was as compressible as a traditional Fermi
liquid.  It  had a longitudinal conductivity $\sigma_{xx}={e^2
\over 8\pi } {q\over k_F}$. This was tested by the damping and
velocity shift of surface acoustic waves by  Willett {\em et al}
(1990,1996) who also found that away from $\nu =1/2$, there was a
resonance in the velocity shift when the wavelength of the surface
acoustic wave coincided with $2R^*$.

HLR identified an overdamped mode in the density-density
correlation with dispersion $\omega \simeq iq^3v(q)$. This mode,
responsible for the compressibility of the system,  also produces
a (logarithmic) divergence in the CF mass by entering its
self-energy. This in turn leads to a gap $E\simeq 1/(p \ln p)$ for
$\nu =p/(2p+1)$ as $p \to \infty $. It has been shown that the
mass divergences do not affect bosonic (e.g., density-density)
correlations\footnote{Kim, Furusaki, Wen, and Lee (1994), Kim,
Lee, Wen, and Stamp (1994), Kim, Lee, and Wen (1995), Stern and
Halperin (1995).}

These successes notwithstanding, the HLR approach had some room
for improvement. It did not allow a clear separation of the LLL
physics, i.e., it did not have a smooth $m\to 0$ limit. The CS
fermion, obtained by attaching two flux tubes to electrons had a
charge of unity (and not $e^*=0$). It did not make contact with
the dipole picture (Read, 1994,1996) based on a wavefunction analysis.

The Extended Hamiltonian Theory (EHT) was devised by us to address
some of these issues. {\em The key idea  is that in order to
discuss the correlation hole that accompanies the electron to form
the CF in tractable form, one must enlarge the Hilbert space to
describe collective charge degrees of freedom and to place a
suitable number of constraints.} This  idea, along with
expressions for the new coordinates, and a certain change of
variables, all  arose in our original tortuous route (Shankar and
Murthy, 1997, Murthy and Shankar, 1998a) from electrons to CF's.
In this review, we spared the reader details of
 the historical route and gave an axiomatic description (Murthy
and Shankar, 2002) of the EHT.

Here we begin with the electronic hamiltonian  Eqn.
(\ref{primordial}), and add for  each electron a new pseudovortex
guiding center coordinate $\bR_v$ (whose algebra corresponding to
charge $-c^2$). Thus the Hilbert space is bigger now and $\bR_v$
has no dynamics.

If we are interested in physics at the cyclotron scale, we can
focus on the kinetic term. The mass $m$ appears here as it should,
in $\omega_0$. The collective coordinate formed out of $\etab_e$
carries the entire Hall current. (This answers the question of who
carries the Hall current at $\nu =\half$ when $e^*=0$.) In fact
they carry the Hall current even at gapped fractions where $e^*\ne
0$, thus substantiating the general belief that the Hall current
is not affected by disorder because it is carried by collective
coordinates. Having identified them as our oscillators, (for which
we have a specific hamiltonian) we have paved the way for a
detailed analysis of the Hall response. We also showed how extract
an effective LLL theory by approximately integrating out the
coordinate $\etab_e$ in our discussion of LL mixing.

 Let us now proceed to  drop higher LL's and work with
\begin{eqnarray}
\barH (\baro) &=& V={1 \over 2} \sum     \bar{\bar{\rho}}(q)\
v(q)e^{-(ql)^2/2}\ \bar{\bar{\rho}} (q)\\ \barro &=& \sum_j \exp
\left[ -i\bq \cdot (\br_j - {\bz \times \Pib_j\over
1+c})\right]\equiv \sum_j\exp \left[-i\bq \cdot \bR_{ej}\right]
\end{eqnarray}
The densities $\barro$ obey the magnetic translation algebra since
$\bR_e$ is the electron guiding center coordinate {\em in the CF
basis}. The advantage of this basis is that the velocity operator
$\Pib$ sees a weaker field $B^*$, leading to a nondegenerate HF
ground state.

The hamiltonian commutes with the operators \beq \barchi = \sum_j
\exp \left[-i\bq \cdot (\br_j +{\bz \times \Pib_j\over
c(1+c)})\right]\equiv \sum_j\exp \left[ -i\bq \cdot
\bR_{vj}\right]. \eeq The pseudovortex densities $\barchi$ thus
form a closed algebra, the symmetry algebra of $H$.

In the first derivation of the theory (Shankar, 1999), the
following constraint naturally emerged: \beq \barchi
|\mbox{physical state} \rangle=0. \eeq In the extended approach
(Murthy and Shankar, 2002), $ \bR_v$ is a cyclic coordinate with
no dynamics. One is free to supplement the theory with the above
constraint since nothing physical depends on the auxiliary
coordinate $\bR_v$. We made this choice so that there would be
just one set of equations to deal with.

We described  two ways to proceed, the choice being dictated by
 what we want to calculate; the conserving
approximation and the shortcut that used the preferred charge.

For situations where the symmetries of $H$ are important one uses
the conserving approximation, i.e., one in which the constraint is
respected at the level of Greens functions. This amounts to
ensuring  gauge invariance. We reviewed the compressibility
paradox where gauge invariance made all the difference. The
paradox concerned  the system at $\nu =\half$ which Halperin and
Stern (1998) argued must be compressible (as had been predicted by
HLR). This result seemed to be at odds with our operator
description in which the charge was dipolar. Halperin and Stern
gave heuristic arguments for how a system of dipoles could still
be compressible if their hamiltonian had $K$-invariance-
invariance of the energy under the shift all momenta, a symmetry
first noted by Haldane, and which appeared in our work as part of
a gauge symmetry. The more detailed analysis of Stern, Halperin,
von Oppen, and Simon (1999) drove the point home. They considered
first a model where the flux tubes were spread over a distance
$1/Q$ and the hamiltonian had $K$-invariance to sufficiently high
order in $Q$ to examine the question of compressibility. The
density-density correlation (now of the dipole-dipole form) was
computed in the RPA (which was exact in the limit $Q \to 0$) and
the factors of $q^2$ in the numerators from the dipoles were
cancelled by the exchange of the overdamped mode. They then showed
that in the actual FQHE problem, $K$-invariance could be built in
if the Landau parameter $F_1=-1$. The conserving approximation was
also employed by Read (1998) within the Pasquier-Haldane (1998)
formalism for $\nu =1$ bosons (to which our theory reduces if we
set $c=1$ or $\nu =\half$). Read (1998)  established that the
fermions interacted with each other as dipoles by exchanging a
transverse collective mode, but that as $q\to 0$, the propagator
for this mode produced enough negative powers of $q$ to overturn
the $q^2$ coming from the dipolar factors from the two ends. These
calculations are important in showing that if constraints are
taken into account, the correct charge of the CF will emerge and
that despite the weaknesses in the wavefunction based arguments
leading to the dipole picture (see Section \ref{dipole-picture}),
the predicted dipole moment is correct in the above sense. Murthy
(2001a) was able to show that in the gapped fractions one could
obtain structure factors in accord with Kohn's (1961) theorem
(${\bar{\bar S}} \simeq q^4$) in a conserving calculation. He also
showed how to compute magnetoexciton dispersions.

In all other situations (where a gap or temperature or both
suppress the deep infrared region $\omega \simeq q^3\ v(q)$)  the
shortcut using $\barrop$ was the weapon of choice. Here we make
the replacement $\baro \to \bar{\bar{\rho}}^p=\baro -c^2\bachi $,
which is allowed in the exact theory. This choice allows one to
employ naive Hartree-Fock calculations that respect Kohn's Theorem
($\bareS \simeq q^4$). As a bonus the CF charge and dipole moment
emerge in the power series expansion of $\baro^p$. It is only in
this sense that the CF can be viewed as the union of an electron
and a correlation hole of charge $-c^2$. With these features built
in at tree level, the usual approximations such as Hartree-Fock
are applicable as long as the large $ql$ region is avoided by the
potential.

The EHT gives a uniform and precise description of the internal
structure of the CF. Whereas in the wavefunction based
description, the Laughlin fractions allowed for a simple picture
of the CF (an electron bound to $2s$ vortices), and the rest of
the Jain series (upon projection) did not, in the EHT the CF is
viewed as an electron plus a pseudovortex. Especially interesting
is the case of $\nu =\half$. The expansion for $\baro^p$ in a
power series shows that it begins with a term that couples to an
external electric field exactly as a dipole moment of strength
$l^2\bz\times \bp$ would. This formula, in operator form, makes no
reference to zeros of the wavefunction or vortices, neither of
which is robust. It does not have the problems of the
wavefunction-based dipole picture. These problems arose upon
antisymmtrization, which we carry out by expressing $\baro^p$ in
second quantized form in terms of fermionic operators. we saw that
the place look for dipoles is not in the wavefunction, but in
correlation functions at high frequencies at and above the CF
Fermi energy.

The operator approach  gives a concrete realization of one of the
primary expectations in FQHE: Once the kinetic energy is quenched
by restricting electrons to the  LLL, it will be resurrected by
interactions and that this low-energy problem will be
characterized by  one common scale, the electron-electron
interaction.   Eqn. (\ref{energy}) which gives $H$ as a quadratic
function of  $\baro$, (or $\baro^p$ if we use the preferred
combination) embodies all these expectations.

The HF approximation to $H(\baro^p)$ was used to compute transport
gaps (Murthy and Shankar, 1999, Shankar, 2001) good to within
$10-20 \%$ for potentials that vanish rapidly with $ql$. It could
be used to explore magnetic transitions from one quantized value
of magnetization to the next (Shankar, 2000, 2001).  In the
absence of disorder it is clear that all physical quantities
pertaining to the FQHE (restricted to the LLL) are functionals of
the potential $v(q)$. For the Zhang-Das Sarma potential which we
use for most of our work, this means a function of $\lambda$. We
saw that given a value of $\lambda$ from experiment, we did not
need several masses for several phenomena, they all came from one
potential (Shankar, 2000, 2001).  This is because the hamiltonian
for the CF's is rather unusual\footnote{Consider for example Eqn.
(\ref{freeh}) for $\nu =\half$.}, and contains the kinetic and
potential terms in a monolithic form. Hidden in it are the various
mass scales $m_a$ and $m_p$ appropriate to various phenomena
(activation, polarization), all functionals of the interaction
$v(q)$.

The operator approach also clarified the question of whether or
not CF's are free. Given the prominent variations in the
magnetoexciton dispersions and the fact that it takes two very
different masses to describe polarization and activation, it is
clear that they are not. But why do they appear to be free for
some magnetic phenomena? Our theory (Shankar, 2000, 2001) shows
that it is an accident coming from rotational invariance and
$d=2$.

The EHT allows us to compute  physical quantities at $T>0$, such
as polarization and relaxation rates for gapless states
(Shankar,2000, 2001). In the experiments of Dementyev {\em at al}
(1999), a $\lambda$ determined from one polarization data point
gives the polarization and relaxation curves for  two tilts and a
range of temperatures. This is to be contrasted with attempts to
fit the data with a mass and interaction pair $(m,J)$, where the
four curves require four nonoverlapping values of these pairs. In
the case of Melinte {\em et al} (2000), Freytag (2001), the
$1/T_1$ predictions (that vary over orders of magnitude) are off
by a factor of $2$, but the polarization data are well described
by a single $\lambda$. Once again the success can be traced back
to the fact that the CF hamiltonian we use is of a nonstandard
form parametrized by a $\lambda$, which in turn determines all the
mass and energy scales relevant to each given process.

When calculating the polarization of gapped states, it turns out
to be essential to take into account spin-waves for spontaneously
polarized cases.  This is done (Murthy, 2000c) by mapping the
low-energy dynamics of the problem on to the continuum quantum
ferromagnet treated in the large-$N$ approximation (Read and
Sachdev, 1996), the parameters of which are extracted from the HF
treatment of the problem in the hamiltonian theory. The results
(Murthy, 2000c) agree extremely well with experiments (Khandelwal
{\it et al}, 1998, Melinte {\it et al}, 2000) up to very high
temperatures. This calculation, like all the others in the
Hamiltonian theory, is in the thermodynamic limit and free from
finite-size effects.

The EHT allows us to calculate the gaps of the Wigner crystal in
terms of CF's (Narevich, Murthy, and Fertig, 2001). These gaps are
off by a factor of $2$ when compared to experiment (Jiang {\it et
al}, 1990, Jiang {\it et al}, 1991). This should be contrasted to
previous approaches in which the gaps were off by two orders of
magnitude. This approach also allows us to consider inhomogeneous
states (Murthy, 2000a) with polarizations not allowed by the CF
theory of homogeneous ground states.

Hopefully we have succeeded in establishing that the hamiltonian
theory of the FQHE is a comprehensive scheme for addressing and
answering a variety of qualitative and quantitative questions, for
gapped and gapless states, at zero and nonzero temperatures.

We have given the reader a taste of what can be done using this
formalism. While many things have been clarified, there are many open
problems where we believe this approach may be fruitfully applied.

One outstanding open problem is that of
computing transport coefficients in the quantum Hall regime from a
microscopic theory. By identifying the collective mode which
carries the Hall current, we have set the stage for a study of
transport in the presence of disorder.

The FQH edge\footnote{For an excellent review, see Kane and Fisher
  (1997).} is another open problem where the hamiltonian approach is
applicable.  After Wen's (1990a,b, 1992) description of the edge as a
chiral Luttinger liquid, other descriptions have appeared based on
wavefunction and field-theoretic approaches\footnote{For
  wavefunction-based approaches, see: Z\"ulicke, U., and A.H.MacDonald
  (1999), Goldman and Tsiper (2001), Mandal and Jain (2001b). For
  field-theoretic descriptions, see: Lee and Wen (1998), Lopez and
  Fradkin (1999), Levitov, Shytov, and Halperin (2001).}. The
field-theoretic descriptions are all effective theories whose
connection to the electron problem has not been rigorously
established. While the wavefunction approaches are microscopic, it
is impractical to calculate time-dependent response functions in
them. As we have seen in the extended formalism, we have an exact
rewriting of the microscopic electron problem, but with the added
advantage of a nondegenerate starting point. The calculation of
edge reconstructions (MacDonald, Yang and Johnson, 1983, Chamon
and Wen, 1994) in CF-HF and the number and dispersions of the edge
collective modes in TDHF seem to be very accessible in the
hamiltonian approach. One of the interesting things to consider in
the edge problem is tunneling\footnote{The theory was described in
a series of
  beautiful papers by Kane and Fisher (1992a,b, 1994).  Recently there
  have appeared some truly remarkable tunneling experiments where the
  data span several orders of magnitude of voltage: Grayson {\it et
    al} (1998), Chang {\it et al} (2001)}.  For this one needs a
description of electron creation and destruction operators within the
CF basis, which is an open problem.

Since there is a gap, the effects of disorder on the Jain series ought
to be describable in some simple approximation, such as the
self-consistent Born approximation for single-particle properties.
Such a treatment seems to capture many of the experimental facts at
$\nu=1$ (see Murthy, 2001b), such as the reduction of the transport gap
due to disorder, the variation of the transport gap as a function of
$E_Z$ (see, for example, Schmeller {\it et al} (1995)), the
polarization at $\nu=1$ \footnote{Tycko {\it et al}
(1995), Barrett {\it et al} (1995), Aifer {\it et al} (1996), Manfra
{\it et al} (1996)}, etc. Such an approach ought to be applicable to
the gapped fractions. The effect of disorder on excitons is also
interesting, and can be treated in a "self-consistent exciton
approximation" (Kallin and Halperin 1985).

A more ambitious problem that might benefit from the hamiltonian
approach is the question of whether the $\nu=\half$ state remains
metallic at $T=0$, when disorder is included. The CS formalism  of
HLR (with its logarithmic corrections) suggests that  metallic
behavior  disappears.  It is known that noninteracting fermions
are always localized in two dimensions, regardless of how weak
disorder is (Abrahams, Anderson, Licciardello, and Ramakrishnan,
1979). By extension, it seems plausible that a Fermi liquid, which
is adiabatically connected to noninteracting fermions, should also
be an insulator on the longest length scales. The $\nu=\half$
system is rather unique, in that the state is produced by
interactions. The Extended hamiltonian has additional symmetries
absent in the zero-field problem.

We have excluded from this review many interesting topics to which
the hamiltonian  is  applicable, such as double-layer
systems\footnote{For some of the early theoretical references see:
Fertig (1989), Wen and Zee (1992, 1993), Ezawa and Iwazaki (1993),
Yang {\it et al} (1994), Moon {\it et al} (1995).  For a recent
review, see Girvin and MacDonald (1997): For recent experiments in
double-layer systems see: Spielman {\it et al} (2001), Kellogg
{\it et al} (2002)}, paired states\footnote{The experiment that
stimulated this subfield was the observation of a plateau in the
Hall resistance at $\nu={5\over2}$: Willett, Eisenstein,
St\"ormer, Tsui, Gossard, and English (1987). For some of the
theoretical references, see: Haldane and Rezayi (1988), Moore and
Read (1991), Greiter, Wen, and Wilczek (1992), Ho (1995), Park,
Melik-Alaverdian, Bonesteel, and Jain (1998), Morf (1998),
Bonesteel (1999), Read and Green (2000). }, etc. These omissions
reflect the double constraints of limited space and our own lack
of expertise.

\begin{acknowledgments}
Over the years, as we have tried to understand the FQHE, we have
acquired valuable insights from countless discussions with many of
our colleagues and gratefully acknowledged them in our various
papers. To mention them all here is unrealistic. We take this
opportunity to thank them all collectively, while apologizing for
the anonymity. This work, performed over a period of six years was
made possible from grants DMR-0071611 (GM), and DMR- 0103639 (RS)
from the National Science Foundation. Without this support, it
would not have been possible for us to pursue this fascinating
subject to our heart's content.
\end{acknowledgments}

\appendix
 \section{Matrix elements\label{matrixelement}}
\label{ME}
 Many of the calculations performed in this paper deal with the
 preferred density $\bar{\bar{\rho}}^p$. In second quantization we
 write it as
 \beq
 \bar{\bar{\rho}}^p({\bf q}) =
 \sum_{m_2n_2;m_1n_1}d^{\dag}_{m_2n_2}d_{m_1n_1}\rho_{m_2n_2;m_1n_1}\label{C1}
 \eeq
 where $d^{\dag}_{m_2n_2}$ creates a particle in the state
 $|m_2\ n_2\rangle$ where $m$ is the angular momentum and $n$
 is the LL index. They are related to the CF cyclotron and guiding
 center coordinates, ${\bf R}$ and $\mbox{\boldmath $\eta $}$  as follows. Let
 \beq
b = {R_x -iR_y\over \sqrt{2l^{*2}}} \ \ \ b^{\dag} = {R_x
+iR_y\over \sqrt{2l^{*2}}} \eeq where $l^* = l /\sqrt{1-c^2}$ is
the CF magnetic length.  These obey the oscillator algebra
 \beq
 \left[ b \ ,b^{\dag} \right] =1
  \eeq
  given
  \beq
  \left[ R_x , R_y \right] =-il^{*2}.
   \eeq
   Similarly we define, in terms of the cyclotron coordinates,
 \beq
a = {\eta_x +i\eta_y\over \sqrt{2l^{*2}}} \ \ \ a^{\dag} = {\eta
_x -i\eta_y\over \sqrt{2l^{*2}}} \eeq which obey the oscillator
algebra
 \beq
 \left[ a \ ,a^{\dag} \right] =1
  \eeq
  given
  \beq
  \left[ \eta_x , \eta_y \right] =il^{*2}.
   \eeq

   The states $|m n\rangle$ are just the tensor products
   \beq
|m n\rangle = {(b^{\dag})^m \over \sqrt{m!}}{(a^{\dag})^n \over
\sqrt{n!}}|00\rangle \eeq where $|00\rangle$ is annihilated by
both $a$ and $b$.

We    will now  show that

\begin{eqnarray} \langle m_2 |e^{-i{\bf q \cdot R}} |m_1\rangle
&=& \sqrt{m_2!\over m_1!} e^{-x/2} \left({-iq_+l^*\over
\sqrt{2}}\right)^{m_1-m_2}\nonumber \\ & &\times \
L_{m_2}^{m_1-m_2}(x) \label{Rmat} \end{eqnarray}
 where
 \beq
 x= q^2l^{*2}/2,\ \ \ \ \ \    q_{\pm} = q_x \pm  iq_y
 \eeq
  $L$ is the associated Laguerre  polynomial, and  $m_1\ge
 m_2$.
 If $m_1<m_2$ one may invoke  the relation
 \beq
\langle m_2 |e^{-i{\bf q \cdot R}} |m_1\rangle = \langle m_1
|e^{+i{\bf q \cdot R}} |m_2\rangle^* .\eeq

Likewise to  establish Eqn. (\ref{Rmat}), consider the coherent
states
 \beq
 |z\rangle = e^{b^{\dag}z}|0\rangle = \sum_{m=0}^{\infty}
 {|m\rangle \over \sqrt{m!}}z^m
 \eeq
 with the inner product
 \beq
 \langle \bar{z}|z \rangle = e^{\bar{z}z}
 \eeq
 First we write from the definitions given above
 \begin{eqnarray}
 \langle \bar{z}|e^{-i{\bf q \cdot R}}|z\rangle \nonumber\\
  &=& \sum_{m_1=0}^{\infty}\sum_{m_2=0}^{\infty}{\bar{z}^{m_2} \over \sqrt{m_2!}}{z^{m_1} \over
  \sqrt{m_1!}}\langle m_2 |e^{-i{\bf q \cdot R}} |m_1\rangle \label{R1} \\
  &\equiv & R(\bar{z},z,{\bf q}).
  \end{eqnarray}
On the other hand
\begin{eqnarray}
\langle \bar{z}|e^{-i{\bf q \cdot R}}|z\rangle &= & \langle
\bar{z}|\exp (-{il^*\over \sqrt{2}}(q_+b^{\dag}+q_-b))|z\rangle
\\
 &=& \langle \bar{z} - {il^*\over \sqrt{2}}q_+|{z} - {il^*\over
\sqrt{2}}q_-\rangle e^{q^2l^{*2}/4}\\ &=& \exp \left[ \bar{z} z -
{il^*\over
\sqrt{2}}(\bar{z}q_-+q_+z)\right]e^{-q^2l^{*2}/4}\label{R2}
\\
 &\equiv &
R(\bar{z},z,{\bf q})
\end{eqnarray}
Comparing Eqns. (\ref{R1}-\ref{R2}) and matching powers of
$\bar{z}^{a}z^b$ we obtain Eqn. (\ref{Rmat}) if we recall\beq
L_{m_2}^{m_1-m_2} (x) = \sum_{t=0}^{m_2} {m_1! \over (m_2- t)!
(m_1-m_2+t)!}{(-1)^t\over t!} x^t \eeq

Likewise, to establish
 \beq \langle n_2 |e^{-i{\bf q}
\cdot \mbox{\boldmath $\eta $}} |n_1\rangle = \sqrt{n_2!\over
n_1!} e^{-x/2} \left({-iq_-l^*\over
\sqrt{2}}\right)^{n_1-n_2}L_{n_2}^{n_1-n_2}(x)\label{etamat}\eeq
(again for $n_1\ge n_2$) we just  need to remember that the
commutation rules of the components of  $\mbox{\boldmath $\eta $}$
have a minus sign relative to those of ${\bf R}$, which  exchanges
the roles of creation and destruction operators and hence $q_+ $
and $q_-$.

Now we consider matrix elements of {$\  $ $\bar{\bar{\rho}},
\bar{\bar{\chi}},\ \bar{\bar{\rho}}^p \ $}. As a first step, let
us express the operators ${\bf R}_e$ and ${\bf R}_v$ in terms of
CF guiding center and vortex coordinates ${\bf R}$ and
$\mbox{\boldmath $\eta $}$.
 We have seen that  in the CF  representation
  \begin{eqnarray}
   {\bf
R}_e &=& {\bf r}-l^2{\hat{\bz}\times {\bf \Pi} \over 1+c}={\bf
R}+\mbox{\boldmath $\eta $} c
\end{eqnarray} if we recall $l^2 = l^{*2}(1-c^2)$.

It can similarly be shown that
  \begin{eqnarray}
   {\bf R}_v &=& {\bf r}+l^2{\hat{\bz}\times {\bf \Pi} \over
c(1+c)}={\bf R}+\mbox{\boldmath $\eta $} /c.
\end{eqnarray}
Thus in first quantization\begin{eqnarray}
 \bar{\bar{\rho}}^p &=& \bar{\bar{\rho}}- c^2  \bar{\bar{\chi}}\\
 \bar{\bar{\rho}}&=& \sum_i  \exp (- i {\bf q \cdot
  R}_i) \exp (- i {\bf q} \cdot
\mbox{\boldmath $\eta $}_ic)\\ \bar{\bar{\chi}}&=& \sum_i \exp (-
i {\bf q \cdot
  R}_i) \exp (- i {\bf q \cdot} \mbox{\boldmath $\eta $}_i/c)  \\
 \end{eqnarray}

 Armed with Eqns. (\ref{Rmat} and \ref{etamat}  ) we may finally
 write for the matrix elements of $\barro$ defined in Eqn.
 (\ref{C1}),
 \begin{eqnarray*}
\lefteqn{ \bar{\bar{\rho}}_{ m_2 n_2;m_1  n_1}=} \\
 & &\sqrt{m_2!\over m_1!} e^{-x/2} \left(
{-iq_+l^*\over \sqrt{2}}\right)^{m_1-m_2}L_{m_2}^{m_1-m_2}(x)
 \\
 & &\bigotimes \left[ \sqrt{n_2!\over n_1!}
\left({-icq_-l^* \over
 \sqrt{2}}\right)^{n_1-n_2}e^{-x c^2/2}L_{n_2}^{n_1-n_2}(xc^2)\right. \\
 & &\left. -c^2 \cdot  f\cdot
\left({-iq_-l^*\over \sqrt{2}c}\right)^{n_1-n_2}
e^{-x/2c^2}L_{n_2}^{n_1-n_2}(x/c^2)\right]\\
 & &\equiv
\rho_{m_2m_1}^{m}\bigotimes \rho_{n_2n_1}^{n}
 \end{eqnarray*}
 Superscripts on $\rho_{m_2m_1}^{m}$ and $\rho_{n_2n_1}^{n}
 $ which will be
   apparent from the subscripts, will usually  be suppressed.

\section{Hall response in the extended picture\label{hallext}}
To compute the DC Hall conductance we just need the zero momentum
component of the current operator  \beq \bJ (0)= {\partial H\over
\partial \bA}_{\bq=0} = {e\over m} \sum_j \bPi_{j}=-{e\over
ml^2}\sum_j \hat{\bz}\times \etab_{ej} .\label{coll}\eeq Note that
the current at $q=0$ depends only on the cyclotron coordinate,
just as it depended only on the oscillator coordinate in the small
treatment. Upon coupling the system to an external potential $\Phi
(\bq)$, $H$ becomes  \begin{eqnarray} H &=& \sum_j
{\etab_{ej}^{2}\over 2ml^4}+e\Phi (\bq)\sum_j e^{-i{\bf q \cdot
r}_j}
\end{eqnarray}
If we now keep the interaction to ${\cal{O}} (\bq)$, and recall
$\br_e = \bR_e + \etab_e$, we obtain \beq H=\sum_j
{\etab_{ej}^{2}\over 2ml^4}+e\Phi (\bq)\sum_j (-i{\bq \cdot
(\bR_{ej} +\etab}_{ej}))+\ldots \eeq If we complete squares on
$\etab_{ej}$, we can read off its   mean value in the presence of
$\Phi$, or the corresponding electric field $i\bq \Phi $. From
this we  obtain a mean current $\bJ (0)$ corresponding to the
right Hall conductance of $ne/B$.

The exercise should make it clear that the oscillator coordinates
are just the collective coordinates formed from $\etab_e$ at small
$q$. If we go to higher orders in $q$ we will find that the
current involves both $\etab_e$ and $\bR_e$. Our extended
formalism allows one to explore corrections due to this mixing in
a small-$q$ expansion.

Note that in this approach the magnetic moment of Simon {\em et
al} need not be put in by hand since $\etab+e$ is still; in the
picture and can respond to a slowly varying magnetic field by a
changing zero-point energy.

The objections of Lee {\em et al}(1999) on the CF Hall conductance
are moot since we do not add resistivities as in Eqn.
(\ref{rhoadd}) but rather  conductivities (of the $\etab_e$  and
CF variables).

\section{\label{HFProof} Proof of Hartree-Fock nature of trial
states} Consider \beq \langle f|H|i\rangle = \langle {\bf
p}|d_fHd^{\dag}_{i}|{\bf p}\rangle \eeq where $|{\bf p}\rangle$
stands for the (ground) state with $p$-filled LL, and $i,f$ label
single-particle excitations on top of this ground state. We want
to show that this matrix element vanishes if $i\ne f$, i.e., the
hamiltonian does not mix these putative  HF particle states. (This
result was established for the small $q$ theory by Murthy, 1999).
The proof, which relies on just the rotational invariance of the
potential, applies with trivial modifications to the hole states,
i.e., to \beq \langle {\bf p}|d_{f}^{\dag}Hd_{i}|{\bf p}\rangle
.\eeq

This matrix element in question  takes the schematic  form \beq
 \langle f|H|i\rangle = \int_q \langle {\bf p}|d_f
 d_{1}^{\dag}d_2d_{3}^{\dag}d_4 d_{i}^{\dag}|{\bf p} \rangle
 \rho_{12}({\bf q})\rho_{34}(-{\bf q})
 \eeq
 where $1$ stands for $m_1n_1$ and so on, and   $\int_q$ stands
 for an integral over {\em a rotationally
 invariant measure}:
 \beq
 \int_q ={1\over2} \int {d^2q\over 4\pi^2}v(q) e^{-q^2l^2/2}.
 \eeq

 Now we use Wick's theorem and perform pairwise contractions on
 the vacuum expectation value,
 bearing mind that
\begin{itemize}
  \item We cannot contract the indices $1$ and $2$ or $3$ and $4$
  since this will require that  $q=0$ at which point the measure
  (which contains the potential) vanishes.

  \item If we contract $i$ and $f$ we already have the desired result.
\end{itemize}

Here is a representative of the contractions we can get:
\begin{equation} \rho_{12}\rho_{34}
\delta_{f1}(1-n_{1}^{F})\delta_{23}(1-n_{2}^{F})\delta_{4i}(1-n_{4}^{F}).
\eeq where $n_{1}^{F}$ is the Fermi function for the LL labeled by
$n_1$ \beq n_{1}^{F}=\theta (p-1-n_1) \eeq and so on. Since $f=1$
and $i=4$, the factor  $(1-n_{2}^{F})(1-n_{4}^{F})=1. $
 The integrand assumes  the form
\begin{eqnarray*}
\lefteqn{\sum_{m_2=0}^{\infty}\sum_{n_2=p}^{\infty}\rho_{f2}({\bf
q}) \rho_{2i}(-{\bf q})=}\\ & & \left[
\sum_{m_2=0}^{\infty}\rho_{m_fm_2}({\bf q})\rho_{m_2m_i}(-{\bf
q})\right] \left[ \sum_{n_2=p}^{\infty}\rho_{n_fn_2}({\bf
q})\rho_{n_2n_i}(-{\bf q})\right]
\\ &=&
\delta_{m_fm_i}\sum_{n_2}q_{-}^{n_i-n_f}F(|q|)
\end{eqnarray*}
where we have also used the fact that $e^{-i{\bf q \cdot R}}\cdot
e^{i{\bf q \cdot R}}=I$ (the identity operator) in doing the sum over
$m_2$, and $F(|q|)$ is some rotationally invariant function.  It
follows that every term in the sum over $n_2$ vanishes unless
$n_f=n_i$ due to the angular integral in ${\bf q}$.

\section{On the conserving nature of TDHF\label{consTDHF}}

We will now verify that  the constraint is a left-eigenvector of
${\cal H}$, as required of the conserving approximation.

To this end we  will need the matrix elements  \eqa
\tilde{\rho}_{n_1n_2}(\bq)
&=&<n_1|e^{-ic\bq\cdot\etab}|n_2>\label{rhomat}\\
\tilde{\chi}_{n_1n_2}(\bq)&=&<n_1|e^{-{i\over
c}\bq\cdot\etab}|n_2> \eea Note that only   the cyclotron part of
the electron and pseudovortex coordinates appear in these
exponentials. An important result we will need is
\begin{eqnarray} \sum_{n}
\tilde{\chi}_{n_1n}(\bq_1)\trho_{nn_2}(\bq_2) &=&<n_1|e^{-{i\over
c}\bq_1\cdot\etab}e^{-ic\bq_2\cdot\etab}|n_2>\\ &=&
e^{-{i\over2}\bQ_1\times\bQ_2} <n_1|e^{-i({\bq_1\over
c}+c\bq_2)\cdot\etab}|n_2> \label{id1a}\\ \bQ&=& \bq
l^*\end{eqnarray} where we have used the completeness of the
states $|n>$ and the commutation rules obeyed by $\etab$. Finally
we separate the exponentials in the reverse order to get the
second useful identity \eqa \sum_{n}
\tchi_{n_1n}(\bq_1)\trho_{nn_2}(\bq_2)\nonumber\\
&=&e^{-i\bQ_1\times\bQ_2}\sum_{n}
\trho_{n_1n}(\bq_2)\tchi_{nn_2}(\bq_1) \label{id2a}\eea Note that
 $\trho$ and $\tchi$ {\it do not commute},
even though $\baro$ and $\bachi$ do.

Let us now right-multiply the putative left-eigenvector $\tchi$ by $\cH$
\begin{eqnarray}\lefteqn{
\sum\limits_{n_1'n_2'}^{}\tchi_{n_1'n_2'}(\bq)\cH(n_1'n_2';n_1n_2;\bq)=}\nonumber
\\ &&(\e(n_1)-\e(n_2))\tchi_{n_1n_2}(\bq)\nonumber\\
&+&\sum\limits_{n_1'n_2'}^{}(N_F(n_2')-N_F(n_1'))
{v(q)\over2\pi(l^*)^2}e^{-q^2l^2/2}
\tchi_{n_1'n_2'}(\bq)\trho_{n_2'n_1'}(-\bq)\trho_{n_1n_2}(\bq)\nonumber\\
&-&\sum\limits_{n_1'n_2'}^{}(N_F(n_2')-N_F(n_1'))\ints
v(s)e^{-{s^2l^2\over2}} \trho_{n_1n_1'}(\bs)\tchi_{n_1'n_2'}(\bq)
\trho_{n_2'n_2}(-\bs)e^{i(l^*)^2\bs\times\bq} \eea

 Let us consider the direct and exchange terms separately. In the
direct term, one $n'$ index can always be summed freely, while the
other is constrained by the Fermi occupation factor $N_F$. The sum
over the free $n'$ gives, according to Eqn. (\ref{id1a}) \eqa
\mbox{direct term} &=&\sum\limits_{n_2'}^{}
N_F(n_2')\trho_{n_1n_2}(\bq) <n_2'|e^{-i({1\over
c}-c)\bq\cdot\etab}|n_2'>\nonumber\\ &-&\sum\limits_{n_1'}^{}
N_F(n_1')\trho_{n_1n_2}(\bq) <n_1'|e^{-i({1\over
c}-c)\bq\cdot\etab}|n_1'> \eea The two terms are immediately seen to
cancel. Now let us turn to the exchange terms, and consider the one
that has the factor $N_F(n_2')$, and a free sum over $n_1'$. In this
term, one can use Eqn. (\ref{id2a}) to exchange the $\trho$ and
$\tchi$ matrix elements to obtain \eqa \mbox{exchange
term}&=&-\sum\limits_{n_2'}^{}N_F(n_2')\sum\limits_{n_1'}
\tchi_{n_1n_1'}(\bq)\times \ints v(s)e^{-{s^2l^2\over2}}
\trho_{n_1'n_2'}(\bs)\trho_{n_2'n_2}(-\bs) \eea Notice that the phase
factor $e^{i(l^*)^2\bs\times\bq}$ has been cancelled by an opposite
phase factor from Eqn. (\ref{id2a}). Now the angular $\bs$ integral
forces $n_1'=n_2$ for a rotationally invariant potential, and the
result contains the Fock energy of the state $n_2$ \eq
\e^{F}(n_2)\tchi_{n_1n_2}(\bq) \ee Similarly, the other exchange term
proportional to $N_F(n_1')$ ends up giving
$-\e^{F}(n_1)\tchi_{n_1n_2}(\bq)$. Due to the peculiar nature of the
Hamiltonian the Hartree energy is a constant independent of the CF-LL
index, and the difference of the Fock energies is the same as the
difference of the full HF energies. Thus, the exchange contributions
cancel the diagonal term $(\e(n_1)-\e(n_2))\tchi_{n_1n_2}(\bq)$, and
$\tchi_{n_1n_2}(\bq)$ is indeed an left eigenvector with zero
eigenvalue for $\cH$.  Also, this property is independent of the form
of $v(q)$ as long as it is rotationally invariant.

\section{Activation gaps\label{gaps}}
           Now we need to find the energy cost of producing a widely
           separated particle-hole (PH) pair. This will be done by evaluating
   \begin{eqnarray}
   \Delta_a &=& \langle {\bf p} + P|H|{\bf p} + P\rangle + \langle {\bf p}+H|H|{\bf p} + H\rangle
  \nonumber \\
  & & -2\langle {\bf p} |H|{\bf p} \rangle\\
   &=& \int_q E(P)+E(H)\\
   \int_q &=&{1\over2} \int {d^2q\over 4\pi^2}v(q) e^{-q^2l^2/2}.
   \end{eqnarray}
   where $P$ denotes a particle added to the state labeled $\mu =
   (n=p,m=0)$ and $H$ denotes a state in which a hole has been
   made in the state $\mu = (n=p-1,m=0)$.
   Let us consider
   \beq
   E(P) =  \langle {\bf p}|d_{\mu}
 d_{1}^{\dag}d_2d_{3}^{\dag}d_4 d^{\dag}_{\mu}|{\bf p} \rangle \rho_{12}\rho_{34}
 \eeq
 In performing the contractions we
\begin{itemize}
  \item Do not make any contractions within $H$. This gets rid of
  $E_0 =\langle {\bf p} |H|{\bf p} \rangle $, the ground state energy.
  \item Do not contract $1$ with  $2$ or $3$ with  $4$ since $v(0)=0$.
\end{itemize}
We end up with
\begin{eqnarray*}
\lefteqn{\int_q \left[ \delta_{\mu
1}\delta_{23}\delta_{4\mu}(1-n_{1}^{F})(1-n_{2}^{F})(1-n_{4}^{F})\right.
} \\ & & -\left. \delta_{\mu
3}\delta_{14}\delta_{2\mu}(1-n_{3}^{F})(n_{4}^{F})(1-n_{2}^{F})
\right]\rho_{12}\rho_{34} \end{eqnarray*}

 Since $4=\mu =1$ in the
first term, we can drop $(1-n_{1}^{F})(1-n_{4}^{F})$ and for
similar reasons $(1-n_{3}^{F})(1-n_{2}^{F})$ in the second giving
us

\begin{eqnarray}
 E(P)& =&  \sum_{m_2=0}^{\infty}\sum_{n_2=p}^{\infty}
 \rho_{\mu 2}({\bf q})\rho_{2\mu}(-{\bf q}) \\ & & -
\sum_{m_2=0}^{\infty}\sum_{n_2=0}^{p-1}
 \rho_{2\mu }({\bf q})
\rho_{\mu 2}(-{\bf q}) .
\end{eqnarray}

Since the sum over $m_2$ is unrestricted, we can use completeness
and
 $e^{-i{\bf q \cdot R}}\cdot
e^{-i{\bf q \cdot R}}=I$ to get rid of the $m$-index altogether.
Thus we end up with \begin{eqnarray} E(P)&=&  \left(
\sum_{n=p}^{\infty}|\rho_{pn}|^2
-\sum_{n=0}^{p-1}|\rho_{pn}|^2\right)\\ & =&  \left[ \langle
p|\rho (q)\rho (-q) |p\rangle -
2\sum_{n=0}^{p-1}|\rho_{pn}|^2\right]
\end{eqnarray}

A similar calculation for the hole state gives (upon dropping the
ground state energy as usual) \beq E(H)=  \left[ -\langle p-1
|\rho (q)\rho (-q) |p-1 \rangle +
2\sum_{n=0}^{p-1}|\rho_{p-1,n}|^2\right]
\end{equation}

where
 \beq
\langle n|
 {\rho} (q)\rho (-q)|n\rangle = \sum_{n'=o}^{\infty}
 |{\rho} (q)_{nn'}|^2.
 \eeq
 Putting all the  pieces together,  we obtain the gap.

\section{Critical fields for magnetic transitions\label{bc}} We need to
calculate \beq E(p-r,r) = \langle {\bf p-r,\ r}|H|\ {\bf p-r,\
r}\rangle \eeq the energy in a state with $p-r$ spin-up LL's and
$r$ spin-down LL's. Since the HF calculation for the spinless case
is very similar, this treatment will be brief. We write \beq H =
\sum_{1234}\int_q d_{1}^{\dag}d_2d_{3}^{\dag}d_4
\rho_{12}\rho_{34} \eeq with the understanding that a label like
$1$ stands for the triplet $(n_1,m_1,s_1)$, $s$ being the spin.
The matrix elements $\rho_{ij}$ are defined by
\begin{eqnarray*}\lefteqn{ \rho_{12}=}\\ && \langle 1|e^{-i{\bf q
\cdot R}}\left[ e^{-i{\bf q} \cdot \mbox{\boldmath $\eta $}c}-c^2
fe^{-i{\bf q \cdot \mbox{\boldmath $\eta $} }/c}\right]|2\rangle
\\ &=&
 \rho_{m_1m_2}\otimes
\rho_{n_1n_2}\otimes \delta_{s_1s_2}\end{eqnarray*}
 and as a result
 \begin{eqnarray*}
 \lefteqn{E(p-r,r)=}
 \\ &&
 \int_q\sum_{n_1n_2s}n_{1}^{F}(s)(1-n_{2}^{F}(s))|\rho_{n_1n_2}|^2\underbrace{\sum_{m}\langle
 m|
 I| m\rangle }_{=n/p}
 \end{eqnarray*}
 where we acknowledge the fact that the occupation factors  $n_{1}^{F}$ and $n_{2}^{F}$ can depend on
 the spin. We have also used the fact that the sum over all values
 of $m$ is the degeneracy of each CF-LL, $n/p$. Carrying out the
 sums over $n_1$ and $n_2$, we obtain

\begin{eqnarray*}
 \lefteqn{E(p-r,r)=}
 \\ &&
{n \over p} \int_q \left[ \sum_{n_1=0}^{p-r-1}\langle n_1|\rho (q)
\rho (-q)|n_1 \rangle
-\sum_{n_1,n_2=0}^{p-r-1}|\rho_{n_1n_2}|^2\right.\\ &&+ \left.
\sum_{n_1=0}^{r-1}\langle n_1|\rho (q) \rho (-q)|n_1 \rangle
-\sum_{n_1,n_2=0}^{r-1}|\rho_{n_1n_2}|^2\right]
 \end{eqnarray*}

It is now straightforward to compute the critical field for the
transition $|{\bf p-r,r}\rangle \to |{\bf p-r-1,r+1}\rangle$ by
invoking \beq E(p-r,r)-E(p-r-1, r+1) = g \left[ {e  \over
2m_e}\right]B^c {n\over p} \eeq

\section{Acronyms and Symbols}

This  paper invokes many symbols and acronyms, not all of which
have been standardized. For the convenience of the reader,  a list
is supplied in Table (\ref{acronyms}).

\begin{table}[b]
\caption{Frequenty used acronyms\label{acronyms}}
\begin{ruledtabular}
\begin{tabular}{|c|c|}
CB & composite bosons \\ CF & composite fermions\\ CQFM &continuum
quantum ferromagnet\\ CS & Chern-Simons\\ EHT & Extended
hamiltonian theory \\ GMP & Girvin, MacDonald and  Platzman\\ HLR
& Halperin, Lee and Read\\
 HF & Hartree-Fock\\ LL & Landau level\\ LLL &lowest Landau
level\\ RPA & random phase approximation\\ TDHF & time-dependent
Hartree-Fock\\ WC & Wigner crystal\\ ZDS & Zhang-Das Sarma
\end{tabular}
\end{ruledtabular}
\end{table}

\newpage

\begin{tabular}{|c|c|}\hline
  Symbol & Significance \\ \hline
  \ & {\bf Widely used  quantities}\\
  $\nu $ & $=p/(2ps+1)$=filling fraction\\
  $2s$ & Number of vortices of flux tubes attached\\
 $p$ & Number of CF LL's\\
  $\hz $ &unit vector along $z$-axis\\
 $c^2$ & $2ps/(2ps+1)$\\
 $m_e \ \ (m) $ & Electron mass in free space (in solid) \\
$ A^*$ \mbox{ or} $B^*$ & Potential or field seen by CF
$A^*=A/(2ps+1)$\\
  ${\bf \Pi}$& Velocity operator for CF, $\bp +e\bA^*$\\
  $g$ & $g$ -factor of electron or CF, taken to be .44\\
  $l$\ \  $(l^*=l/\sqrt{1-c^2})$ & electron (CF)  magnetic length \\
\ $\bR_e \ \mbox{or}\  \bR_v \ \mbox{or}\ \bR $\  & Electron,
pseudovortex, or CF guiding center coordinate\ \ \\
  $\etab_e \ \mbox{or} \ \etab$ & Electron or CF cyclotron
  coordinate\\
  $\barH$ & Hamiltonian in the LLL\\
  $\bar{\bar{\rho}} ({\bf q})$ & Electron density in the LLL \\
  $\bar{\bar{\chi }}({\bf q})$ & Constraint\\
  $\bar{\bar{\rho}}^p ({\bf q})$ &$\bar{\bar{\rho}} ({\bf q})- c^2
  \bar{\bar{\chi}} ({\bf q})$\  (preferred charge)\\

  \ & {\bf Quantities related to gaps}\\
$\Delta_{a,p}$& Activation or polarization gap\\ $m_{a,p}^{(2s)}$&
Defined by $\Delta_{a,p} =eB^*/(m_{a,p}^{(2s)}$\\
  $\delta$  & $ \Delta/(e^2/\varepsilon l)$\\
  $\lambda$ & Defined by  $v_{ZDS}(q) = 2\pi e^2  e^{-ql\lambda}/q$\\
\ & {\bf Magnetic quantities}\\
  $S$ & Number of spin up minus down CFs \\
  $E(S)$ & Ground state energy density \\
  ${\cal E}_{\pm}(k)$ & Hartree Fock energy for up/down spin at momentum $k$ \\
  $|{\bf p-r,r}\rangle$ &  CF state with  $p-r$ LL's spin up   and $r$ down.\\
  \ & {\bf Matrix elements}\\
$\rho_{n_1n_2}$& single-particle matrix element of
  $\bar{\bar{\rho}}^p$  between  LL's
   $n_1$ and $n_2$\\
 $\tchi_{nn'}$ & $\langle n|\exp {-i\bq \cdot \etab /c   }|n'\rangle$\\
   \hline
\end{tabular}

\end{document}